\documentclass[12pt,letterpaper]{article}
\pdfoutput=1

\usepackage{tipa}
\usepackage{amsmath,amssymb,calc, amsthm,bbm, epsfig,psfrag, mathtools,comment}
\usepackage{mathrsfs}
\usepackage{needspace}
\usepackage{graphicx, enumerate}
\usepackage{float}
\usepackage{comment}
 \usepackage[all,cmtip]{xy}
\usepackage[numbers,sort&compress]{natbib}
\usepackage[dvipsnames]{xcolor}
\usepackage{tikz}
\usepackage[dvipsnames]{xcolor}
\definecolor{light-gray}{gray}{0.9}
\usepackage{pgfplots}
\pgfplotsset{compat=1.15}
\usepgfplotslibrary{fillbetween}
\usepackage{tikz-cd}
\usepackage{physics}
\usepackage{tensor}
\usepackage{enumitem}
\usepackage[compat=1.1.0]{tikz-feynman}
\usepackage[utf8]{inputenc} 
\definecolor{azure}{rgb}{0.0, 0.5, 1.0}
\definecolor{darkblue}{rgb}{0.15,0.35,0.7}
\definecolor{reddish}{rgb}{0.65, 0.2, 0.2}
\definecolor{brandeisblue}{rgb}{0.0, 0.44, 1.0}
\definecolor{ceruleanblue}{rgb}{0.16, 0.32, 0.75}
\definecolor{indigo(dye)}{rgb}{0.0, 0.25, 0.42}
\usepackage[linktocpage=true]{hyperref}
\hypersetup{
colorlinks=true,
citecolor=ceruleanblue,
linkcolor=ceruleanblue,
urlcolor=ceruleanblue,
pdfauthor={},
pdftitle={},
pdfsubject={}
}

\usepackage{tikz}
\tikzset{sines/.style={
        thick,
        line join=round, 
        draw=black, 
        decorate, 
        decoration={complete sines, amplitude=1mm,
        segment length=2mm}
    }}
    
\tikzset{esines/.style={
        thick,
        line join=round, 
        decorate, 
        decoration={complete sines, amplitude=1mm,
        segment length=2mm}
    }}

\usetikzlibrary{arrows.meta}

\usetikzlibrary{decorations}

\tikzset{/pgf/decoration/.cd,
    number of sines/.initial=10,
    angle step/.initial=20,
}
\newdimen\tmpdimen
\pgfdeclaredecoration{complete sines}{initial}
{
    \state{initial}[
        width=+0pt,
        next state=move,
        persistent precomputation={
            \pgfmathparse{\pgfkeysvalueof{/pgf/decoration/angle step}}%
            \let\anglestep=\pgfmathresult%
            \let\currentangle=\pgfmathresult%
            \pgfmathsetlengthmacro{\pointsperanglestep}%
                {(\pgfdecoratedremainingdistance/\pgfkeysvalueof{/pgf/decoration/number of sines})/360*\anglestep}%
        }] {}
    \state{move}[width=+\pointsperanglestep, next state=draw]{
        \pgfpathmoveto{\pgfpointorigin}
    }
    \state{draw}[width=+\pointsperanglestep, switch if less than=1.25*\pointsperanglestep to final, % <- bit of a hack
        persistent postcomputation={
        \pgfmathparse{mod(\currentangle+\anglestep, 360)}%
        \let\currentangle=\pgfmathresult%
    }]{%
        \pgfmathsin{+\currentangle}%
        \tmpdimen=\pgfdecorationsegmentamplitude%
        \tmpdimen=\pgfmathresult\tmpdimen%
        \divide\tmpdimen by2\relax%
        \pgfpathlineto{\pgfqpoint{0pt}{\tmpdimen}}%
    }
    \state{final}{
        \ifdim\pgfdecoratedremainingdistance>0pt\relax
            \pgfpathlineto{\pgfpointdecoratedpathlast}
        \fi
   }
}
\newcommand\sgn{\text{sgn}}

\definecolor{indigo(dye)}{rgb}{0.0, 0.25, 0.42}

\newcommand{\overbar}[1]{\mkern 1.5mu\overline{\mkern-1.5mu#1\mkern-1.5mu}\mkern 1.5mu}

\newcommand\wb{{\overbar{w}}}

\def\TT{{T\overbar{T}}}

\newcommand{\Abar}{\overbar{A}}

\usepackage{amsthm}

\usepackage{cleveref}
\usepackage{bm}
\crefname{lem}{lemma}{lemmas}
\crefname{thm}{theorem}{theorems}
\crefname{cor}{corollary}{corollaries}
\crefname{rem}{remark}{remarks}
\crefname{prop}{proposition}{propositions}

\makeatletter
\renewcommand\section{\@startsection {section}{1}{\z@}%
                               {-3.5ex \@plus -1ex \@minus -.2ex}%nn
                               {2.3ex \@plus.2ex}%
                               {\normalfont\large\bfseries}}
\renewcommand\subsection{\@startsection{subsection}{2}{\z@}%
                                 {-3.25ex\@plus -1ex \@minus -.2ex}%
                                 {1.5ex \@plus .2ex}%
                                 {\normalfont\bfseries}}
\makeatother

\let\non\nonumber

\newfont{\goth}{ygoth.tfm scaled 1200}                   % gothic font (usual)
\renewcommand{\thesection}{\arabic{section}}

\numberwithin{equation}{section}

\newcommand{\ul}{\underline}

\begin{document}
\begin{titlepage}

\begin{center}

%{\today}
%\today
\hfill         \phantom{xxx}  %EFI--20-5
\vskip 2 cm {\Large \bf \scalebox{1}{Flows in the Space of Interacting Chiral Boson Theories}}

\vskip 1.25 cm {\bf Stephen Ebert,$^1$ Christian Ferko,$^{2}$ Cian Luke Martin,$^{3}$ and \\
Gabriele Tartaglino-Mazzucchelli${}^{3}$} \non\\

\vskip 0.2 cm
{\it $^1$ Mani L. Bhaumik Institute for Theoretical Physics,\\\scalebox{0.99}{Department of Physics \& Astronomy,   
University of California, Los Angeles, CA 90095, USA}}

\vskip 0.2 cm
 {\it $^2$ Center for Quantum Mathematics and Physics (QMAP), \\ Department of Physics \& Astronomy,  University of California, Davis, CA 95616, USA}

\vskip 0.2 cm
 {\it $^3$ School of Mathematics \& Physics, University of Queensland, \\ St Lucia, Brisbane, Queensland 4072, Australia}

\end{center}
\vskip 1.5 cm

\begin{abstract}

We study interacting theories of $N$ left-moving and $\overbar{N}$ right-moving Floreanini-Jackiw bosons in two dimensions. A parameterized family of such theories is shown to enjoy (non-manifest) Lorentz invariance if and only if its Lagrangian obeys a flow equation driven by a function of the energy-momentum tensor. We discuss the canonical quantization of such theories along classical stress tensor flows, focusing on the case of the root-$\TT$ deformation, where we obtain perturbative results for the deformed spectrum in a certain large-momentum limit. In the special case $N = \overbar{N}$, we consider the quantum effective action for the root-$\TT$-deformed theory by expanding around a general classical background, and we find that the one-loop contribution vanishes for backgrounds with constant scalar gradients. Our analysis can also be interpreted via dual $U(1)$ Chern-Simons theories in three dimensions, which might be used to describe deformations of charged $\mathrm{AdS}_3$ black holes or quantum Hall systems.

\baselineskip=18pt

\end{abstract}

\end{titlepage}

\tableofcontents
%\newpage 

\section{Introduction} \label{Intro}

In physics, we are frequently interested in parameterized families of classical or quantum field theories. The tangent vectors to these families often have an interpretation as operators within a given theory. One familiar example appears in the study of conformal field theories, which may possess certain exactly marginal operators. Deforming a CFT by a marginal operator generates motion on the conformal manifold, which is one such family of theories.

Another simple one-parameter family generated from any quantum field theory is the well-known renormalization group flow. We can interpret this as a curve of theories labeled by an energy scale $\mu$.
For a CFT, this curve degenerates to a single point, but for other QFTs, one finds an infinite family of theories connecting two RG fixed points at the UV and IR ends of this flow. The operator that plays the role of the tangent vector to this curve is the trace of the energy-momentum tensor, which generates scale transformations.

The renormalization group example is especially useful because it is \emph{universal}: any translation-invariant field theory admits an energy-momentum tensor $T_{\mu \nu}$, so we may always deform by the trace $\tensor{T}{^\mu_\mu}$ to flow toward the infrared. It is natural to explore other deformations constructed from the stress tensor, which are also universal. These stress tensor deformations generate a larger class of flows, which includes the renormalization group flow as a special case, but which also includes other famous examples such as the $\TT$ deformation of two-dimensional quantum field theories \cite{Zamolodchikov:2004ce, Smirnov:2016lqw,Cavaglia:2016oda}.

Even at the classical level, stress tensor flows often give rise to interesting parameterized families of theories. For instance, consider classical theories of a single Abelian gauge field $A_\mu$ whose Lagrangians depend on the field strength $F_{\mu \nu}$ but not its derivatives. Construct the parameterized family which contains the Maxwell theory, $\mathcal{L} = - \frac{1}{4} F_{\mu \nu} F^{\mu \nu}$, and all other theories that can be reached from the Maxwell theory by deformations involving the energy-momentum tensor. This family is precisely the collection of theories of non-linear electrodynamics which are invariant under electric-magnetic duality rotations \cite{Ferko:2023wyi}, which is of interest in its own right.\footnote{Strictly speaking, there are some isolated points in this space such as the Bialynicki-Birula theory \cite{Bialynicki-Birula:1992rcm} which are not connected to Maxwell, so to be precise we should say that the family generated in this way gives one connected component in the space of duality-invariant theories.}

Another example concerns theories of a two-form gauge potential $A_{\mu \nu}$ with a self-dual three-form field strength $F_{\mu \nu \rho}$ in six spacetime dimensions. Any family of such theories -- e.g., the collection of interacting chiral tensor theories which describe the worldvolume theory on an M5-brane, labeled by a parameter $T$ that controls the tension of the brane -- also obeys a stress tensor flow equation \cite{Ferko:2024zth}. We say that both $4d$ theories of duality-invariant electrodynamics and $6d$ chiral tensor theories are closed under stress tensor flows, in the sense that deforming any member of one of these classes of theories by a Lorentz scalar constructed from $T_{\mu \nu}$ produces another member of the same class.

In this paper, we will investigate another space of theories, which is also closed under deformations involving the energy-momentum tensor. The theories that we consider here describe the dynamics of a collection of $N$ chiral and $\overbar{N}$ anti-chiral bosonic fields in two spacetime dimensions. Such theories of several chiral bosons appear in many contexts, such as in the T-duality symmetric formulation of worldsheet string theory \cite{Tseytlin:1990nb,Tseytlin:1990va}. The simplest member of this class of theories, with $N = 1$ and $\overbar{N} = 0$, is the theory of a single chiral boson which is described by the Floreanini-Jackiw action \cite{Floreanini:1987as}, namely
\begin{align}\label{fj_action}
    S_{\text{FJ}} = \frac{1}{2} \int d^2 x \, \left( \partial_t \phi \partial_\theta \phi - \partial_\theta \phi \partial_\theta \phi \right) \, .
\end{align}
Here we work in a $2d$ spacetime with coordinates $(t, \theta)$. As is well-known, it is not straightforward to write a manifestly Lorentz-invariant Lagrangian for a field that obeys a chirality (or self-duality) constraint. One approach, which we will follow in this work, is to sacrifice manifest Lorentz invariance and work with actions of the form (\ref{fj_action}) that explicitly single out a preferred time coordinate $t$; we will then need to impose that the theory enjoy a non-manifest Lorentz symmetry. Another strategy is to introduce one or more auxiliary fields to restore manifest Lorentz invariance, which is the tactic used to describe chiral tensor theories in six dimensions using, e.g., the Pasti-Sorokin-Tonin (PST) formulation \cite{Pasti:1995tn,Pasti:1996vs,Pasti:1997gx} (and later extended to higher dimensions \cite{Buratti:2019guq}). A related technique was used to present a manifestly Lorentz invariant description of the Floreanini-Jackiw action in \cite{Townsend:2019koy}.

For a single chiral (or anti-chiral) boson, it is known that no continuous deformation of the free theory to include Lorentz-invariant self-interactions is possible \cite{Buratti:2019guq,Bandos:2020hgy} (although see \cite{Avetisyan:2022zza} for such an interacting model which is not continuously connected to the free theory). In this work, we will give a new interpretation of this fact: all Lorentz-invariant interacting chiral boson theories are generated from stress tensor deformations, but (\ref{fj_action}) is a fixed point of all such flows, and, therefore, there is no way to deform it to include interactions. However, for a theory with $N \geq 1$ chiral and $\overbar{N} \geq 1$ anti-chiral bosons, such self-interactions are possible, and it is natural to describe them with an interaction function $V ( \partial_\theta \phi^i , \partial_\theta \overbar{\phi}^{\overline{i}} )$ that depends on the spatial derivatives of the fields:
\begin{align}\label{interaction_function}
    S_{\text{int}} = \frac{1}{2} \int d^2 x \, \left( \partial_t \phi^i \partial_\theta \phi^i - \partial_t \overbar{\phi}^{\overline{i}} \partial_\theta \overbar{\phi}^{\overline{i}} - V ( \partial_\theta \phi^i , \partial_\theta \overbar{\phi}^{\overline{i}} ) \right) \, .
\end{align}
In this expression, $i = 1 , \ldots , N$ runs over the chiral fields and $\overline{i} = 1 , \ldots , \overbar{N}$ labels the anti-chiral fields. We will be primarily interested in theories that are invariant under the $O ( N ) \times O ( \overbar{N} )$ symmetry rotating the chiral and anti-chiral bosons among themselves, although we will give some results that do not make this assumption; we will see that it is also possible to promote (\ref{interaction_function}) to include a target-space metric $G_{ij}$, $\overbar{G}_{\overline{i} \overline{j}}$ for the bosons, or couplings to an antisymmetric tensor field $B_{ij}$, $\overbar{B}_{\overline{i} \overline{j}}$, (which in general breaks $O ( N ) \times O ( \overbar{N} )$) without significantly changing our analysis. Because the Lagrangian appearing in (\ref{interaction_function}) is first-order in time derivatives, the function $V$ can also be interpreted as the Hamiltonian of the model. This structure is similar to that of the PST description of a $6d$ chiral tensor theory, after gauge-fixing the auxiliary field $v_\mu$ of this formalism to the value $v_\mu = \tensor{\delta}{^0_\mu}$, whose action is
\begin{align}
    S_{\text{PST, gauge-fixed}} = \int d^6 x \, \left( \frac{1}{4} B_{ij} \partial_0 A^{ij} - \mathcal{H} ( s, p ) \right) \, .
\end{align}
Here $s = \frac{1}{4} B^{ij} B^{kl} \delta_{ik} \delta_{jl}$ and $p = \sqrt{p^i p_i}$, where $p_i = \frac{1}{8} \varepsilon_{i j k l m} B^{jk} B^{lm}$, are two $SO(5)$-invariant quantities constructed from the ``magnetic field'' $B_{ij}$, where $E^{ij}$ and $B^{ij}$ are related to the fundamental field $F_3 = d A_2$ as $E^{ij} = F^{i j 0}$, $B^{ij} = \widetilde{F}^{i j 0}$, and $\widetilde{F}$ denotes the Hodge dual of $F$. This gauge-fixed form of the PST action is closely related to the Perry-Schwarz formalism \cite{Perry:1996mk}. In our two-dimensional example, the role of the magnetic components $B^{ij}$ of the three-form field strength is played by the spatial derivatives $\partial_\theta \phi^i$ and $\partial_\theta \overbar{\phi}^{\overline{i}}$ of the bosons.

Although we will not consider other formulations of chiral boson theories in this work, let us briefly mention that several other approaches have been used to describe such models. One presentation, due to Sen \cite{Sen:2015nph,Sen:2019qit}, introduces an additional ``spectator'' field which decouples from the dynamics; $\TT$ flows within this formalism have been studied in \cite{Chakrabarti:2020dhv,Chakrabarti:2022lnn,Chakrabarti:2023czz}.\footnote{The latter analysis also illuminates a surprising connection between the solvability of $\TT$-like deformations and that of another deformation of quantum mechanics involving a $\cosh ( p )$ kinetic term \cite{Grassi:2018bci}.} Another presentation introduced by Mkrtchyan includes an additional auxiliary scalar field $R$ and reduces to the PST form of the chiral boson action after integrating out $R$ \cite{Mkrtchyan:2019opf}. See \cite{Arvanitakis:2022bnr} for a comparison of some of these formulations and the realization of chiral bosons via a $3d$ Chern-Simons theory. Finally, a notable presentation by Siegel \cite{Siegel:1983es} expresses the chiral boson action in terms of a symmetric and traceless auxiliary tensor field $\lambda^{\alpha \beta}$:
\begin{equation}
\begin{aligned}\label{eq:Siegel}
 S_{\text{Siegel}}&=-\frac{1}{4} \int d^2x \bigg[ \partial_\alpha \phi \partial^\alpha \phi + \lambda^{\alpha \beta} \left( \partial_\alpha \phi - \epsilon_{\alpha \sigma} \partial^\sigma \phi \right) \left( \partial_\beta \phi - \epsilon_{\beta \rho} \partial^\rho \phi \right)\bigg]
 \\&=\int dt d\theta \bigg[ \frac{1}{4} \left(\partial_t \phi \partial_t \phi  - \partial_\theta \phi \partial_\theta \phi \right) + \frac{\lambda^{01} - \lambda^{00}}{2} \left(\partial_t \phi - \partial_\theta \phi \right)^2\bigg]\,.
\end{aligned}
\end{equation}
Siegel's action \eqref{eq:Siegel} is \emph{classically} equivalent to the Floreanini-Jackiw action \eqref{fj_action} assuming one can gauge the two independent components of $\lambda^{\alpha \beta}$ to $(\lambda^{00}, \lambda^{01}) =( \frac{1}{4}, - \frac{1}{4})$  \cite{Bernstein:1988zd}. For applications extending Siegel's action to gravity and string theory, see \cite{Gates:1987sy,Bellucci:1988uv,Kuzenko:1990mk,Bellucci:1991id,Kuzenko:1991ew,Gates:1991am}. The study of chiral bosons and other self-dual fields has a long history, and we refer the reader to an incomplete sampling \cite{Deser:1976iy,Marcus:1982yu,Henneaux:1987hz,Henneaux:1988gg,Schwarz:1993vs,Mezincescu:2022hnb,Evnin:2022kqn} of earlier work, and references therein, for other results.

Our motivation for studying this class of interacting chiral boson theories in this work is twofold. The first reason is purely classical: we would like to characterize the space of all such interacting theories, each of which is determined by an interaction function $V$, which enjoy non-manifest Lorentz invariance. As we will see, this condition will require that the function $V$ satisfy a certain partial differential equation which is very similar to those that appear in the cases of $4d$ duality-invariant electrodynamics \cite{Ferko:2023wyi} and $6d$ chiral tensor theories \cite{Ferko:2024zth}. The space of solutions to this partial differential equation is intimately connected to stress tensor flows. More precisely, given any parameterized family of Lorentz-invariant theories with interaction functions $V^{(\lambda)}$ labeled by a parameter $\lambda$, we will show that $\partial_\lambda V^{(\lambda)}$ can always be written as a function of the stress tensor $T_{\mu \nu}^{(\lambda)}$ of the theory at the same value of the parameter $\lambda$. Conversely, any flow equation of the form
\begin{align}\label{classical_flow}
    \partial_\lambda V^{(\lambda)} = f \left( T_{\mu \nu}^{(\lambda)} , \lambda \right) \, , \qquad \lim_{\lambda \to 0} V^{(\lambda)} = V^{(0)} \, ,
\end{align}
along with a Lorentz-invariant initial condition $V^{(0)}$, gives rise to a one-parameter family of Lorentz-invariant theories. Therefore, families of Lorentz-invariant interacting chiral boson theories are in one-to-one correspondence with stress tensor flows. These statements are the precise $2d$ analogs of the $4d$ and $6d$ results in \cite{Ferko:2023wyi} and \cite{Ferko:2024zth}.

The second motivation for this study concerns quantization. The general form (\ref{interaction_function}) of an interacting theory is convenient for canonical quantization, since the dependence on time derivatives is fixed and thus the definition of the conjugate momenta is unaffected by the interaction function. One can study the quantization of theories in this class in a uniform way, at least for cases that admit a controlled perturbative expansion which makes calculations tractable. When we consider the quantization of a one-parameter family of theories defined by interaction functions $V^{(\lambda)}$ that satisfy a differential equation of the form (\ref{classical_flow}), we will say that we are studying ``quantization along the classical flow.''

We will be especially interested in quantization along the flow driven by the function 
\begin{align}\label{root_TT_flow}
    \partial_\gamma V^{(\gamma)} = \mathcal{R} \left[ T_{\mu \nu} \right] = \frac{1}{\sqrt{2}} \sqrt{ T^{\mu \nu} T_{\mu \nu} - \frac{1}{2} \left( \tensor{T}{^\mu_\mu} \right)^2 } \, , 
\end{align}
where we suppress the dependence of $T_{\mu \nu}$ on the flow parameter $\gamma$. This non-analytic combination $\mathcal{R}$ is the two-dimensional root-$\TT$ operator \cite{Ferko:2022cix}, which is the unique marginal combination of stress tensors that defines a classical flow equation which commutes with the irrelevant $\TT$ flow in $2d$. The root-$\TT$ deformation shares some of the remarkable properties of the $\TT$ deformation, such as preserving classical integrability in many examples \cite{Borsato:2022tmu,Ferko:2024ali} and admitting a holographic interpretation in terms of modified boundary conditions for $\mathrm{AdS}_3$ gravity \cite{Ebert:2023tih,Tian:2024vln}. It also plays a role in classical flows for $3d$ gauge theories \cite{Ferko:2023sps} and has connections to $\text{BMS}_3$ symmetry and ultra/non-relativistic limits of $2d$ CFTs \cite{Rodriguez:2021tcz,Bagchi:2022nvj,Tempo:2022ndz}.

Another motivation for studying this operator is that the corresponding commuting $\TT$-like and root-$\TT$-like flows in four spacetime dimensions, with the initial condition given by the free Maxwell Lagrangian, were shown in \cite{Conti:2018jho,Babaei-Aghbolagh:2022uij,Ferko:2022iru,Conti:2022egv,Ferko:2023ruw} to produce an interesting family of gauge theories referred to as ModMax-Born-Infeld, which was first written down in \cite{Bandos:2020hgy}. This family depends on two parameters $\lambda$ and $\gamma$. When $\gamma$ is taken to zero, the theory reduces to the $4d$ Born-Infeld model which gives an effective description of the gauge dynamics on a D3-brane. As $\lambda \to 0$, one recovers the so-called Modified Maxwell or ModMax theory, which is the unique conformally invariant and electric-magnetic duality-invariant extension of the Maxwell theory \cite{Bandos:2020jsw}. This theory can be supersymmetrized \cite{Bandos:2021rqy,Kuzenko:2021cvx,Kuzenko:2021pqm}, or deformed to include higher derivative interactions \cite{Kuzenko:2024zra}, it admits a family of Carrollian analogues \cite{Chen:2024vho},
and the entire class of ModMax-Born-Infeld theories can be lifted to a similar family of $6d$ chiral tensor theories \cite{Bandos:2020hgy} which also satisfies commuting stress tensor flow equations \cite{Ferko:2024zth}.\footnote{See \cite{Deger:2024jfh} for a construction of solutions of the $6d$ ModMax-like chiral tensor theory coupled to gravity.} For an introduction to theories of non-linear electrodynamics, see the lecture notes \cite{Sorokin:2021tge}.

Although several classical aspects of the ModMax theory (and its ModMax-Born-Infeld extension) have been studied \cite{Dassy:2021ulu,Nastase:2021uvc,Escobar:2021mpx,Lechner:2022qhb,Neves:2022jqq}, the quantization of this model appears to be more subtle because the Lagrangian is non-analytic around $F_{\mu \nu} = 0$. One strategy is to perform perturbative quantization of this theory around a non-zero background for the field strength \cite{LukeMartin:2024gsb}.\footnote{Another approach would be to use heat kernel techniques. We are grateful to Sergei Kuzenko and Dmitri Sorokin for discussions on this topic and for informing us of their unpublished results. See also \cite{pinelliThesis} for a Master's thesis which computes the one-loop effective action for ModMax using such techniques.}
Another approach is to look for lower-dimensional analogs of the ModMax theory, which one might hope are simpler to quantize. The most extreme case is to dimensionally reduce the Modified Maxwell theory all the way down to $(0+1)$ spacetime dimensions, which yields a theory of particle mechanics known as the ModMax oscillator \cite{Garcia:2022wad,Ferko:2023ozb} that can be quantized exactly \cite{Ferko:2023iha}. An intermediate case is to reduce ModMax from $4d$ to $2d$, which was done in \cite{Conti:2022egv}, and this reduction yields precisely the same theory that one obtains by deforming a collection of free scalars by the $2d$ root-$\TT$ flow \cite{Ferko:2022cix,Babaei-Aghbolagh:2022leo}. This ``Modified Scalar'' theory is the model whose quantization we consider in the present work.

For one non-chiral boson, or one left-moving and one-right moving chiral boson, the Modified Scalar theory collapses to a free massless scalar with a re-scaled kinetic term, but for multiple scalars, the theory is non-trivial. As we will see later, the Modified Scalar theory with a general number of scalars may also be related to a free theory by a series of more complicated, non-local field redefinitions; similar field redefinitions, and related non-local ``dressed'' operators, have also played a role in the study of $\TT$ flows \cite{Kruthoff:2020hsi,Guica:2020uhm,Kraus:2021cwf,Ebert:2022cle,Kraus:2022mnu}.

One of our goals in studying the quantization of this model is to test a flow equation for certain energies in a root-$\TT$ deformed CFT, which was obtained via a holographic analysis in \cite{Ebert:2023tih}. Under some assumptions, this equation predicts that the deformed energy $E_\gamma$ associated with a seed CFT state that has undeformed energy $E_0$ and momentum $P_0$ is
\begin{align}\label{zero_mode_formula}
    E_\gamma = E_0 \cosh ( \gamma ) + \sqrt{ E_0^2 - P_0^2 } \sinh ( \gamma ) \, .
\end{align}
This formula was derived for states dual to BTZ black holes in $\mathrm{AdS}_3$ with mass $M \geq 0$ and spin $| J | \leq M$, which correspond to \emph{constant} stress tensor backgrounds. We will, therefore, refer to the flow equation (\ref{zero_mode_formula}) as the ``zero-mode energy formula''  since it applies to states of a CFT on a cylinder whose stress tensors are constant along the circular direction (that is, the formula applies to the zero mode of the stress tensor). It would be quite unusual if this energy formula held \emph{universally}, even for states whose stress tensors are spatially varying. And indeed, we will see explicitly in this work that the zero-mode energy formula fails for states with such spatial gradients. One might therefore think of (\ref{zero_mode_formula}) as the first term in a gradient expansion, which is corrected by terms that depend on derivatives $\partial T$.\footnote{The idea of performing such a gradient expansion is philosophically similar to the strategy adopted in hydrodynamics or the fluid-gravity correspondence \cite{Bhattacharyya:2007vjd} (see \cite{Rangamani:2009xk} for a review).}

The key ingredient in our check of the energy formula (\ref{zero_mode_formula}), which allows us to resolve the square root and perform a perturbative analysis, is to consider a certain large-momentum limit and expand in powers of $\frac{1}{p}$. Although this approach involves a specific choice of background around which to expand, one could expand around \emph{any} field configuration for which the gradients of the scalars are non-vanishing, since the combination of stress tensors (\ref{root_TT_flow}) which appears in the classical Lagrangian for the Modified Scalar theory is only non-analytic around zero-energy configurations. We will also present a related analysis which involves expanding around a general classical background for $N = \overbar{N}$, in which case the equal number of chiral and anti-chiral bosons can be assembled into a manifestly Lorentz invariant theory of $N$ non-chiral bosons, and compute loop corrections to the Modified Scalar action. This offers further insight into the perturbative quantization of this model.

This paper is organized as follows. In section \ref{sec:classical}, we compute the stress tensor for a generic interacting chiral boson theory and study classical properties of flows driven by functions of $T_{\mu \nu}$, such as preservation of the Lorentz invariance condition. We then give a complementary perspective on such chiral boson theories in section \ref{sec:cs}, interpreting them as the boundary duals to $U(1)$ Chern-Simons gauge theories, and we show that deformations such as root-$\TT$ can be implemented using certain modified boundary conditions for the bulk gauge fields. In section \ref{sec:quantization}, we review general machinery for the canonical quantization of first-order systems like (\ref{interaction_function}) along classical stress tensor flows using a mode expansion; we then specialize to quantization along the root-$\TT$ flow and study the cases of $\left( N , \overbar{N} \right) = ( 1, 1 )$ and $\left( N , \overbar{N} \right) = ( 2, 1 )$ in detail. In section \ref{sec:cian}, we perform a diagrammatic analysis of quantum corrections along the root-$\TT$ flow for a deformed theory of $N = \overbar{N}$ \emph{non-chiral} bosons, using the background field method. Finally, section \ref{sec:Conclusion&Outlook1} summarizes our results and outlines some interesting future directions. An order-by-order analysis for more general stress tensor flows is presented in appendix \ref{app:TTn}, and the computational steps used to evaluate certain Feynman diagrams in dimensional regularization have been relegated to appendix \ref{app:feynman}.

\section{Classical Stress Tensor Flows for Chiral Boson Theories}\label{sec:classical}

In this section, we will discuss some generalities about classical deformations of interacting chiral boson theories which are driven by functions of the energy-momentum tensor. Quite generally, we refer to any differential equation for the Lagrangian which takes the form
\begin{align}\label{stress_tensor_flow_defn}
    \frac{\partial \mathcal{L}^{(\lambda)}}{\partial \lambda} = f \left( T_{\alpha \beta}^{(\lambda)} , \lambda \right) \, , 
\end{align}
along with an initial condition $\mathcal{L}^{(\lambda = 0)} = \mathcal{L}^{(0)}$, as a stress tensor flow. We emphasize that the function $f$ is a Lorentz scalar constructed from the Hilbert stress tensor associated with the Lagrangian $\mathcal{L}^{(\lambda)}$, and not with the undeformed theory $\mathcal{L}^{(0)}$. For theories that can be coupled to gravity using only the metric tensor $g_{\alpha \beta}$, the stress tensor is given by
\begin{align}\label{hilbert_defn_standard}
    T_{\alpha \beta} = - \frac{2}{\sqrt{-g}} \frac{\delta S}{\delta g^{\alpha \beta}} = - 2 \frac{\partial \mathcal{L}}{\partial g^{\alpha \beta}} + g_{\alpha \beta} \mathcal{L} \, .
\end{align}
However, for theories involving fermions or the chiral bosons of interest in this work, the standard definition (\ref{hilbert_defn_standard}) is not sufficient. We will instead need to work in a tetrad formalism, introducing vielbein fields (or frame fields) $\tensor{E}{^a_\alpha}$ so that
\begin{align}
    g_{\alpha \beta} = \tensor{E}{^a_\alpha} \tensor{E}{^b_\beta} \eta_{ab} \, .
\end{align}
We will use Greek symbols such as $\alpha$ and $\beta$ to refer to curved\footnote{We use the term ``curved'' for spacetime indices, even when we set the spacetime metric $g_{\alpha \beta}$ to the flat Minkowski metric $\eta_{\alpha \beta}$, to distinguish them from ``flat'' tangent space indices like those on $\eta_{ab}$.} indices in the two-dimensional spacetime with metric $g_{\alpha \beta}$ on which our fields are defined, in contrast with early Latin letters like $a$ and $b$ which refer to the flat tangent-space indices that are raised and lowered with the Minkowski metric $\eta_{ab}$. These are not to be confused with the lowercase middle Latin symbols like $i, j$ which are used to index the chiral scalars $\phi^i$, or their antichiral variants $\overbar{i}$, $\overbar{j}$ which are decorated with a bar and label the anti-chiral scalars $\overbar{\phi}^{\overline{i}}$.

We also define $E = \det \left( \tensor{E}{^a_\alpha} \right) = \sqrt{|g|}$. Because this determinant is non-vanishing, the matrix $\tensor{E}{^a_\alpha}$ has an inverse, which we refer to as the inverse vielbein and write as $\tensor{E}{^\alpha_a}$. This inverse frame field obeys
\begin{align}
    \tensor{E}{^a_\alpha} \tensor{E}{^\alpha_b} = \tensor{\delta}{^a_b} \, , \qquad \tensor{E}{^\alpha_a} \tensor{E}{^a_\beta} = \tensor{\delta}{^\alpha_\beta} \, ,
\end{align}
and similarly
\begin{align}
    \tensor{E}{^\alpha_a} \tensor{E}{^\beta_b} g_{\alpha \beta} = \eta_{ab} \, .
\end{align}
Within the tetrad formalism, the appropriate generalization of the Hilbert stress tensor with one curved and one flat index is defined by
\begin{align}\label{stress_tensor_flat_curved}
    \tensor{T}{_\beta^a} = - \frac{1}{E} \frac{\delta S}{\delta \tensor{E}{^\beta_a}} \, .
\end{align}
All tangent space indices can be converted to spacetime indices, or vice-versa, by contracting with vielbeins or inverse vielbeins as needed. For instance, the conventional stress tensor with two curved indices is then
\begin{align}
    T_{\alpha \beta} = \tensor{T}{_\alpha^a} \tensor{E}{^\gamma_a} g_{\gamma \beta} \, .
\end{align}
The tetrad formalism will allow us to compute the energy-momentum tensor and define stress tensor flows for an arbitrary interacting chiral boson theory of the type in equation (\ref{interaction_function}). We will perform the coupling to vielbeins in such a way that the stress tensor is automatically symmetric, $T_{\alpha \beta} = T_{\beta \alpha}$, but this is not sufficient to guarantee that the theory is invariant under boosts; for a generic choice of the interaction function $V$, the theory is \emph{not} Lorentz-invariant. In this work, we will be primarily interested in theories which \emph{do} enjoy Lorentz invariance, although this Lorentz symmetry will not be manifest within this formalism. Therefore, we will now pause to discuss the non-manifest Lorentz invariance of these models, including the conditions this imposes upon the interaction function $V$ and the connection between Lorentz symmetry and stress tensor flows.

\subsection{Lorentz invariance}\label{sec:lorentz}

We begin by reviewing one way to see the non-manifest Lorentz invariance of the simplest theory within the class of interest, the Floreanini-Jackiw action describing a single chiral boson. Although this is a well-known story, the discussion will fix our notation and set the stage for the analysis of Lorentz invariance with more general interaction functions.

\subsubsection*{\ul{\it One free chiral boson}}

Much like the electric-magnetic duality invariance of the $4d$ Maxwell theory, which is a symmetry of the equations of motion but not of the action itself, the Lorentz symmetry of the chiral boson theories we study here will be easier to understand at the level of the equations of motion. We illustrate this simple principle beginning with the action (\ref{fj_action}), which we rewrite for convenience:
\begin{equation}
\label{eq:FJR}
    S= \frac{1}{2} \int d^2x \left( \dot{\phi} \phi' - \phi^{\prime 2} \right) \, .
\end{equation}
Here, we have defined
\begin{align}
    \dot{\phi} = \partial_t \phi = \frac{\partial \phi}{\partial x^0} \, ,  \quad \phi' = \partial_\theta \phi = \frac{\partial \phi}{\partial x^1} \, , 
\end{align}
to ease notation. Now consider an infinitesimal Lorentz boost $\tensor{\Lambda}{^\alpha_\beta} = \tensor{\delta}{^\alpha_\beta} + \tensor{\omega}{^\alpha_\beta}$ with parameter $\omega_{0 1} = - \omega_{10} = \epsilon$. In this section, we work in Lorentzian signature with spacetime metric $\eta_{\alpha \beta} = \left[ \begin{smallmatrix} -1 & 0 \\ 0 & 1 \end{smallmatrix} \right]$. The change in the components of $\partial^\alpha \phi$ is
\begin{align}
    \delta \left( \partial^\alpha \phi \right) = \tensor{\omega}{^\alpha_\beta} \partial^\beta \phi \, ,
\end{align}
and thus the components of the covector $\partial_\alpha \phi = ( \dot{\phi} , \phi' )$ transform as
\begin{equation}
\label{eq:Iboost}
    \dot{\phi} \rightarrow \dot{\phi} + \epsilon \phi'\,, \quad \phi' \rightarrow \phi' + \epsilon \dot{\phi} \, .
\end{equation}
The change in the action \eqref{eq:FJR} is therefore

\begin{equation}
    \delta S = \frac{\epsilon}{2} \int d^2x \left( \dot{\phi} - \phi' \right)^2 + \mathcal{O}(\epsilon^2) \, .
\end{equation}
This is not an off-shell total derivative, so it is not manifest that this transformation is a symmetry of the theory. However, this property is more transparent if we work directly with the equations of motion. The Euler-Lagrange equation associated with (\ref{eq:FJR}) is
\begin{equation}
    \dot{\phi}' - \phi'' = 0 \, ,
\end{equation}
where $\dot{\phi}' = \partial_t \partial_\theta \phi$. This equation of motion can be expressed as $\partial_\theta \left( \dot{\phi} - \phi' \right) = 0$, which means that the quantity $\dot{\phi} - \phi'$ is independent of the spatial coordinate $\theta$:
\begin{align}\label{eom_no_gauge}
    \dot{\phi} - \phi' = f ( t ) \, .
\end{align}
The time-dependent function $f ( t )$ can be thought of as a choice of gauge, which is not physically meaningful. Indeed, suppose that we transform the function $\phi$ by
\begin{align}\label{delta_phi_gauge}
    \delta \phi = h ( t ) 
\end{align}
for a general time-dependent function $h$. Then $\delta \dot{\phi} = \dot{h}$ and $\delta \phi' = 0$, so the change in the Floreanini-Jackiw action is
\begin{align}
    \delta S = \frac{1}{2} \int d^2 x \, \left( \dot{h} \phi' \right) = \frac{1}{2} \int d^2 x \, \partial_\theta \left( \dot{h} \phi \right) \, , 
\end{align}
which is an integral of a total spatial derivative, and thus the action is unchanged. Therefore, given any solution to the equations of motion which takes the form (\ref{eom_no_gauge}), we are always free to perform a gauge transformation (\ref{delta_phi_gauge}) with
\begin{align}
    h ( t ) = \int^{t} f ( t' ) \, dt' \, ,  \qquad \dot{h} ( t ) = f ( t ) \, , 
\end{align}
which has the effect of eliminating the function $f ( t )$ on the right side of (\ref{eom_no_gauge}), and thus brings the equation of motion to the form
\begin{align}\label{eom_gauge_choice}
    \dot{\phi} - \phi' = 0 \, .
\end{align}
We will always work in the gauge (\ref{eom_gauge_choice}) in what follows. If we write equation (\ref{eom_gauge_choice}) as
\begin{align}
    \mathcal{E} ( \dot{\phi} , \phi' ) = 0 \, , \qquad \mathcal{E} = \dot{\phi} - \phi' \, , 
\end{align}
then acting with a Lorentz transformation on this quantity $\mathcal{E}$ gives
\begin{align}
    \delta \mathcal{E} = \delta \left( \dot{\phi} - \phi' \right) = - \epsilon \left( \dot{\phi} - \phi' \right) = - \epsilon \mathcal{E} \, .
\end{align}
That is, the variation of the equation of motion is proportional to the equation of motion itself. This means that, on the mass shell, the equations of motion are invariant under Lorentz transformations, which we write as
\begin{align}
    \delta \mathcal{E} \simeq 0 \, , 
\end{align}
where the symbol $\simeq$ means ``equal when the fields satisfy their equations of motion.'' This is sufficient for the theory to enjoy Lorentz invariance.

From this simple exercise, we see that the Floreanini-Jackiw theory of 
a single chiral boson does indeed exhibit non-manifest Lorentz invariance. This discussion also motivates a couple of definitions. We say that any function $\mathcal{O}$ of the fields and their derivatives is a \emph{Lorentz-invariant function} if $\delta \mathcal{O} \simeq 0$, that is, if the quantity $\mathcal{O}$ is invariant under Lorentz transformations when the fields satisfy their equations of motion. Likewise, we say that a Lagrangian $\mathcal{L}$ defines a \emph{Lorentz-invariant theory} if the Euler-Lagrange equations associated with $\mathcal{L}$ can be written as $\mathcal{E} = 0$ where $\mathcal{E}$ is a Lorentz-invariant function. 

\subsubsection*{\ul{\it Multiple interacting bosons}}

We now promote the action to depend on $N$ chiral bosons $\phi^i$ and $\overbar{N}$ anti-chiral bosons $\overbar{\phi}^{\overline{i}}$. A general theory with interactions that depend on spatial derivatives of the fields is\footnote{In this paper, we do not consider higher-derivative interactions.}
\begin{equation}\label{interacting_again}
    S = \int d^2x \left( \frac{1}{2} (\dot{\phi}^i \phi^{\prime \, i} - \dot{\overbar{\phi}} {}^{\overline{i}} \, \overbar{\phi}^{\prime \, \overline{i}}  ) - V ( \phi', \overbar{\phi}' ) \right) \, ,
\end{equation}
where we suppress indices on the fields in the argument of the interaction function $V$. Following the notation of the $N = 1$ analysis above, we can write the equations of motion for this model as a collection of equations $\mathcal{E}^i = 0$ and $\overbar{\mathcal{E}}^{\overline{i}} = 0$, where
\begin{align}\label{interacting_eom}
    \mathcal{E}^i = \dot{\phi}^i - \frac{\partial V}{\partial \phi^{\prime \, i}} \, , \qquad \overbar{\mathcal{E}}^{\overline{i}} = \dot{\overbar{\phi}}{}^{\overline{i}} + \frac{\partial V}{\partial \overbar{\phi}^{\overline{i}}} \, .
\end{align}
Note that we do not distinguish between upstairs and downstairs $i, j$ and $\overline{i}$, $\overline{j}$ indices on the scalars, instead choosing index placement for typographical convenience. In expressing the equations of motion as the vanishing of the quantities (\ref{interacting_eom}), we have also implicitly chosen the analog of the gauge $h(t) = 0$, as in the discussion around equation (\ref{eom_gauge_choice}) for the case of one chiral boson.

Let us again consider a Lorentz boost parameterized by $\omega_{01} = - \omega_{10} = \epsilon$. All of the fields transform in the same way as before:
\begin{equation}
    \dot{\phi}^i \rightarrow \dot{\phi}^i + \epsilon \phi^{\prime \, i} \, , \quad \phi^{\prime \, i} \rightarrow \phi^{\prime \, i} + \epsilon \dot{\phi}^i \, , \quad \dot{\overbar{\phi}}{}^{\overline{i}} \rightarrow \dot{\overbar{\phi}}{}^{\overline{i}} + \epsilon \overbar{\phi}^{\prime \, \overline{i}} \,, \quad \overbar{\phi}^{\prime \, \overline{i}} \rightarrow \overbar{\phi}^{\prime \, \overline{i}} + \epsilon \dot{\overbar{\phi}}{}^{\overline{i}} \, .
\end{equation}
We now ask: under what conditions on the interaction function $V$ will the action (\ref{interacting_again}) define a Lorentz-invariant theory, which means that $\delta \mathcal{E}^i \simeq 0$ and $\delta \overbar{\mathcal{E}}^{\overline{i}} \simeq 0$ under this Lorentz transformation? The variation of the chiral equations of motion is
\begin{align}
    \delta \mathcal{E}^i &= \delta \dot{\phi}^i - \frac{\partial^2 V}{\partial \phi^{\prime \, i} \, \partial \phi^{\prime \, j}} \, \delta \phi^{\prime \, j } - \frac{\partial^2 V}{\partial \phi^{\prime \, i} \, \partial \overbar{\phi}^{\prime \, \overline{j}}} \, \delta \overbar{\phi}^{\prime \, \overline{j} } \nonumber \\
    &= \epsilon \phi^{\prime \, i} - \epsilon V_{i j} \dot{\phi}^j - \epsilon V_{i \overline{j}}  \dot{\overbar{\phi}}{}^{\overline{j}} \nonumber \\
    &\simeq \epsilon \left[ \phi^{\prime \, i} - V_{ij} V_j + V_{i \overline{j}} V_{\overline{j}} \right] \, ,
\end{align}
where in the second step we have introduced the notation
\begin{align}
    V_i = \frac{\partial V}{\partial \phi^{\prime \, i}} \, , \quad V_{ij} = \frac{\partial^2 V}{\partial \phi^{\prime \, i} \, \partial \phi^{\prime \, j}} \, , \quad V_{i \overline{j}} = \frac{\partial^2 V}{\partial \phi^{\prime \, i} \, \partial \overbar{\phi}^{\prime \, \overline{j}}} \, , 
\end{align}
and so on, and in the third line, we have replaced the time derivatives $\dot{\phi}^i$, $\dot{\overbar{\phi}}{}^{\overline{j}}$ using the equations of motion and therefore used the on-shell equality symbol $\simeq$. An identical calculation for the anti-chiral equations of motion gives
\begin{align}
    \delta \overbar{\mathcal{E}}^{\overline{i}} \simeq \epsilon \left[ \overbar{\phi}^{\prime \, \overline{i}} - V_{\overline{i} \overline{j}} V_{\overline{j}} + V_{\overline{i} j} V_j \right] \, .
\end{align}
Therefore, for the quantities $\mathcal{E}^i$ and $\overbar{\mathcal{E}}^{\overline{i}}$ to be Lorentz-invariant functions, we must impose the two conditions
\begin{align}
    \phi^{\prime \, i } + V_{i \overline{j}} V_{\overline{j}} = V_{ij} V_j \, , \qquad \overbar{\phi}^{\prime \, \overline{i}} + V_{\overline{i} j} V_j = V_{\overline{i} \overline{j}} V_{\overline{j}} \, .
\end{align}
It is convenient to write these two equations in terms of the derivatives of products,
\begin{align}\label{lorentz_intermediate}
    \phi^{\prime \, i } + \frac{1}{2} \partial_i \left( V_{\overline{j}} V_{\overline{j}} \right) = \frac{1}{2} \partial_i \left( V_j V_j \right) \, , \qquad \overbar{\phi}^{\prime \, \overline{i}} + \frac{1}{2} \partial_{\overline{i}} \left( V_j V_j \right) = \frac{1}{2} \partial_{\overline{i}} \left( V_{\overline{j}} V_{\overline{j}} \right) \, ,
\end{align}
where the repeated $j$, $\overline{j}$ indices are summed and where $\partial_i = \frac{\partial}{\partial \phi^{\prime \, i}}$, $\partial_{\overline{i}} = \frac{\partial}{\partial \overbar{\phi}^{\prime \, \overline{i}}}$.

We can now integrate the first of the equations (\ref{lorentz_intermediate}) with respect to $\phi^{\prime \, i}$ and the second with respect to $\overbar{\phi}^{\prime \, \overline{i}}$ to find
\begin{align}\label{lorentz_intermediate_two}
    \left( \phi^{\prime \, i} \right)^2 + V_{\overline{j}} V_{\overline{j}} = V_{j} V_{j} + C^i ( \phi^{\prime \, k \neq i} , \overbar{\phi}^{\prime \, \overline{k}} ) \, , \qquad \left( \overbar{\phi}^{\prime \, \overline{i}} \right)^2 + V_j V_j = V_{\overline{j}} V_{\overline{j}} + \overbar{C}^{\overline{i}} ( \phi^{\prime \, k} , \overbar{\phi}^{\prime \, \overline{k} \neq \overline{i}} ) \, .
\end{align}
Here we have introduced two integration constants, $C^i$ which is independent of $\phi^{\prime \, i}$ and $\overbar{C}^{\overline{i}}$ which is independent of $\overbar{\phi}^{\prime \, \overline{i}}$. Also note that equation (\ref{lorentz_intermediate_two}) holds separately for each fixed $i$ and $\overline{i}$; the quantity $\left( \phi^{\prime \, i} \right)^2$ is the square of one such fixed $\phi^{\prime \, i}$, and is not summed on $i$. We can fix these integration constants by noting that the choice of interaction function
\begin{align}\label{free_interaction}
    V ( \phi , \overbar{\phi} ) = \frac{1}{2} \left( \phi^{\prime \, j} \phi^{\prime \, j} + \overbar{\phi}^{\prime \, \overline{j}} \overbar{\phi}^{\prime \, \overline{j}} \right) \, ,
\end{align}
which is just a sum of non-interacting chiral and anti-chiral bosons, must necessarily satisfy the Lorentz-invariance condition. This will be true if we choose
\begin{align}
    C^i = \overbar{\phi}^{\prime \, \overline{j}} \overbar{\phi}^{\prime \, \overline{j}} \, + \, \sum_{k \neq i} \phi^{\prime \, k} \phi^{\prime \, k} , \qquad \overbar{C}^{\overline{i}} = \phi^{\prime \, j} \phi^{\prime \, j} \, + \, \sum_{\overline{k} \neq \overline{i}} \overbar{\phi}^{\prime \, \overline{k}} \overbar{\phi}^{\prime \, \overline{k}} \, ,
\end{align}
which means that the two equations in (\ref{lorentz_intermediate_two}) are proportional to one another, and we are left with the single condition
\begin{align}\label{lorentz_intermediate_three}
    \phi^{\prime \, j} \phi^{\prime \, j} - \overbar{\phi}^{\prime \, \overline{j}} \overbar{\phi}^{\prime \, \overline{j}} = V_j V_j - V_{\overline{j}} V_{\overline{j}} \, , 
\end{align}
for Lorentz invariance. Suppose that we now further assume that the interaction function is invariant under $O ( N )$ rotations of the $N$ chiral fields and $O ( \overbar{N} )$ rotations of the $\overbar{N}$ anti-chiral fields. This means that we can parameterize $V$ as a function of the two invariants\footnote{The invariant $S$ should not be confused with the action $S = \int d^2 x \, \mathcal{L}$; we trust that the reader can distinguish between the two based on context.}
\begin{align}\label{S_and_P_invariants}
    S = \frac{1}{2} \left( \phi^{\prime \, j} \phi^{\prime \, j} + \overbar{\phi}^{\prime \, \overline{j}} \overbar{\phi}^{\prime \, \overline{j}} \right) \, , \qquad P = \frac{1}{2} \left( \phi^{\prime \, j} \phi^{\prime \, j} - \overbar{\phi}^{\prime \, \overline{j}} \overbar{\phi}^{\prime \, \overline{j}} \right) \, .
\end{align}
Note that, for the theory defined by the free interaction function (\ref{free_interaction}), the quantities $S$ and $P$ represent the total Hamiltonian density and momentum density, respectively. In terms of these variables, the condition (\ref{lorentz_intermediate_three}) can be written as
\begin{align}\label{lorentz_pde_final}
    V_S^2 + \frac{2 S}{P} V_S V_P + V_P^2 = 1 \, .
\end{align}
Partial differential equations of the schematic form (\ref{lorentz_pde_final}) have appeared in many contexts. Most directly relevant for this analysis, precisely the same differential equation appears as the condition for Lorentz invariance of the phase space actions for theories of self-dual electrodynamics in $d = 4$ or for chiral tensor theories in $d = 6$; see, for instance, sections 2.2 and 2.3 of \cite{Bandos:2020hgy} for these two cases, respectively. Our condition (\ref{lorentz_pde_final}) is merely the $2d$ version of these results, in the case where one considers arbitrary numbers of chiral and anti-chiral bosons. Note that, in the case $\overbar{N} = 0$ which describes only chiral bosons, the two invariants (\ref{S_and_P_invariants}) collapse to
\begin{align}
    S = P \, , 
\end{align}
so that $V$ is a function of one variable, and the constraint (\ref{lorentz_intermediate_three}) simplifies to
\begin{align}
    \phi^{\prime \, j} \phi^{\prime \, j} = V_j V_j \, , 
\end{align}
or in terms of the variable $S = \frac{1}{2} \phi^{\prime \, j} \phi^{\prime \, j}$,
\begin{align}\label{only_S}
    V_S = 1 \, .
\end{align}
This means that the only solution is the free case, $V = S = \frac{1}{2} \phi^{\prime \, j} \phi^{\prime \, j}$, in accordance with known results. The same conclusion holds for only anti-chiral bosons, $N = 0$ but $\overbar{N} > 0$.

A similar partial differential equation, which differs only by signs, occurs as the condition for a Lagrangian for $4d$ non-linear electrodynamics to have equations of motion that are invariant under electric-magnetic duality rotations. In this case, the appropriate PDE reads
\begin{align}\label{duality_pde}
    \mathcal{L}_S^2 - \frac{2 S}{P} \mathcal{L}_S \mathcal{L}_P - \mathcal{L}_P^2 = 1 \, , 
\end{align}
where now $S = - \frac{1}{4} F_{\mu \nu} F^{\mu \nu}$ and $P = - \frac{1}{4} F_{\mu \nu} \widetilde{F}^{\mu \nu}$ are the two independent Lorentz scalars that can be constructed from the field strength $F_{\mu \nu}$, and $\widetilde{F}_{\mu \nu}$ denotes the Hodge dual of $F_{\mu \nu}$. This version of the differential equation, with the signs as in (\ref{duality_pde}), also appears as the condition for a certain class of non-linear sigma models in $d = 2$ to have equations of motion which are equivalent to the flatness of a Lax connection which takes a prescribed form \cite{Borsato:2022tmu} (see equations (7.3) - (7.5) of \cite{Ferko:2023wyi} for the definitions of $S$ and $P$ in this case).

In either presentation, with the choice of signs in (\ref{lorentz_pde_final}) or the one in (\ref{duality_pde}), this differential equation has many solutions besides the free one. For instance, equation (\ref{lorentz_pde_final}) admits the two-parameter family of solutions
\begin{align}\label{scalar_modified_nambu_goto}
    V ( S, P ; \gamma, \lambda ) = \frac{1}{\lambda} \left( \sqrt{ 1 + 2 \lambda \left( \cosh ( \gamma ) S + \sinh ( \gamma ) \sqrt{ S^2 - P^2 } \right) + \lambda^2 P^2 } - 1 \right) \, .
\end{align}
This family of interaction functions is the $2d$ chiral boson analog of the two-parameter family of $4d$ ModMax-Born-Infeld gauge theories, which we mentioned in the introduction. As in the $4d$ case, the function $V$ of equation (\ref{scalar_modified_nambu_goto}) satisfies two commuting flow equations which relate $\partial_\lambda V$ and $\partial_\gamma V$ to an irrelevant $\TT$-like and a marginal root-$\TT$-like operator built from the energy-momentum tensor of the model, respectively:
\begin{align}\label{two_nice_flows}
    \frac{\partial V}{\partial \lambda} = - \mathcal{O}_{TT} = - \frac{1}{4} \left( T^{\alpha \beta} T_{\alpha \beta} - \left( \tensor{T}{^\alpha_\alpha} \right)^2 \right) \, , \qquad \frac{\partial V}{\partial \gamma} = - \mathcal{R} = - \frac{1}{\sqrt{2}} \sqrt{ T^{\alpha \beta} T_{\alpha \beta} - \frac{1}{2} \left( \tensor{T}{^\alpha_\alpha} \right)^2 } \, .
\end{align}
This example illustrates that, at least in this case, solutions to the differential equation (\ref{lorentz_pde_final}) can be obtained by deforming the interaction function by Lorentz-invariant quantities, such as Lorentz scalars constructed from $T_{\mu \nu}$. This statement applies quite generally to any deformation of $V$ by a Lorentz-invariant function, as we describe next.

\subsubsection*{\ul{\it Lorentz-invariant functions}}

In the preceding discussion, we derived a condition on the function $V$ (equation (\ref{lorentz_pde_final})) which guarantees that this interaction function describes a Lorentz-invariant theory. By definition, this means that the equations of motion $\mathcal{E}^i$, $\overbar{\mathcal{E}}^i$ are Lorentz-invariant functions. One might ask, more generally, given an arbitrary function $\mathcal{O} ( S, P )$ which depends on the two combinations $S$ and $P$ defined in (\ref{S_and_P_invariants}), under what conditions is $\mathcal{O}$ a Lorentz-invariant function? That is, for which operators $\mathcal{O}$ is $\delta \mathcal{O} \simeq 0$, where $\delta$ is a Lorentz transformation?

This question can be answered using a similar calculation as the one above. One has
\begin{align}
    \delta \mathcal{O} ( S, P ) &= \mathcal{O}_S \delta S + \mathcal{O}_P \delta P \nonumber \\
    &= \mathcal{O}_S \left( \phi^{\prime \, j} \, \delta \phi^{\prime \, j}  + \overbar{\phi}^{\prime \, \overline{j}} \, \delta \overbar{\phi}^{\prime \, \overline{j}} \right) + \mathcal{O}_P \left( \phi^{\prime \, j} \delta \phi^{\prime \, j}  - \overbar{\phi}^{\prime \, \overline{j}} \, \delta \overbar{\phi}^{\prime \, \overline{j}} \right) \, ,
\end{align}
where subscripts represent partial derivatives with respect to the argument. On-shell, one has the variations
\begin{align}
    \delta \phi^{\prime \, j} = \epsilon \dot{\phi}^{j} \simeq \epsilon V_j \, , \qquad \delta \overbar{\phi}^{\prime \, j} = \epsilon \dot{\overbar{\phi}}{}^{\overline{j}} \simeq - V_{\overline{j}} \, , 
\end{align}
and thus one finds
\begin{align}
    \delta \mathcal{O} \simeq \epsilon \mathcal{O}_S \left( \phi^{\prime \, j} V_j - \overbar{\phi}^{\prime \, \overline{j}} V_{\overline{j}} \right) + \epsilon \mathcal{O}_P \left( \phi^{\prime \, j} V_j + \overbar{\phi}^{\prime \, \overline{j}} V_{\overline{j}} \right) \, .
\end{align}
Expressing the derivatives of $V$ in terms of $V_S$ and $V_P$ using
\begin{align}
    V_j = \left( V_S + V_P \right) \phi^{\prime \, j} \, , \qquad V_{\overline{j}} = \left( V_S - V_P \right) \overbar{\phi}^{\prime \, \overline{j}} \, , 
\end{align}
we conclude that $\delta \mathcal{O} \simeq 0$ if and only if
\begin{align}\label{lorentz_invariant_function}
    V_S \mathcal{O}_S + \frac{S}{P} \left( V_S \mathcal{O}_P + V_P \mathcal{O}_S \right) + V_P \mathcal{O}_P = 0 \, .
\end{align}
It is easy to see that the condition (\ref{lorentz_invariant_function}) is identical to the constraint that one finds by expanding the Lorentz-invariance condition (\ref{lorentz_pde_final}) for a perturbed interaction function
\begin{align}
    V ( S, P ) \to V ( S, P ) + \lambda \mathcal{O} ( S , P ) \, , 
\end{align}
assuming that $V$ itself satisfies the Lorentz-invariance condition, and then demanding that the deformed interaction function preserve this condition (\ref{lorentz_pde_final}) to leading order in $\lambda$.

We conclude that linearized Lorentz-preserving deformations of a boost-invariant theory of chiral bosons, described by an interaction function $V$, are in one-to-one correspondence with Lorentz-invariant functions $\mathcal{O}$ within this same theory defined by $V$. Again, this result is the $2d$ analog of the corresponding statements about linearized deformations which preserve electric-magnetic duality invariance in $4d$ \cite{Ferko:2023wyi} or PST gauge invariance in $6d$ \cite{Ferko:2024zth}. As in those contexts, this extends to an all-orders result: given a one-parameter family of interaction functions $V 
( \lambda )$ with an initial condition $V_0 = V ( \lambda = 0 )$ which satisfies (\ref{lorentz_pde_final}), the entire family of functions $V ( \lambda )$ satisfies the Lorentz invariance condition if and only if
\begin{align}\label{all_orders_flow}
    \frac{\partial V(\lambda)}{\partial \lambda} = \mathcal{O}^{(\lambda)} \, , 
\end{align}
where at each value of $\lambda$, the function $\mathcal{O}^{(\lambda)}$ obeys the constraint (\ref{lorentz_invariant_function}) with respect to the interaction function $V ( \lambda )$ at the same value of $\lambda$. 

There are several ways to prove this claim, which we will not present in detail since they are similar to the $4d$ and $6d$ cases. One strategy is to first argue that any such family of Lorentz-invariant functions $\mathcal{O}^{(\lambda)}$ can be expressed in terms of Lorentz scalars constructed from $T_{\mu \nu}^{(\lambda)}$, as we will show shortly, and then to show that an all-orders flow of the form (\ref{all_orders_flow}) driven by a function of the stress tensor preserves the Lorentz-invariance condition, by following an inductive argument like that in appendix A.1 of \cite{Ferko:2023wyi}.

\subsection{Stress tensor for general interacting theory}

We now turn to the computation of the energy-momentum tensor for a generic member of our class of chiral boson theories. Contractions built from this stress tensor, such as $\tensor{T}{^\mu_\mu}$ and $T^{\mu \nu} T_{\mu \nu}$, are canonical examples of the Lorentz-invariant functions which yield Lorentz-preserving deformations (\ref{all_orders_flow}) of the interaction function -- and, in fact, \emph{any} such deformation can be expressed in terms of such stress tensor scalars, as we will see.

In order to calculate the stress tensor defined in (\ref{stress_tensor_flat_curved}), we will couple a general theory of chiral bosons to gravity in the vielbein formulation following the approach of \cite{Bastianelli:1989cu}, which demonstrated how to perform this coupling for the standard Floreanini-Jackiw boson with interaction function $V ( S, P ) = S$. In the case of a general interaction function, the corresponding Lagrangian including the vielbein couplings takes the form
\begin{align}\label{general_class_vielbeins}
    \mathcal{L} = \frac{1}{2} \left( G_{ij} \dot{\phi}^i \phi^{\prime \, j} - \overbar{G}^{\overline{i} \overline{j}} \dot{\overbar{\phi}}{}^{\overline{i}} \overline{\phi}^{\prime j} \right) - \left( E_\theta^- E_t^+ + E_t^- E_\theta^+ \right)  P - E V ( S, P ) + \mathcal{L}_{\text{top}} \, , 
\end{align}
where now $S$ and $P$ are coupled to the frame fields as
\begin{align}
    S = - \frac{1}{4 E_\theta^- E_\theta^+} \left( G_{ij} \phi^{\prime \, i} \phi^{\prime \, j} + \overbar{G}_{\overline{i} \overline{j}} \overbar{\phi}^{\prime \overline{i}} \overbar{\phi}^{\prime \overline{j}} \right) \, , \qquad P = - \frac{1}{4 E_\theta^- E_\theta^+} \left( G_{ij} \phi^{\prime \, i} \phi^{\prime \, j} - \overbar{G}_{\overline{i} \overline{j}} \overbar{\phi}^{\prime \overline{i}} \overbar{\phi}^{\prime \overline{j}} \right) \, .
\end{align}
A few remarks are in order. We work in light-cone coordinates $x^a = x^{\pm}$ for the tangent space indices, so the vielbeins and inverse vielbeins carry one $( + , - )$ index and one $(t, \theta)$ index. After varying with respect to the vielbeins, we will set them to their flat-space values 
\begin{align}\label{flat_vielbeins}
    E^+_t = - E^+_\theta = E^-_\theta = E^-_t = \frac{1}{\sqrt{2}} \, ,
\end{align}
at the end of the calculation, which is appropriate for the light-cone tangent space metric $\eta_{ab} = \left[ \begin{smallmatrix} 0 & - 1 \\ - 1 & 0 \end{smallmatrix} \right]$. We have also introduced general target-space metrics $G_{ij} ( \phi )$ and $\overbar{G}_{\overline{i} \overline{j}} ( \overbar{\phi} )$ for the chiral and anti-chiral bosons, which does not affect the computation of the stress tensor. In equation (\ref{general_class_vielbeins}), we have allowed for the inclusion of a general topological term $\mathcal{L}_{\text{top}}$, which does not couple to the frame fields and which therefore drops out of the computation of $T_{\mu \nu}$. An example of such a topological term is a coupling to a target-space antisymmetric tensor field $B_{ij}$, $\overbar{B}_{\overline{i} \overline{j}}$. In manifestly Lorentz-invariant notation, which is perhaps more familiar, such a coupling would be expressed as $B_{ij} \epsilon^{\alpha \beta} \partial_\alpha \phi^i \partial_\beta \phi^j$, and is independent of the metric.

Note that, in the special case $G_{ij} = \delta_{ij}$, $\overbar{G}_{\overline{i} \overline{j}} = \delta_{\overline{i} \overline{j}}$, $\mathcal{L}_{\text{top}} = 0 $, and with the vielbeins equal to their flat-space values (\ref{flat_vielbeins}), the Lagrangian (\ref{general_class_vielbeins}) reduces to
\begin{equation}\label{interacting_again_again}
    \mathcal{L} = \frac{1}{2} (\dot{\phi}^i \phi^{\prime \, i} - \dot{\overbar{\phi}} {}^{\overline{i}} \, \overbar{\phi}^{\prime \, \overline{i}}  ) - V ( S, P  )  \, ,
\end{equation}
which agrees with (\ref{interacting_again_again}), and the quantities $S$ and $P$ become
\begin{align}\label{S_and_P_invariants_again}
    S = \frac{1}{2} \left( \phi^{\prime \, j} \phi^{\prime \, j} + \overbar{\phi}^{\prime \, \overline{j}} \overbar{\phi}^{\prime \, \overline{j}} \right) \, , \qquad P = \frac{1}{2} \left( \phi^{\prime \, j} \phi^{\prime \, j} - \overbar{\phi}^{\prime \, \overline{j}} \overbar{\phi}^{\prime \, \overline{j}} \right) \, ,
\end{align}
which agrees with (\ref{S_and_P_invariants}). 

It may come as a surprise that the kinetic terms in (\ref{general_class_vielbeins}), which involve $\dot{\phi}^i \phi^{\prime \, j}$ and $\dot{\overbar{\phi}}{}^{\overline{i}} \overline{\phi} {}^{\prime \overline{j}} $, are independent of the vielbeins, and do not even include a factor of $E$ which plays the role of $\sqrt{ g }$ that usually accompanies any scalar within a spacetime integral. This is a consequence of the specific method for coupling the chiral boson to gravity developed in \cite{Bastianelli:1989cu}, which first introduces an unconstrained bosonic field and then incorporates auxiliary fields $P$ and $b$ which enforce the chirality constraint. This combined system is then coupled to gravity, and then integrating out the auxiliary fields $P$ and $b$ has the effect of eliminating the factor of $E$ that normally multiplies the kinetic term. We will see in section \ref{sec:cs} that the absence of vielbein dependence in these terms has a natural interpretation in the dual Chern-Simons description of chiral boson theories.

We can now explicitly perform the variation with respect to the vielbeins to compute the stress tensor $\tensor{T}{_a^\beta}$, as defined in equation (\ref{stress_tensor_flat_curved}), or more usefully, the version $T_{\alpha \beta}$ with two spacetime indices:
\begin{align}\label{general_stress_tensor}
    T_{tt} &= V ( S, P ) \, , \nonumber \\
    T_{t \theta} &= - P = T_{\theta t} \, , \nonumber \\
    T_{\theta \theta} &= - V + 2 \left( S V_S + P V_P \right) \, .
\end{align}
Note that the off-diagonal terms of $T_{\alpha \beta}$ are therefore identical and both proportional to $P$, which has the interpretation of the momentum along the $\theta$ circle. This is a consequence of the way we have coupled to the vielbeins in the second term of (\ref{general_class_vielbeins}), which is proportional to $P$ but which vanishes in the flat-space limit.

In principle, one could consider more general couplings of these chiral boson theories to vielbeins, which would lead to stress tensors that may not be symmetric and which are related to (\ref{general_stress_tensor}) by an improvement transformation. However, we find the choice of coupling that we have made here to be physically motivated for the problem of studying flow equations of the form (\ref{all_orders_flow}) which are connected to the free interaction function (\ref{free_interaction}). For instance, in the quantum theory, the momentum along a circle of radius $R$ is quantized in units of $\frac{1}{R}$, and, therefore, cannot flow with any deformation parameterized by a continuous $\lambda$. The coupling to frame fields which leads to (\ref{general_stress_tensor}) makes this manifest, even at the level of the classical stress tensor, since for any interaction function $V ( S, P )$, the linear momentum along the circle is fixed to its value $T_{t \theta} = - P$ in the free theory.

The trace of the stress tensor,
\begin{align}\label{trT}
    \Tr ( T ) = \tensor{T}{^\alpha_\alpha} = - 2 \left( V - S V_S - P V_P \right) \, , 
\end{align}
vanishes if the interaction function $V$ is a homogeneous function of degree $1$ in its arguments, which is equivalent to the scale invariance of the theory as expected. The other Lorentz invariant that one can construct from the stress tensor is
\begin{align}\label{trT2}
    \Tr ( T^2 ) = T^{\mu \nu} T_{\mu \nu} = V^2 - 2 P^2 + \left( V - 2 \left( S V_S + P V_P \right) \right)^2 \, .
\end{align}
One can check by explicit computation that the two invariants (\ref{trT}) and (\ref{trT2}) each satisfy the condition (\ref{lorentz_invariant_function}), assuming that the interaction function $V$ itself obeys the condition (\ref{lorentz_pde_final}). In fact, more is true: given either of these two Lorentz-invariant functions $\tensor{T}{^\mu_\mu}$ and $T^{\mu \nu} T_{\mu \nu}$, locally and away from exceptional points, we can implicitly express any other Lorentz-invariant function $f$ in terms of this stress tensor invariant. To see this, let $f ( S, P )$ and $g ( S, P )$ be any two functions that satisfy the Lorentz-invariance condition (\ref{lorentz_invariant_function}). Consider the Jacobian for the change of variables from $(S, P)$ to $(f, g)$, namely
\begin{align}
    J = \begin{bmatrix} f_S & f_P \\ g_S & g_P \end{bmatrix} \, ,
\end{align}
and, in particular, its determinant,
\begin{align}\label{jac_det}
    \det ( J ) = f_S g_P - f_P g_S \, .
\end{align}
Since $f$ and $g$ each satisfy equation (\ref{lorentz_invariant_function}), we can solve this equation to express one of the partial derivatives of each function in terms of the other. For instance, we can choose
\begin{align}
    f_S = - \frac{f_P \left( P V_P + S V_S \right) }{S V_P + P V_S} \, , \qquad g_S = - \frac{g_P \left( P V_P + S V_S \right) }{S V_P + P V_S} \, .
\end{align}
Substituting these into the determinant (\ref{jac_det}), we find
\begin{align}
    \det ( J ) =  - \frac{f_P g_P \left( P V_P + S V_S \right) }{S V_P + P V_S} + \frac{f_P g_P \left( P V_P + S V_S \right) }{S V_P + P V_S} = 0 \, . 
\end{align}
Because $\det ( J ) = 0$, this change of variables is singular, which means that there exists a functional relation of the form $F ( f, g ) = 0$. By the implicit function theorem, under some assumptions on the derivatives of $F$, we can locally express $f ( S, P )$ as a function of $g ( S, P )$, or vice-versa. Thus, ignoring exceptional points, any pair of Lorentz-invariant functions are functionally dependent. Since the quantities $T^{\mu \nu} T_{\mu \nu}$ and $\tensor{T}{^\mu_\mu}$ are examples of such invariant quantities, it follows that any other Lorentz-invariant function -- again, away from singular points, and excluding trivial examples such as the case where one of the functions is a constant -- can be expressed as a function of the stress tensor.

Combining this conclusion with the previous statement around equation (\ref{all_orders_flow}), it also follows that, given any parameterized family of interaction functions $V ( \lambda )$ for Lorentz-invariant theories, one can write
\begin{align}
    \frac{\partial V ( \lambda )}{\partial \lambda} = \mathcal{O}^{(\lambda)} \equiv f ( T_{\mu \nu}^{(\lambda)} , \lambda ) \, ,
\end{align}
where in the last step we have used that $\mathcal{O}^{(\lambda)}$ can be implicitly expressed as a function of the stress tensor, given that this $\mathcal{O}^{(\lambda)}$ satisfies the Lorentz-invariance condition (\ref{lorentz_invariant_function}).

Therefore, the stress tensor flows that we have introduced in equation (\ref{stress_tensor_flow_defn}) are quite generic: any family of Lorentz-invariant interaction functions obeys a differential equation of this form, and conversely, any such flow equation (along with a Lorentz-invariant initial condition) defines a family of Lorentz-invariant interacting chiral boson theories.

Interesting examples of such flows are the ones defined in equation (\ref{two_nice_flows}), which are driven by the operators $\mathcal{O}_{\TT}$ and $\mathcal{R}$. We can express these two operators in terms of the interaction function $V$ and its derivatives using the general results (\ref{trT}) and (\ref{trT2}):
\begin{align}\label{simplified_TT_and_R}
    \mathcal{O}_{\TT} = V \left( S V_S + P V_P \right) - \frac{1}{2} \left( V^2 + P^2 \right) \, , \qquad \mathcal{R} = \sqrt{ \left( S V_S + P V_P + P \right) \left( S V_S + P V_P - P \right) } \, .
\end{align}
One can check directly that the two-parameter family of interaction functions (\ref{scalar_modified_nambu_goto}) solves the flow equations driven by the two operators given in (\ref{simplified_TT_and_R}).\footnote{When $\gamma = 0$, one recovers the theory of $\TT$-deformed Floreanini-Jackiw bosons, which also appears in the boundary graviton action for AdS$_3$ gravity at a finite radial cutoff; see equation (3.70) of \cite{Ebert:2022cle}.} The root-$\TT$ flow equation can also be solved in more generality. Suppose we begin from the flow equation
\begin{align}\label{root_TT_flow_V}
    \frac{\partial V ( \gamma )}{\partial \gamma} = - \mathcal{R} = - \sqrt{ \left( S V_S + P V_P + P \right) \left( S V_S + P V_P - P \right) } \, , 
\end{align}
and we furthermore assume that the function $V$ satisfies the Lorentz-invariance condition (\ref{lorentz_pde_final}) everywhere along the flow (which it is guaranteed to do, assuming the initial condition is Lorentz-invariant). Then the general solution to the differential equation (\ref{root_TT_flow_V}) with initial condition $V ( \gamma = 0 , S , P ) = V_0 ( S, P )$ is
\begin{align}\label{general_solution_root_TT}
    V ( \gamma, S , P ) = V_0 \left( \cosh ( \gamma ) S + \sinh ( \gamma ) \sqrt{ S^2 - P^2 } , P \right) \, .
\end{align}
That is, we simply replace all occurrences of the variable $S$ in the initial condition $V_0 ( S, P )$ with the combination $\cosh ( \gamma ) S + \sinh ( \gamma ) \sqrt{ S^2 - P^2 }$, while leaving all occurrences of $P$ unchanged. The result is a solution to (\ref{root_TT_flow_V}) with the correct initial condition at $\gamma = 0$.

Let us point out that the formulas (\ref{general_stress_tensor}) for the stress tensor components are valid when $N \geq 1$ and $\overbar{N} \geq 1$. In the case of all chiral bosons ($\overbar{N} = 0$), or all anti-chiral bosons ($N = 0$), the two invariants $S$ and $P$ become proportional to one another, so some of the structures in the Lagrangian collapse. For instance, for a theory of all chiral bosons, we have $S = P$ and the components of the stress tensor are
\begin{align}
    T_{tt} &= V ( S ) \, , \nonumber \\
    T_{t \theta} &= - S = T_{\theta t} \, , \nonumber \\
    T_{\theta \theta} &= - V ( S ) + 2 S V' ( S ) \, .
\end{align}
We have seen that the only solution to the Lorentz-invariance condition (\ref{lorentz_pde_final}) for all chiral bosons is $V = S$, and the stress tensor for this theory is
\begin{align}
    T_{\alpha \beta} = \frac{1}{2} \phi^{\prime j} \phi^{\prime j} \begin{bmatrix} 1 & -1 \\ -1 & 1 \end{bmatrix} \, .
\end{align}
Here one has $\tensor{T}{^\alpha_\alpha} = 0$ and $T^{\alpha \beta} T_{\alpha \beta} = 0$. The same conclusion holds for all anti-chiral bosons, where we have $S = -P$ rather than $S = P$, but again one finds $\Tr ( T ) = 0 = \Tr ( T^2 )$. For either of these scenarios, since both Lorentz scalars constructed from the stress tensor are vanishing, the theory is a fixed point of all Lorentz-preserving stress tensor deformations.\footnote{Another way to see this is by considering complex coordinates $(w, \wb)$, with $T = T_{ww}$ and $\overbar{T} = T_{\wb \wb}$. A theory of all chiral bosons has $\overbar{T} = 0$ and a theory of all anti-chiral bosons has $T = 0$. In either case, the product $\TT$ vanishes, and the trace vanishes by conformal invariance, so any Lorentz-preserving stress tensor flow is trivial. Of course, one could generate non-trivial interacting models by breaking Lorentz invariance and studying, for example, $f(T)$ (or $f(\overbar{T})$) flows, but we will not pursue this option here.}

This gives another way to understand the fact that there is no way to introduce Lorentz-invariant interactions as a continuous deformation of a free theory involving only chiral bosons, or only anti-chiral bosons. Indeed, if a family of such interacting theories did exist, they would necessarily satisfy a stress tensor flow equation. But no such flow can exist which includes the free theory $V = S$, as this theory is left invariant by any stress tensor deformation. Since a theory of only chiral bosons has the Hamiltonian $\mathcal{H} = S = P$, one can view it as a $2d$ version of the $4d$ theory of Bialynicki-Birula electrodynamics, which is also a fixed point of all stress tensor flows.

\subsection{Self-duality and chirality}\label{sec:sd_and_chiral}

To conclude this section, we will point out one additional feature of the chiral boson models considered here. Although this property is trivially satisfied for any interacting chiral boson theory, regardless of the interaction function $V ( S, P )$, the analogous property for theories in the dual Chern-Simons description will play an important role in the next section.

Suppose that we begin with a general action of the form that we have been considering, which we repeat here for convenience:
\begin{equation}\label{interacting_start_selfduality}
    S = \int d^2x \left( \frac{1}{2} (\dot{\phi}^i \phi^{\prime \, i} - \dot{\overbar{\phi}} {}^{\overline{i}} \, \overbar{\phi}^{\prime \, \overline{i}}  ) - V ( S, P ) \right) \, .
\end{equation}
We would like to exchange the gradients $\partial_\alpha \phi^i = ( \dot{\phi}^i , \phi^{\prime i} )$ of the scalar fields for a vector field $A_\alpha = ( A_0, A_1 )$, and likewise for the anti-chiral scalars. To do this, we introduce a collection of Lagrange multiplier fields $\lambda^{i \alpha}$ and $\overbar{\lambda}^{\overline{i} \alpha}$, and write the equivalent action
\begin{equation}\label{equivalent_interacting}
    S = \int d^2x \left( \frac{1}{2} \left( A_0^i A_1^i - \overbar{A}_0^{\overline{i}} \overbar{A}_1^{\overline{i}} \right) - V ( S_A , P_A ) + \frac{1}{2} \lambda^{\alpha i} ( A_\alpha^i - \partial_\alpha \phi^i ) - \frac{1}{2} \overbar{\lambda}^{\alpha \overline{i}} ( \overbar{A}_\alpha^{\overline{i}} - \partial_\alpha \overbar{\phi}^{\overline{i}} ) \right) \, .
\end{equation}
Here the variables $S_A$ and $P_A$ are defined by replacing instances of $\phi^{\prime i}$ with $A_1^i$ and replacing $\overbar{\phi}^{\prime \overline{i}}$ with $\Abar^{\overline{i}}_1$:
\begin{align}
    S_A = \frac{1}{2} \left( A_1^i A_1^i + \overbar{A}_1^{\overline{i}} \overbar{A}_1^{\overline{i}} \right) \, , \qquad P_A = \frac{1}{2} \left( A_1^i A_1^i - \overbar{A}_1^{\overline{i}} \overbar{A}_1^{\overline{i}} \right) \, .
\end{align}
If one integrates out the auxiliary fields $\lambda^{\alpha i}$ and $\overbar{\lambda}^{\alpha \overline{i}}$ in the action (\ref{equivalent_interacting}), these fields simply act as Lagrange multipliers which set $A_\alpha^i = \partial_\alpha \phi^i$ and $\overbar{A}_\alpha^{\overline{i}} = \partial_\alpha \overbar{\phi}^{\overline{i}}$, and the action then reduces to (\ref{interacting_start_selfduality}). 

However, suppose that we wish to proceed in the opposite direction, instead integrating out the fields $A_\alpha^i$ and $\overbar{A}_\alpha^{\overline{i}}$. To do this, we vary the action with respect to the fields $A_\alpha^i$ and $\overbar{A}_\alpha^{\overline{i}}$ to obtain their equations of motion, whose solutions take the form
\begin{align}\label{A_eom_soln}
    A_0^i &= - \lambda^{1 i} - 2 ( V_{S_A} + V_{P_A} ) \lambda^{0 i}  \, , \qquad A_1^i = - \lambda^{0 i} \, , \nonumber \\
    \overbar{A}_0^{\overline{i}} &= - \overbar{\lambda}^{1 \overline{i}} + 2 ( V_{S_A} - V_{P_A} ) \overbar{\lambda}^{0 \overline{i}} \, , \qquad \overbar{A}_1^{\overline{i}} = - \overbar{\lambda}^{0 \overline{i}}   \, .
\end{align}
Integrating out $A_\alpha^i$ and $\overbar{A}_\alpha^{\overline{i}}$ by replacing them with their on-shell values (\ref{A_eom_soln}) then gives
\begin{align}
    S = \int d^2 x \, \left( \frac{1}{2} \left( \overbar{\lambda}^{0 \overline{i}} \overbar{\lambda}^{1 \overline{i}} - \lambda^{0 i} \lambda^{1 i} \right) - V ( S_\lambda, P_\lambda ) + \frac{1}{2} \left( \phi^i \partial_\alpha \lambda^{\alpha i} - \overbar{\phi}^{\overline{i}} \partial_\alpha \overbar{\lambda}^{\alpha \overline{i}} \right) \right)  \, ,
\end{align}
 where we have integrated by parts to move the derivatives on the final two terms, and where now $S_\lambda$ and $P_\lambda$ are defined as
\begin{align}\label{S_lambda_P_lambda}
    S_\lambda = \frac{1}{2} \left( \lambda^{0 i} \lambda^{0 i} + \overbar{\lambda}^{0 \overline{i}} \overbar{\lambda}^{0 \overline{i}} \right) \, , \qquad P_\lambda = \frac{1}{2} \left( \lambda^{0 i} \lambda^{0 i} - \overbar{\lambda}^{0 \overline{i}} \overbar{\lambda}^{0 \overline{i}} \right) \, .
\end{align}
Note that (\ref{S_lambda_P_lambda}) involve the \emph{time} components of the $\lambda$ fields, rather than the spatial components. We see that the fields $\phi^i$ and $\overbar{\phi}^{\overline{i}}$ act as Lagrange multipliers to enforce the constraints
\begin{align}
    \partial_\alpha \lambda^{\alpha i} = 0 = \partial_\alpha \overbar{\lambda}^{\alpha \overline{i}} \, , 
\end{align}
which admit the general solutions
\begin{align}
    \lambda^{\alpha i} = \epsilon^{\alpha \beta} \partial_\beta \psi^i \, , \qquad \overbar{\lambda}^{\alpha \overline{i}} = \epsilon^{\alpha \beta} \partial_\beta \overbar{\psi}^{\overline{i}} \, ,
\end{align}
for some scalar fields $\psi^i$, $\overbar{\psi}^{\overline{i}}$. Here we use the conventions $\epsilon^{0 1} = 1$, so
\begin{align}\label{lambda_to_psi}
    \lambda^{0 i} = \partial_x \psi^i = \psi^{\prime i} \, , \quad \lambda^{1 i} = - \partial_t \psi^i = - \dot{\psi}^i \, , \quad \overbar{\lambda}^{0 \overline{i}} = \partial_x \overbar{\psi}^{\overline{i}} = \overbar{\psi}^{\prime \overline{i}} \, , \quad \overbar{\lambda}^{1 \overline{i}} = - \partial_t \overbar{\psi}^{\overline{i}} = - \dot{\overbar{\psi}}{}^{\overline{i}} \, .
\end{align}
After integrating out $\phi^i$ and $\overbar{\phi}^{\overline{i}}$ and replacing $\lambda^{\alpha i}$, $\overbar{\lambda}^{\alpha \overline{i}}$ in favor of $\psi^i$, $\overbar{\psi}^{\overline{i}}$, we arrive at the dual form of the action
\begin{align}\label{psi_action}
    S = \int d^2x \left( \frac{1}{2} (\dot{\psi}^i \psi^{\prime \, i} - \dot{\overbar{\psi}} {}^{\overline{i}} \, \overbar{\psi}^{\prime \, \overline{i}}  ) - V ( S_\psi , P_\psi ) \right) \, ,
\end{align}
where, according to the map in equation (\ref{lambda_to_psi}), the dualization has replaced time components with space components in the definition of the $S$ and $P$ variables,
\begin{align}\label{S_and_P_invariants_psi}
    S_\psi = \frac{1}{2} \left( \psi^{\prime \, i} \psi^{\prime \, i} + \overbar{\psi}^{\prime \, \overline{i}} \overbar{\psi}^{\prime \, \overline{i}} \right) \, , \qquad P_\psi = \frac{1}{2} \left( \psi^{\prime \, i} \psi^{\prime \, i} - \overbar{\psi}^{\prime \, \overline{i}} \overbar{\psi}^{\prime \, \overline{i}} \right) \, .
\end{align}
The result (\ref{psi_action}) is in fact identical to our starting point (\ref{interacting_start_selfduality}). Therefore, any interacting chiral boson theory is ``self-dual'' in the sense that the theory is left invariant under the process of introducing auxiliary fields and then integrating out to express the theory in terms of the ``dual'' $\psi$ variables rather than the original $\phi$ variables.

Versions of this simple argument are well-known in various contexts. The observation that the standard Floreanini-Jackiw action with $V ( S, P ) = S$ exhibits this self-duality appeared in \cite{Miao:1999pr}, which we have simply generalized to the interacting case. Similar manipulations also appear, for instance, when discussing T-duality in string theory from the worldsheet point of view.

However, we would like to emphasize two aspects of this observation. The first is that, unlike Lorentz invariance -- which only holds for interaction functions which satisfy the differential equation (\ref{lorentz_pde_final}) -- this self-duality holds for \emph{any} system of interacting chiral bosons, regardless of the form of $V ( S, P )$. We will therefore take the view that the property of self-duality should be part of the \emph{definition} of a theory of chiral bosons. Since we have seen that any chiral boson theory enjoys self-duality in the sense described here when presented in the Floreanini-Jackiw formulation, we will demand that any other description of chiral bosons should also have a corresponding self-duality property. That is, we will take self-duality as a necessary condition for a theory to describe chiral degrees of freedom.

The second observation is that, if one rewrites the action (\ref{equivalent_interacting}) as
\begin{align}
    S &= \int d^2x \left( \mathcal{L}_A + \frac{1}{2} \lambda^{\alpha i} ( A_\alpha^i - \partial_\alpha \phi^i ) - \frac{1}{2} \overbar{\lambda}^{\alpha \overline{i}} ( \overbar{A}_\alpha^{\overline{i}} - \partial_\alpha \overbar{\phi}^{\overline{i}} ) \right) \, , \nonumber \\
    \mathcal{L}_A &= \frac{1}{2} \left( A_0^i A_1^i - \overbar{A}_0^{\overline{i}} \overbar{A}_1^{\overline{i}} \right) - V ( S_A , P_A ) \, , 
\end{align}
then the equations of motion for the fields $A_\alpha^i$ and $\overbar{A}_\alpha^{\overline{i}}$ are
\begin{align}\label{generalized_legendre}
    \lambda^{\alpha i} = - 2 \frac{\partial \mathcal{L}_A}{\partial A_\alpha^i} \, , \qquad \overbar{\lambda}^{\alpha \overline{i}} = 2 \frac{\partial \mathcal{L}_A}{\partial \Abar_\alpha^{\overline{i}}} \, .
\end{align}
Therefore, in a sense, one can think of the fields $\lambda$, $\overbar{\lambda}$ as the duals (or conjugates) of the fields $A$ and $- \overbar{A}$. Since the fields $A_\alpha^i = \partial_\alpha \phi^i$ and $\overbar{A}_\alpha^{\overline{i}} = \partial_\alpha \overbar{\phi}^{\overline{i}}$ are given by derivatives of a scalar field on-shell, one can also view the relations (\ref{generalized_legendre}) as a sort of Legendre transform. From this perspective, the self-duality of chiral boson models is the statement that such theories are invariant under a Legendre transform, or that one is free to rotate the fields $A_\alpha$, $\overbar{A}_\alpha$ into their duals $\lambda_\alpha$ and $- \overbar{\lambda}_\alpha$. This is very similar to the structure of theories of duality-invariant nonlinear electrodynamics in four dimensions, which are invariant under rotations mixing the field strength $F_{\mu \nu}$ with a certain dual field strength tensor $G_{\mu \nu}$. We will review this structure in more detail around equation (\ref{duality_rotations}) in the next section.

\section{Deformations of Dual Chern-Simons Theories}\label{sec:cs}

The chiral boson theories that we have considered in section \ref{sec:classical} often arise as the edge modes, or boundary duals, associated with the dynamics of Chern-Simons gauge fields in $3d$ bulk theories \cite{Witten:1996hc,ELITZUR1989108,Belov:2006jd}. For instance, physical descriptions of a quantum Hall droplet often involve a gauge field defined on a disk whose circular boundary supports edge modes modeled by chiral bosons \cite{PhysRevB.25.2185,PhysRevLett.64.220,PhysRevLett.64.216}.
Another example is found in $\mathrm{AdS}_3$ holography, where a collection of $U(1)$ Chern-Simons gauge fields in the bulk are dual to a corresponding collection of chiral currents in the $2d$ boundary. The addition of such bulk Chern-Simons terms to the action for $\mathrm{AdS}_3$ gravity allows BTZ black hole solutions to carry $U(1)$ charges \cite{Kraus:2006nb,Kraus:2006wn,Moussa:2008sj}.

In this section, we will show how stress tensor deformations of $2d$ chiral boson theories can be interpreted from the perspective of $3d$ bulk Chern-Simons gauge theories. We will begin by making some preliminary observations about the behavior of such $3d$ Chern-Simons theories in the presence of general boundary terms.

\subsection{$U(1)$ Chern-Simons theories with general boundary terms}\label{sec:cs_bdry}

Throughout this section, we will consider gauge theories defined on a bulk spacetime manifold $\mathcal{M}_3$ with boundary $\partial \mathcal{M}_3$. We will not specify whether $\partial \mathcal{M}_3$ is a true physical boundary or a conformal boundary, since our results apply uniformly in both cases.

Let us give a concrete example for each of these two cases to keep in mind as applications. In the former case, with a physical boundary, an example is furnished by the spacetime manifold $\mathcal{M}_3 = \mathcal{H}_2^+ \times \mathbb{R}_t$, where
\begin{align}
    \mathcal{H}_2^+ = \left\{ ( x, y ) \mid x, y \in \mathbb{R} \, , \, y \geq 0 \right\}
\end{align}
is the upper half-plane, viewed as a spatial manifold, and the factor of $\mathbb{R}_t$ represents a non-compact time direction. In this case, the boundary is $\partial \mathcal{M}_3 = \mathbb{R}_x \times \mathbb{R}_t$, where $\mathbb{R}_x$ is the spatial boundary $\partial \mathcal{H}_2^+ = \left\{ ( x, 0 ) \mid x \in \mathbb{R} \right\}$ and $\mathbb{R}_t$ is again the time direction.

An example of the latter case, with a conformal boundary, is a three-dimensional negatively curved bulk manifold $\mathcal{M}_3$, which asymptotically approaches an $\mathrm{AdS}_3$ spacetime that is characterized by a length scale $\ell_{\text{AdS}_3}$. The metric on $\mathcal{M}_3$ plays almost no role in this example, since the bulk Chern-Simons action is topological, but it is convenient to use the structure of the metric to characterize the conformal boundary $\partial \mathcal{M}_3$. The most general asymptotically $\mathrm{AdS}_3$ metric can be written in the form of a Fefferman-Graham expansion
\begin{align}\label{fg_expansion}
    ds^2 = \frac{\ell_{\text{AdS}_3}^2}{4 \rho^2} \, d \rho^2 + \left( \frac{g_{\alpha \beta}^{(0)} ( x^\gamma ) }{\rho} + g_{\alpha \beta}^{(2)} ( x^\gamma ) + \rho g_{\alpha \beta}^{(4)} ( x^\gamma ) \right) \, dx^\alpha \, d x^\beta \, .
\end{align}
The important point about this asymptotic form is that it determines a conformal boundary $\partial \mathcal{M}_3$ for our spacetime, located near $\rho = 0$, which has a boundary metric $g_{\alpha \beta}^{(0)} ( x^\gamma )$ determined by the leading term in the expansion (\ref{fg_expansion}). Here $\rho$ has the interpretation of a bulk radial coordinate whereas $x^\alpha$ label the two coordinates on the conformal boundary.

From now onwards, we will not distinguish between the two qualitatively different cases above, using the notation $\partial \mathcal{M}_3$ for either type of boundary. We will describe the $2d$ boundary in Euclidean signature using coordinates $x^\alpha = (w, \overbar{w})$ and the flat metric
\begin{align}
    ds^2 = g_{\alpha \beta} \, dx^\alpha \, dx^\beta = dw \, d \overbar{w} \, .
\end{align}
Although this signature and coordinate choice differs from the ones used in section \ref{sec:classical}, it allows for easier comparison with the holographic analysis of the root-$\TT$ deformation in \cite{Ebert:2023tih}. We will also use the convention that
\begin{align}
    \sqrt{ g } = \sqrt{ \det \left( \begin{bmatrix} 0 & \frac{1}{2} \\ \frac{1}{2} & 0 \end{bmatrix} \right) } = \frac{i}{2} \, , 
\end{align}
which will introduce some unfamiliar factors of $i$ in various places.

Our primary interest is to study the dynamics of Abelian gauge fields defined on the bulk manifold $\mathcal{M}_3$. Consider a collection of $U(1)$ gauge fields $A_i$, $i = 1 , \ldots , N$, and $\overbar{A}_{\overline{i}}$, $\overline{i} = 1 , \ldots , \overbar{N}$. Of course, the standard kinetic term for such gauge fields is the Maxwell term $F_{\alpha \beta}^i F^{\alpha \beta}_i$ where $F^i = d A^i$ is the field strength associated with the gauge field $F^i$. However, as we are in three spacetime dimensions, it is also possible to write down a Chern-Simons term which takes the form $A_i \wedge d A^i$ for the gauge fields $A^i$. The Maxwell term involves two derivatives and two factors of $A_i$, whereas the Chern-Simons term has only a single derivative and two factors of $A_i$. Therefore, by power counting, we see that the infrared behavior of the theory will be dominated by the Chern-Simons terms.

This motivates us to study the gauge theory with purely Chern-Simons couplings for the gauge fields $A_i$ and $\overbar{A}_{\overline{i}}$, which we parameterize as
\begin{align}\label{eq:EuclideanU(1)AdS3}
    I_{\text{CS}} = \frac{i}{8 \pi} \int \left( k^{ij} A_i \wedge dA_j - \overbar{k}^{\overline{i} \overline{j}} \overbar{A}_{\overline{i}} \wedge d \overbar{A}_{\overline{j}} \right) \, , 
\end{align}
where $k^{ij}$ and $\overbar{k}^{\overline{i} \overline{j}}$ are constant matrices which we assume are symmetric and have positive eigenvalues.\footnote{Throughout this section we will use the symbol $I$ rather than $S$ for actions to emphasize that we are in Euclidean signature.} These matrices will play the role of the metrics $G_{ij}$ and $\overbar{G}_{\overline{i} \overline{j}}$ of section \ref{sec:classical}. 

In addition to the Chern-Simons term (\ref{eq:EuclideanU(1)AdS3}), one can add a boundary term of the form
\begin{align}\label{boundary_term}
    I_{\text{bdry}} = - \frac{1}{8 \pi} \int_{\partial \mathcal{M}_3} d^2 x \, \sqrt{g} \, \mathcal{L}_{\text{bdry}} \left( A_{i \alpha} , \overbar{A}_{\overline{i} \alpha} \right) \, ,
\end{align}
where $\mathcal{L}_{\text{bdry}}$ is a Lorentz scalar constructed from the quantities $A_{i \alpha}$, $\overbar{A}_{\overline{i} \alpha}$, which are the restrictions of the three-dimensional gauge fields to the boundary $\partial \mathcal{M}_3$. The full description of the theory is then given by the combined action
\begin{align}
    I = I_{\text{CS}} + I_{\text{bdry}} \, .
\end{align}
The standard choice of boundary term is the one which corresponds to the free interaction function $V ( \phi , \overbar{\phi} )$ given in equation (\ref{free_interaction}), and is written as
\begin{equation}\label{free_CS_bdry}
    I_{\text{bdry}} = - \frac{1}{16 \pi} \int_{\partial \mathcal{M}_3} d^2 x \, \sqrt{g} \, g^{\alpha \beta} \left( k^{ij} A_{i \alpha} A_{j \beta} + \overbar{k}^{\overline{i} \overline{j}} \overbar{A}_{\overline{i} \alpha} \overbar{A}_{\overline{j} \beta} \right) \, .
\end{equation}
However, in this section we will be interested in studying more general choices of boundary term, especially those which arise by deformations of the conventional boundary term (\ref{free_CS_bdry}).

It may seem strange that one can write down a general boundary term (\ref{boundary_term}) which is an arbitrary function of the variables $A_{i \alpha}$ and $\Abar_{\overline{i} \alpha}$, or after assuming Lorentz invariance and $O ( N ) \times O ( \overbar{N} )$ symmetry under rotations of the gauge fields, an arbitrary function of the two combinations
\begin{align}
    S = \frac{1}{2} \left( k^{ij} A_i^\alpha A^{j}_{\alpha} + \overbar{k}^{\overline{i} \overline{j}} \overbar{A}_{\overline{i}}^\alpha \overbar{A}^{\overline{j}}_{\alpha} \right) \, , \qquad P = \frac{1}{2} \left(  k^{ij}  A_i^\alpha A^{j}_{\alpha} - \overbar{k}^{\overline{i} \overline{j}} \overbar{A}_{\overline{i}}^\alpha \overbar{A}^{\overline{j}}_{\alpha} \right) \, .
\end{align}
Any such boundary term $\mathcal{L}_{\text{bdry}} ( S , P )$ is manifestly compatible with boundary Lorentz invariance. This is in contrast with the analysis of section \ref{sec:classical}, where only interaction functions $V ( S, P )$ which obey the differential equation (\ref{lorentz_pde_final}) yield Lorentz-invariant theories.

The resolution to this tension is that the Floreanini-Jackiw and Chern-Simons descriptions of Lorentz-invariant chiral boson theories make different aspects of the models manifest. In the Floreanini-Jackiw description of section \ref{sec:classical}, it is manifest that the bosons $\phi^i$ are chiral since the theory is automatically self-dual (which we take as part of the definition of chirality) as we saw in section \ref{sec:sd_and_chiral}. However, it is not manifest that the Floreanini-Jackiw equations of motion respect Lorentz invariance, and requiring boost symmetry imposes a condition on $V ( S, P )$. Conversely, in the Chern-Simons description, it is manifest that the boundary theory enjoys Lorentz invariance since $\mathcal{L}_{\text{bdry}}$ is a Lorentz scalar. However, it is not manifest that the theory describes \emph{chiral} edge modes, which in particular requires that the theory be self-dual under the appropriate notion of duality transformation. Demanding chirality, or self-duality, will yield a constraint on $\mathcal{L}_{\text{bdry}}$, to be given in equation (\ref{CS_chirality_constraint}).

An analogy with electrodynamics is apt. Suppose that one wishes to describe a theory of an Abelian gauge field $A_\mu$ in four spacetime dimensions, whose Lagrangian $\mathcal{L}$ depends on the field strength $F_{\mu \nu}$ but not its derivatives. We assume that the equations of motion of this theory are invariant under both Lorentz transformations and under electric-magnetic duality rotations $\delta_\theta$ which act as
\begin{align}\label{duality_rotations}
    \delta_\theta F_{\mu \nu} = \theta G_{\mu \nu} ( F ) \, ,
\end{align}
where $G_{\mu\nu} = - \frac{1}{2} \varepsilon_{\mu\nu\rho\tau} \widetilde{G}^{\rho\tau}$ is the Hodge dual of $\widetilde{G}_{\mu \nu}$, which is itself defined as
\begin{align}
\widetilde{G}_{\mu \nu} = 2 \frac{\partial \mathcal{L}}{\partial F^{\mu \nu}} \, .
\end{align}
One option for describing such a theory is by giving the Lagrangian $\mathcal{L}$ itself. As the Lagrangian is a Lorentz scalar, this description makes Lorentz invariance manifest. However, invariance under duality rotations (\ref{duality_rotations}) is not automatic, and requires that the Lagrangian satisfy the differential equation (\ref{duality_pde}). 

Another option is to describe the theory in terms of its Hamiltonian $\mathcal{H} ( \vec{D}, \vec{B} )$, where $\vec{D} = \frac{\partial \mathcal{L}}{\partial \vec{E}}$ is the electric displacement. In these variables, the duality transformation (\ref{duality_rotations}) acts as an $SO(2)$ rotation which mixes the vectors $\vec{D}$ and $\vec{B}$. The most general duality-invariant Hamiltonian can be written as a function of the two variables
\begin{align}
    s = \frac{1}{2} \left( | \vec{D} |^2 + | \vec{B} |^2 \right) \, , \qquad p = | \vec{D} \times \vec{B} | \, .
\end{align}
These quantities $s$ and $p$ are invariant under $SO(3)$ rotations of the spatial coordinates and under duality rotations, so any Hamiltonian $\mathcal{H} ( s, p )$ is manifestly duality-invariant. However, because the canonical formulation has singled out a time direction as special, Lorentz invariance is no longer manifest. Imposing boost symmetry requires that the Hamiltonian satisfy the differential equation
\begin{align}
    \mathcal{H}_s^2 + \frac{2 s}{p} \mathcal{H}_s \mathcal{H}_p + \mathcal{H}_p^2 = 1 \, .
\end{align}
The upshot is that, in the electrodynamics example, either Lorentz invariance or duality invariance can be made manifest, and then imposing a partial differential equation will ensure that the remaining non-manifest symmetry will be respected.

In the chiral boson version of this story, the Floreanini-Jackiw formulation is analogous to the Hamiltonian presentation of $4d$ duality-invariant electrodynamics, since any theory of Floreanini-Jackiw bosons is automatically self-dual although Lorentz invariance is not manifest. The Chern-Simons presentation, on the other hand, is analogous to the Lagrangian description, since Lorentz invariance is manifest but self-duality is not guaranteed.

To understand the condition which must be imposed upon the Chern-Simons boundary term to ensure self-duality, which is the subject of section \ref{sec:cs_self_duality}, it will first be useful to study the currents obtained from varying the boundary gauge fields.

\subsubsection*{\ul{\it Boundary currents}}

Quite generically, we expect that gauge fields couple to conserved currents. In the case of the $3d$ Chern-Simons theory, although we have not coupled the bulk gauge fields to any sources in $\mathcal{M}_3$, the variation of the on-shell action localizes to a boundary term, so we can therefore define boundary currents that live in $\partial \mathcal{M}_3$. We normalize these currents as
\begin{align}\label{currents_defn}
    J^\alpha_i = - \frac{2 \pi i}{\sqrt{g}} \frac{\delta I}{\delta A_\alpha^i} \Big\vert_{\text{on-shell}} \, , \qquad \overbar{J}^\alpha_{\overline{i}} = - \frac{2 \pi i}{\sqrt{g}}  \frac{\delta I}{\delta \Abar_\alpha^{\overline{i}}} \Big\vert_{\text{on-shell}} \, .
\end{align}
We would like to compute these currents in a Chern-Simons theory with a general boundary term that is an arbitrary function of the $O(N) \times O ( \overbar{N} )$ invariant combinations $S$ and $P$. To do this, we consider a general variation of the action. The Chern-Simons term varies as
\begin{align}
    \delta I_{\text{CS}} &= \frac{i}{8 \pi} \int_{\mathcal{M}_3} \left( k^{ij} \left( \delta A_i \wedge d A_j + A_i \wedge d \delta A_j \right) - \overbar{k}^{\overline{i} \overline{j}} \left( \delta \Abar_{\overline{i}} \wedge d \Abar_{\overline{j}} + \Abar_{\overline{i}} \wedge d \delta \Abar_{\overline{j}} \right) \right) \nonumber \\
    &= \frac{i}{4 \pi} \int_{\mathcal{M}_3} \left( k^{ij}  \delta A_i \wedge d A_j - \overbar{k}^{\overline{i} \overline{j}} \delta \Abar_{\overline{i}} \wedge d \Abar_{\overline{j}}\right) - \frac{i}{8 \pi} \int_{\mathcal{M}_3} d \left( k^{ij} A_i \wedge \delta A_j - \overbar{k}^{\overline{i} \overline{j}} \Abar_{\overline{i}} \wedge \delta \Abar_{\overline{j}} \right) \, .
\end{align}
The first term vanishes after imposing the bulk equations of motion $d A_j = 0 = d \Abar_{\overline{j}}$, while the second term localizes to a boundary contribution,
\begin{align}
    \delta I_{\text{CS}} \Big\vert_{\text{on-shell}} = - \frac{i}{8 \pi} \int_{\partial \mathcal{M}_3} \left( k^{ij} A_\alpha^i \, \delta A_\beta^j - \overbar{k}^{\overline{i} \overline{j}} \Abar^{\overline{i}}_{\alpha} \, \delta \Abar^{\overline{j}}_{\beta} \right) \, dx^\alpha \wedge dx^\beta \, .
\end{align}
Since we are assuming that $\mathcal{L}_{\text{bdry}}$ takes the form
\begin{align}
    \mathcal{L}_{\text{bdry}} = f ( S, P ) \, ,
\end{align}
the variation of the boundary term can be written as
\begin{align}
    \delta I_{\text{bdry}} = - \frac{1}{8 \pi} \int_{\partial \mathcal{M}_3} \sqrt{g} \left( \left( f_S + f_P \right) k^{ij} A_i^\alpha \delta A_{j \alpha} + \left( f_S - f_P \right) \overbar{k}^{\overline{i} \overline{j}} \Abar_{\overline{i}}^\alpha \delta \Abar_{\overline{j} \alpha} \right) \, .
\end{align}
In coordinates $(w, \wb)$, after raising the indices using $A^w_i = 2 A_{\wb i}$ and $A^\wb_i = 2 A_{w i}$, the variation of the combined action is then
\begin{align}\label{CS_variation_total_action}\hspace{-10pt}
    \delta I \Big\vert_{\text{on-shell}} &= - \frac{i}{8 \pi} \int_{\partial \mathcal{M}_3} \left( k^{ij} \left( A_w^i \, \delta A_{\wb}^j - A_{\wb}^i \, \delta A_w^j  \right) - \overbar{k}^{\overline{i} \overline{j}} \left( \Abar^{\overline{i}}_{w} \, \delta \Abar^{\overline{j}}_{\wb} - \Abar^{\overline{i}}_{\wb} \, \delta \Abar^{\overline{j}}_{w} \right) \right) \nonumber \\
    &- \frac{1}{4 \pi} \int_{\partial \mathcal{M}_3} \sqrt{g} \left( k^{ij} \left( f_S + f_P \right) \left( A_{\wb}^{i} \delta A^{j}_{w} + A_{w}^{i} \delta A^{j}_{\wb}  \right) + \overbar{k}^{\overline{i} \overline{j}} \left( f_S - f_P \right) \left( \Abar^{\overline{i}}_{\wb} \delta \Abar^{\overline{j}}_{w} + \Abar^{\overline{i}}_{w} \delta \Abar^{\overline{j}}_{\wb} \right) \right) \, .
\end{align}
Using $\sqrt{g} = \frac{i}{2}$, we can therefore read off the currents (\ref{currents_defn}),
\begin{align}\label{currents_wwbar}
    J^w_i = \frac{i}{2} k^{ij} \left( f_S + f_P - 1 \right) A_{\wb}^j \, , \qquad J^{\wb}_i = \frac{i}{2} k^{ij} \left( f_S + f_P + 1 \right) A_w^j \, , \nonumber \\
    \overbar{J}^w_{\overline{i}} = \frac{i}{2} \overbar{k}^{\overline{i} \overline{j}} \left( f_S - f_P + 1 \right) \Abar_{\wb}^{\overline{j}} \, , \qquad \overbar{J}^{\wb}_{\overline{i}} = \frac{i}{2} \overbar{k}^{\overline{i} \overline{j}} \left( f_S - f_P - 1 \right) \Abar_w^{\overline{j}} \, .
\end{align}
These can also be written more covariantly as
\begin{align}
    J_{\alpha i} &= \frac{i}{4} k^{ij} \left( g_{\alpha \beta} ( f_S + f_P ) + \frac{1}{2} \epsilon_{\alpha \beta} \right) A^\beta_j \, , \nonumber \\
    \overbar{J}^\alpha_{\overline{i}} &= \frac{i}{4} \overbar{k}^{\overline{i} \overline{j}} \left( g_{\alpha \beta} ( f_S - f_P ) - \frac{1}{2} \epsilon_{\alpha \beta} \right) \overbar{A}^\beta_{\overline{j}} \, ,
\end{align}
which agrees with (\ref{currents_wwbar}) for $g_{w \wb} = \frac{1}{2}$, $\epsilon_{w \wb} = 1 = - \epsilon_{\wb w}$.

We note that variation of the total on-shell action has two qualitatively different contributions. The terms in the first line of (\ref{CS_variation_total_action}) are ``universal'' in the sense that they are present for any Chern-Simons theory and do not depend on the details of the boundary term $f ( S, P )$. These universal terms are also independent of the boundary metric, since they come from the integral of a $2$-form. In contrast, the terms on the second line of (\ref{CS_variation_total_action}) are ``model-dependent'' as they make explicit reference to the choice of boundary term $f ( S, P )$. Furthermore, these terms are metric-dependent and include an overall factor of $\sqrt{ g }$.

These two types of terms are analogous to those in the Lagrangian (\ref{general_class_vielbeins}) which couples a generic chiral boson theory to gravity. In that setting, the role of the ``universal'' and metric-independent contributions is played by the kinetic terms $G_{ij} \dot{\phi}^i \phi^{\prime \, j}$ and $\overbar{G}_{\overline{i} \overline{j}} \dot{\overbar{\phi}}{}^{\overline{i}} \overline{\phi}^{\prime j}$, which as we explained below equation (\ref{S_and_P_invariants_again}), do not include a factor of $E$. The Chern-Simons perspective gives another way to understand the metric-independence of these terms, since they may be viewed as the duals of contributions which arise from a topological bulk term. Similarly, the remaining metric-dependent and interaction-function-dependent terms in (\ref{general_class_vielbeins}) can be viewed as the analogs of the second line of (\ref{CS_variation_total_action}).

The expressions for the $J_i^\alpha$ and $\overbar{J}_{\overline{i}}^\alpha$ also determines the boundary conditions on the gauge fields which we impose in order to have a well-defined variational principle. In general, the on-shell variation of the action can be written as
\begin{align}\label{delta_I_os}
    \delta I \Big\vert_{\text{on-shell}} \sim \int_{\partial \mathcal{M}_3} \left( J_i^\alpha \, \delta A^{i}_{\alpha} + \overbar{J}_{\overline{i}}^{\alpha} \, \delta \Abar^{\overline{i}}_{\alpha} \right) \, .
\end{align}
We must ensure that the quantity (\ref{delta_I_os}) vanishes to have a good variational principle. To do this, we impose boundary conditions which hold fixed some particular combination of the boundary gauge fields $A^{i}_{\alpha}$ and $\Abar^{\overline{i}}_{\alpha}$. Schematically, this constraint takes the form
\begin{align}\label{F_bdry_conditions}
    F \left( A^{i}_{\alpha} \right) = 0 \, , \qquad \overbar{F} \left( \Abar^{\overline{i}}_{\alpha} \right) = 0 \, ,
\end{align}
where the precise form of the functions $F$ and $\overbar{F}$ depend on the case under consideration. In particular, this means that the allowed variations of the gauge fields must be constrained to satisfy the equations
\begin{align}\label{boundary_delta_A_constraints}
    \frac{\partial F}{\partial A^i_\alpha} \delta A^i_\alpha = 0 \, , \qquad \frac{\partial \overbar{F}}{\partial \Abar^{\overline{i}}_\alpha} \delta \Abar^{\overline{i}}_{\alpha} = 0 \, .
\end{align}
For instance, if both of the boundary variations $\delta A^i_w$ and $\delta A^i_{\wb}$ are non-zero, the constraints (\ref{boundary_delta_A_constraints}) can in principle be inverted to express one of these two boundary variations in terms of the other. This means that only one combination of the boundary gauge fields is free to fluctuate, while the other is held fixed. This is in agreement with the general expectation that imposing Dirichlet boundary conditions on \emph{both} components of the gauge field is too strong, and one would not find smooth solutions to the equations of motion for generic choices of the fixed gauge fields.

We also note that these boundary conditions will restrict the class of bulk gauge transformations that are permissible. A general gauge transformation $A^i \to A^i + d \Lambda^i$, $\overbar{A}^{\overline{i}} \to \overbar{A}^{\overline{i}} + d \overbar{\Lambda}^{\overline{i}}$ in the bulk leads to a variation of the Chern-Simons term which takes the form
\begin{align}
    \delta I_{\text{CS}} = \frac{i}{8 \pi} \int_{\partial \mathcal{M}_3} \left( k^{ij} d A^i \wedge \Lambda^j - \overbar{k}^{\overline{i} \overline{j}} d \overbar{A}^{\overline{i}} \wedge \overbar{\Lambda}^{\overline{j}} \right) \, ,
\end{align}
which, for general choices of the gauge parameters, will not be compatible with our choice of boundary conditions. We must therefore allow only a subclass of bulk gauge transformations which preserve the desired boundary conditions. Physically, one can think of this restriction as giving rise to physical degrees of freedom on the boundary.
% \footnote{Imposing a gauge symmetry means that some field configurations become identified as physically equivalent, which reduces the size of the phase space. Conversely, restricting the class of allowed gauge transformations means that there are \emph{fewer} identifications, so some previously-equivalent physical states are now inequivalent; it follows that the size of the physical phase space has increased, so new degrees of freedom have been added. Here, the new degrees of freedom are the boundary chiral bosons.}

To give a specific example illustrating the general observations above, let us consider the standard boundary term $f = S$. In this case, evaluating the currents (\ref{currents_wwbar}) gives
\begin{align}\label{holo_currents}
    J_{\wb}^{i} &= 0 \, , \quad J_{w}^i = \frac{i}{2} k^{ij} A^j_w \, , \quad \overbar{J}_{\wb}^{\overline{i}} = \frac{i}{2} \overbar{k}^{\overline{i} \overline{j}} \Abar^{\overline{j}}_{\wb} \, , \quad \overbar{J}_{w}^{\overline{i}} = 0 \, .
\end{align}
Therefore, with the conventional boundary term, the currents $J_\alpha^i$ are purely holomorphic and the currents $\overbar{J}^{\overline{i}}_\alpha$ are purely anti-holomorphic. The variation of the on-shell action is
\begin{align}\label{variation_std_term}
    \delta I \Big\vert_{\text{on-shell}} \sim \int_{\partial \mathcal{M}_3} \left( J^i_w \, \delta A^i_{\wb} + \overbar{J}^{\overline{i}}_{\wb} \, \delta \Abar^{\overline{i}}_{w} \right) \, .
\end{align}
The variation (\ref{variation_std_term}) vanishes if we require that $\delta A^i_{\wb} = 0$ and $\delta \Abar^{\overline{i}}_w = 0$, which is equivalent to imposing Dirichlet boundary conditions on the components $A^i_{\wb}$ and $\Abar^{\overline{i}}_w$ at the boundary $\partial \mathcal{M}_3$. For instance, one can demand that these components are both set to zero, which corresponds to the choice of functions $F$, $\overbar{F}$ in (\ref{F_bdry_conditions}) given by
\begin{align}
    F \left( A^{i}_{\alpha} \right) = A^i_{\wb} =  0 \, , \qquad \overbar{F} \left( \Abar^{\overline{i}}_{\alpha} \right) = \Abar^{\overline{i}}_{w} = 0 \, .
\end{align}
We must then allow only bulk gauge transformations which do not change the values of $A^i_{\wb}$ and $\Abar^{\overline{i}}_{w}$ on the boundary, and this restriction gives rise to boundary degrees of freedom. To see why these degrees of freedom are chiral, it is convenient to think of the holomorphic currents as $J_w^i = \partial \varphi^i$ and the anti-holomorphic currents as $\overbar{J}_{\wb}^{\overline{i}} = \overbar{\partial} \varphi^i$, where the $\varphi^i$ are $c = 1$ free bosons. Then it is clear that the $J_w^i$ play the role of the left-moving chiral half of a non-chiral boson, and the $\overbar{J}_{\wb}^{\overline{i}}$ act as the right-moving anti-chiral parts.

\subsection{Self-duality condition for Chern-Simons theories}\label{sec:cs_self_duality}

Let us now consider the question of self-duality for Chern-Simons theories. As we argued in section \ref{sec:sd_and_chiral}, self-duality should be viewed as a necessary condition to impose on the theory so that it describes chiral degrees of freedom. In the Floreanini-Jackiw description, self-duality meant that we could express the action either in terms of the original variables $A_\alpha^{i} = \partial_\alpha \phi^{i}$, or in terms of the dual variables $\lambda_\alpha^i = \epsilon_{\alpha \beta} \partial^\beta \psi^i$. 
The relationship between $A_\alpha$ and $\lambda_\alpha$, as expressed around equation (\ref{generalized_legendre}) is very similar to the relationship between the boundary Chern-Simons gauge field $A_\alpha$ and the corresponding current. Let us compare them side-by-side. In section \ref{sec:sd_and_chiral}, we had the relations
\begin{align}\label{lambda_to_J}
    \lambda^{\alpha i} = - 2 \frac{\partial \mathcal{L}}{\partial A_\alpha^i} \, , \qquad \overbar{\lambda}^{\alpha \overline{i}} = 2 \frac{\partial \mathcal{L}}{\partial \overbar{A}_\alpha^{\overline{i}}} \, ,
\end{align}
where in this formula the symbol $A_\alpha$ refers to the vector field appearing in the action (\ref{equivalent_interacting}). In the Chern-Simons setting, we instead have the schematic relations 
\begin{align}\label{J_CS_analogy}
    J^{\alpha i} &= - \frac{2 \pi i}{\sqrt{g}} \frac{\delta I}{\delta A_\alpha^i} \Big\vert_{\text{on-shell}} = - \frac{2 \pi i}{\sqrt{g}} \frac{\partial \mathcal{L}_{\text{on-shell}}}{\partial A_\alpha^i} \, , \nonumber \\
    \overbar{J}^{\alpha \overline{i}} &=  - \frac{2 \pi i}{\sqrt{g}}  \frac{\delta I}{\delta \Abar_\alpha^{\overline{i}}} \Big\vert_{\text{on-shell}} = - \frac{2 \pi i}{\sqrt{g}}  \frac{\partial \mathcal{L}_{\text{on-shell}}}{\partial \overbar{A}_\alpha^{\overline{i}}} \, ,
\end{align}
where now the symbol $A_\alpha$ refers to the boundary Chern-Simons gauge field.\footnote{In equation (\ref{J_CS_analogy}), the partial derivatives of the Lagrangian $\mathcal{L}_{\text{on-shell}}$ are understood to be defined as the integrands of corresponding variations of the on-shell action in the middle expression of each line.} Insofar as the gauge field acts as a good proxy for the gradient of the Floreanini-Jackiw bosons, this suggests that the role of the dual variable $\lambda_\alpha^i$ is now played by
\begin{align}
    \lambda_\alpha^i = - \frac{1}{2} J_\alpha^i \, , \qquad \overbar{\lambda}_\alpha^{\overline{i}} = \frac{1}{2} \overbar{J}_\alpha^{\overline{i}} \, , 
\end{align}
where the sign difference is due to the relative sign in (\ref{generalized_legendre}), which itself originates from the difference in signs between the kinetic terms for chiral and anti-chiral bosons.

This analogy leads us to propose a notion of self-duality for Chern-Simons theories. We will phrase this condition via an infinitesimal transformation, rather than a finite one. That is, in section \ref{sec:sd_and_chiral}, the duality transformation was a $\mathbb{Z}_2$ action which replaced the fields $A_\alpha$ with the fields $\lambda_\alpha$. In the present context, we will instead propose a continuous transformation which infinitesimally rotates the fields $A_\alpha^i$, $\Abar_\alpha^{\overline{i}}$ into their duals $J_\alpha^i$, $- \overbar{J}_\alpha^{\overline{i}}$.

We say that a Chern-Simons theory with boundary term $f ( S, P )$ is \emph{self-dual} if the on-shell variation of the action identically vanishes under the transformation
\begin{align}\label{cs_sd_trans}
    \delta A_\alpha^i = \epsilon J_\alpha^i \, , \qquad \delta \overbar{A}_\alpha^{\overline{i}} = - \epsilon \overbar{J}_\alpha^{\overline{i}} \, .
\end{align}
To see why this is the right notion of self-duality, let us find the condition on the boundary term $f ( S, P )$ under which the transformation (\ref{cs_sd_trans}) is a symmetry. By equation (\ref{delta_I_os}), under this variation the change in the on-shell action is
\begin{align}
    \delta I \Big\vert_{\text{on-shell}} &\sim \int_{\partial \mathcal{M}_3} \left( J_i^\alpha \, \delta A^{i}_{\alpha} + \overbar{J}_{\overline{i}}^{\alpha} \, \delta \Abar^{\overline{i}}_{\alpha} \right) \, \nonumber \\
    &= \epsilon \int_{\partial \mathcal{M}_3} \left( J_i^\alpha \, J^{i}_{\alpha} - \overbar{J}_{\overline{i}}^{\alpha} \, \overbar{J}^{\overline{i}}_{\alpha} \right) \, ,
\end{align}
so the rotation (\ref{cs_sd_trans}) is a symmetry if and only if
\begin{align}\label{JJ_is_JbJb}
    J^i_w J^i_{\wb} - \overbar{J}^{\overline{i}}_w \overbar{J}^{\overline{i}}_\wb = 0 \, .
\end{align}
Using the general expression (\ref{currents_wwbar}) for the currents, and expressing the condition in terms of $S$ and $P$, we find that (\ref{JJ_is_JbJb}) is equivalent to the condition
\begin{align}\label{CS_chirality_constraint}
    f_S^2 + \frac{2 S}{P} f_S f_P + f_P^2 = 1 \, .
\end{align}
Remarkably, the Chern-Simons boundary term is self-dual if and only if it satisfies precisely the same differential equation (\ref{lorentz_pde_final}) which the Floreanini-Jackiw interaction function $V(S, P)$ must satisfy in order to guarantee Lorentz invariance. Because of the identical structure of the constraints on $f ( S , P)$ and $V(S,P)$, some of our observations from section \ref{sec:classical} can be immediately translated to analogous statements in the Chern-Simons setting.

For instance, if $\overbar{N} = 0$ so that the theory features only a collection of unbarred gauge fields $A_\alpha^i$ but no barred fields $\overbar{A}_{\alpha}^{\overline{i}}$, the two invariants collapse as $S = P$ and the only solution to the constraint (\ref{CS_chirality_constraint}) is $f ( S, P ) = S$. This is consistent with the comments around equation (\ref{only_S}) in the Floreanini-Jackiw formulation, namely that no Lorentz-invariant interactions are possible for a system of purely chiral (or purely anti-chiral) bosons. Here we are seeing the Chern-Simons counterpart of this statement: although we can write down any boundary term $f(S)$ that we like, and still respect boundary Lorentz invariance, only the choice $f(S) = S$ will respect chirality (or self-duality) of the boundary theory.

In the remainder of this section, we will view the differential equation (\ref{CS_chirality_constraint}) as a consistency condition which a boundary term $f ( S, P )$ must satisfy to describe chiral bosons. One can also understand this constraint as an analog of electric-magnetic duality invariance for $3d$ Chern-Simons theories. Of course, the conventional form of electric-magnetic duality is inapplicable for $3d$ gauge theories, since the Hodge dual of a two-form field strength $F_2$ in three spacetime dimensions is a one-form, which is interpreted as the field strength of a dual scalar rather than a dual $1$-form. However, demanding invariance under the duality rotation (\ref{cs_sd_trans}) is closely related to imposing invariance under the rotations (\ref{duality_rotations}); in both cases, the symmetry exchanges the field appearing in the Lagrangian with a certain ``dual'' that is defined via the derivative of the Lagrangian with respect to this field.

\subsubsection*{\ul{\it Linear and non-linear self-duality constraints for currents}}

One typically describes a free chiral $p$-form field in $2p$ dimensions, where $p$ is odd, as a form which satisfies a linear Hodge self-duality constraint. For instance, a free chiral $3$-form field $F_3$ in six dimensions obeys $\ast F_3 = F_3$. Likewise, the Floreanini-Jackiw bosons $\phi^i$, $\overbar{\phi}^{\overline{i}}$ with free interaction function $V ( S, P ) = S$ are self-dual and anti-self-dual, respectively. Introducing interactions for such $p$-forms then modifies this constraint to a non-linear self-duality condition, which can be viewed as determining the self-dual part of the $p$-form as a function of the anti-self-dual part, or vice-versa.

We would now like to see how these self-duality constraints can be understood from the Chern-Simons description of chiral bosons. Since we are working in a two-dimensional Euclidean spacetime, the appropriate self-duality conditions for a one-form are \emph{imaginary} self-duality or anti-self-duality, since the definition of the Hodge star,
\begin{align}
    \left( \ast V \right)_\beta = \sqrt{ g } V^\alpha \epsilon_{\alpha \beta} \, ,
\end{align}
includes a factor of $\frac{i}{2}$ from the measure $\sqrt{ g }$. With these conventions, the dual of a general one-form $V_\alpha$ with components $V_w$, $V_\wb$ is
\begin{align}
    \left( \ast V \right)_\alpha = \left( - i V_w , i V_\wb \right) \, .
\end{align}
Thus a holomorphic one-form $V_\alpha = ( V_w, 0 )$ obeys an imaginary anti-self-duality condition
\begin{align}
    \ast V = - i V \, , 
\end{align}
whereas a purely anti-holomorphic one-form $V_\alpha = ( 0, V_\wb )$ is imaginary-self-dual,
\begin{align}
    \ast V = i V \, .
\end{align}
We therefore see that all of the currents $J_i$ and $\overbar{J}_i$ of equation (\ref{holo_currents}), which correspond to the standard boundary term $f(S,P) = S$,  satisfy $\ast J_i = - i J_i$ and $\ast \overbar{J}_i = + i \overbar{J}_i$. This can also be expressed by defining the projectors onto imaginary-self-dual and imaginary-anti-self-dual parts of a one-form,
\begin{align}
    P_{\pm} = \frac{1}{2} \left( 1 \mp i \ast \right) \, .
\end{align}
In terms of these projectors, the fact that the $J^\alpha_i$ are purely holomorphic can be expressed as $P_- J^\alpha_i = J^\alpha_i$, and the fact that the $\overbar{J}^{\alpha}_{\overline{i}}$ are purely anti-holomorphic is equivalent to the statement that $P_+ \overbar{J}^{\alpha}_{\overline{i}} = \overbar{J}^{\alpha}_{\overline{i}}$. Therefore, by adding the boundary term $f ( S, P ) = S$ to the Chern-Simons action, we obtain chiral currents which obey a \emph{linear} self-duality condition. This is the image of the usual statement that free chiral $p$ forms in $2p$ dimensions, for $p$ odd, obey linear self-duality constraints.

Next we would like to understand how a more general boundary term gives rise to a \emph{non-linear} self-duality constraint, which corresponds to an interacting system of boundary chiral bosons. In this case, rather than obeying the standard chirality constraints
\begin{align}
    P_- J_i^\alpha = J_i^\alpha \, , \qquad P_+ \overbar{J}_{\overline{i}}^\alpha = \overbar{J}_{\overline{i}}^\alpha \, ,
\end{align}
which correspond to (linear) Hodge imaginary-self-duality or imaginary-anti-self-duality,
\begin{align}\label{linear_sd}
    \ast J_i = - i J_i \, , \qquad \ast \overbar{J}_{\overline{i}} = i \overbar{J}_{\overline{i}} \, ,
\end{align}
the currents will satisfy more general, non-linear or twisted self-duality conditions, each characterized by an operator $\mathcal{T}^{(i)}$ or $\overbar{\mathcal{T}}^{(i)}$:
\begin{align}
    \ast J_i = \mathcal{T}^{(i)} J_i \, , \qquad \ast \overbar{J}_{\overline{i}} = \overbar{\mathcal{T}}^{(\overline{i})} \overbar{J}_{\overline{i}} \, .
\end{align}
In the case where $\mathcal{T}^{(i)} = - i \, \mathbb{I}$ and $\overbar{\mathcal{T}}^{(\overline{i})} = i \, \mathbb{I}$ are both proportional to the identity operator $\mathbb{I}$, this reduces to the standard chirality condition (\ref{linear_sd}). In the more general case we allow $\mathcal{T}^{(i)}$, $\overbar{\mathcal{T}}^{(i)}$ to be non-trivial operators which can depend on the fields.

Twisted self-dual boundary conditions characterized by operators of this form have been considered in \cite{Severa:2016prq,Arvanitakis:2022bnr}, primarily in the setting of non-Abelian Chern-Simons theories. In the Abelian case, which is the focus of this work, no non-trivial operator $\mathcal{T}$ exists for a system obeying (\ref{CS_chirality_constraint}) with either $N = 0$ or  $\overbar{N} = 0$ (i.e. a self-dual theory which only describes fields $\overbar{A}_{\overline{i}}^\alpha$ but no $A_i^\alpha$, or with only  $A_i^\alpha$  but none of the $\overbar{A}_{\overline{i}}^\alpha$, respectively). This is again related to the statement, which we have seen in section \ref{sec:lorentz}, that there are no possible Lorentz-invariant interactions for a system of purely chiral (or purely anti-chiral) bosons.\footnote{Alternatively, this is because there are no solutions to the self-duality equation (\ref{CS_chirality_constraint}) besides the trivial solution $f ( S, P ) = S$ when either $N = 0$ or $\overbar{N} = 0$.} However, in a theory which features both chiral and anti-chiral bosons -- or both $A_i^\alpha$ and $\overline{A}_{\overline{i}}^\alpha$, from the Chern-Simons perspective -- such Lorentz-invariant interactions are possible, which manifests as the existence of allowable operators $\mathcal{T}$ besides the identity.

It is easy to see that, for a general boundary term $\mathcal{L}_{\rm bdry} =f( S, P )$, the currents
\begin{align}
    J_w^i = \frac{i}{4} k^{ij} \left( f_S + f_P + 1 \right) A_w^j \, , \qquad J_{\wb}^i = \frac{i}{4} k^{ij} \left( f_S + f_P - 1 \right) A_\wb^j \, ,
\end{align}
satisfy the non-linear self-duality condition
\begin{align}
    \left( \ast J^i \right)_\alpha &= \tensor{\left( \mathcal{T}^{(i)} \right)}{_\alpha^\beta} J^i_\beta \, \nonumber , \\
    \mathcal{T}^{(i)} &= - i \begin{bmatrix} 1 & 0 \\ - \frac{2 \tensor{k}{^i_j} A_{\wb}^j }{\tensor{k}{^i_k} A_w^k} \frac{f_S + f_P - 1}{f_S + f_P + 1} & 1 \end{bmatrix} \, .
\end{align}
This expression gives the components of the matrix $\mathcal{T}^{(i)}$ with respect to its Lorentz indices $\alpha, \beta = w, \wb$, where $i$ is a fixed internal index. When $f_S = 1$ and $f_P = 0$, we see that $\mathcal{T}^{(i)}$ reduces to $-i \, \mathbb{I}$, which expresses the usual imaginary-anti-self-duality constraint. 

Similarly, the general currents
\begin{align}
    \overbar{J}^{\overline{i}}_w = \frac{i}{4} \overbar{k}^{\overline{i} \overline{j}} \left( f_S - f_P - 1 \right) \Abar_w^{\overline{j}} \, , \qquad \overbar{J}^{\overline{i}}_\wb = \frac{i}{4} \overbar{k}^{\overline{i} \overline{j}} \left( f_S - f_P + 1 \right) \Abar_{\wb}^{\overline{j}} \, ,
\end{align}
satisfy the non-linear self-duality condition
\begin{align}
    \left( \ast \overbar{J} \right)^{\overline{i}}_\alpha &= \tensor{\left( \overbar{\mathcal{T}}^{(\overline{i})} \right)}{_\alpha^\beta} \overbar{J}^{\overline{i}}_\beta \, , \nonumber \\
    \overbar{\mathcal{T}}^{(\overline{i})} &= i \begin{bmatrix} 1 & - \frac{2 \tensor{\overline{k}}{^{\overline{i}}_{\overline{j}}} \Abar_w^{\overline{j}} ( 1 + f_P - f_S ) }{\tensor{\overline{k}}{^{\overline{i}}_{\overline{k}}} \Abar_\wb^{\overline{k}} ( -1 + f_P - f_S ) } \\ 0 & 1 \end{bmatrix} \, .
\end{align}
Likewise, when $f_S = 1$ and $f_P = 0$, we see that $\overbar{\mathcal{T}}^{(\overline{i})} = i \, \mathbb{I}$ so this reduces to the usual imaginary-self-duality condition $\ast \overbar{J}^{\overline{i}} = i \overbar{J}^{\overline{i}}$.

We should point out that, in other studies of twisted self-duality in Chern-Simons theories such as \cite{Severa:2016prq,Arvanitakis:2022bnr}, the twisting operator $\mathcal{T}$ commutes with the Hodge star operation. As a result, acting with the Hodge star operator on each side of the twisted self-duality constraint $\ast J = \mathcal{T} J$, one has
\begin{align}
    \ast \ast J = \ast \mathcal{T} J = \mathcal{T} \ast J = \mathcal{T}^2 J \, .
\end{align}
Since the Hodge star is an anti-involution, $\ast \ast = - \mathbb{I}$, in two Euclidean dimensions, one therefore arrives at the constraint
\begin{align}\label{anti_involution}
    \mathcal{T}^2 = - \mathbb{I} \, .
\end{align}
In Lorentzian signature, this would instead give the constraint $\mathcal{T}^2 = \mathbb{I}$.

However, in our case the twisting operators $\mathcal{T}^{(i)}$ and $\overbar{\mathcal{T}}^{(\overline{i})}$ have non-trivial structure in their Lorentz indices and therefore do not commute with the Hodge star. This is why, in our case, these twisting operators do not satisfy an anti-involutive constraint like (\ref{anti_involution}).

One can now proceed as in the linear case and define projection operators
\begin{align}\label{four_projection_operators}
    P_+^{(i)} &= \begin{bmatrix} 0 & 0 \\ - \frac{\tensor{k}{^i_j} A_{\wb}^j }{\tensor{k}{^i_k} A_w^k} \frac{f_S + f_P - 1}{f_S + f_P + 1} & 1 \end{bmatrix} \, , \qquad \overbar{P}_{+}^{(\overline{i})} = \begin{bmatrix} 0 & \frac{ \tensor{\overline{k}}{^{\overline{i}}_{\overline{j}}} \Abar_w^{\overline{j}} ( 1 + f_P - f_S ) }{\tensor{\overline{k}}{^{\overline{i}}_{\overline{k}}} \Abar_\wb^{\overline{k}} ( -1 + f_P - f_S ) } \\ 0 & 1 \end{bmatrix}  \, , \nonumber \\
    P_-^{(i)} &=  \begin{bmatrix} 1 & 0 \\ \frac{\tensor{k}{^i_j} A_{\wb}^j }{\tensor{k}{^i_k} A_w^k} \frac{f_S + f_P - 1}{f_S + f_P + 1} & 0 \end{bmatrix} \, , \qquad \overbar{P}_-^{(\overline{i})} = \begin{bmatrix} 1 & - \frac{\tensor{\overline{k}}{^{\overline{i}}_{\overline{j}}} \Abar_w^{\overline{j}} ( 1 + f_P - f_S ) }{\tensor{\overline{k}}{^{\overline{i}}_{\overline{k}}} \Abar_\wb^{\overline{k}} ( -1 + f_P - f_S ) } \\ 0 & 0 \end{bmatrix} \, ,
\end{align}
which satisfy the expected properties of orthogonal projectors,
\begin{align}
    \left( P_\pm^{(i)} \right)^2 = P_\pm^{(i)}  \, , \quad \left( \overbar{P}_{\pm}^{(\overline{i})} \right)^2 = \overbar{P}_{\pm}^{(\overline{i})} \, , \quad P_\pm^{(i)} P_\mp^{(i)} = 0 = \overbar{P}_{\pm}^{(\overline{i})} \overbar{P}_{\mp}^{(\overline{i})}   \, , 
\end{align}
along with the chirality conditions
\begin{align}
    P_-^{(i)} J^i = J^i \, , \quad P_+^{(i)} J^i = 0 \, , \quad \overbar{P}_{+}^{(\overline{i})} \overbar{J}^{\overline{i}} = \overbar{J}^{\overline{i}} \, , \quad \overbar{P}_{-}^{(\overline{i})} \overbar{J}^{\overline{i}} = 0 \, .
\end{align}
Therefore, even in the interacting case, one can view the currents as satisfying an appropriate non-linear self-duality constraint. This expresses, in Chern-Simons language, the equations of motion (\ref{interacting_eom}) for interacting Floreanini-Jackiw bosons.

We should point out that this construction has now produced two separate pairs of projection operators $P_{\pm}^{(i)}$, $\overbar{P}_{\pm}^{(\overline{i})}$ for each fixed choice of indices $i, \overline{i}$, or equivalently, two separate twist operators $\mathcal{T}^{(i)}$ and $\overbar{\mathcal{T}}^{(i)}$. This is in contrast with the linear-self duality constraint, which is described by only two projectors $P_{\pm} = \frac{1}{2} \left( 1 \mp i \ast \right)$, where
\begin{align}\label{linear_projectors}
    P_+ = \overbar{P}_+ \, \qquad P_- = \overbar{P}_- \, .
\end{align}
In the linear case, there are relations that cause these four operators to collapse to just two independent projectors, and it is clear that these operators project onto one-dimensional eigenspaces which represent physically opposite chiralities.

In the non-linear case, there are also relations (albeit more complicated ones) between the two twist operators. For instance, one can see that $\mathcal{T}^{(i)}$ can be obtained from $\overbar{\mathcal{T}}^{(i)}$ by simultaneously transposing the matrix in its Lorentz indices and interchanging all barred and unbarred quantities. That is, one exchanges
\begin{align}
    k^{ij} \longleftrightarrow \overbar{k}^{\overline{i} \overline{j}} \, , \quad A^i \longleftrightarrow \overbar{A}^{\overline{i}} \, \quad w \longleftrightarrow \wb \, ,
\end{align}
which also has the effect of sending $P \to -P$ (and thus $f_P \to - f_P$). This relation holds regardless of the choice of boundary term. When the function $f ( S, P )$ satisfies the self-duality condition (\ref{CS_chirality_constraint}) necessary to describe chiral modes, there are further constraints between the twist operators. To see one such constraint, we can rewrite (\ref{CS_chirality_constraint}) as
\begin{align}
    J^i \wedge \ast J^i = \overbar{J}^{\overline{i}} \wedge \ast \overbar{J}^{\overline{i}} \, .
\end{align}
Since $\ast J^i = \mathcal{T}^{(i)} J^i$ and $\ast \overbar{J}^{\overline{i}} = \overbar{\mathcal{T}}^{(\overline{i})} \overbar{J}^{\overline{i}}$, this relation can also be expressed as
\begin{align}\label{wedge_constraint}
    J^i \wedge \mathcal{T}^{(i)} J^i =  \overbar{J}^{\overline{i}} \wedge \overbar{\mathcal{T}}^{(\overline{i})} \overbar{J}^{\overline{i}} \, .
\end{align}
Equation (\ref{wedge_constraint}) is a consequence of the fact that, when the boundary term obeys the self-duality constraint, the chiral and anti-chiral twist operators are ``compatible'' in a sense which generalizes the statements that $\mathcal{T}^{(i)} = - \overbar{\mathcal{T}}^{(\overline{i})}$, or that the projection operators satisfy (\ref{linear_projectors}), in the linear case.

\subsection{Current deformations of boundary terms}\label{sec:cs_defs}

We will now consider flow equations which modify the boundary term $\mathcal{L}_{\text{bdry}}$ of a bulk Chern-Simons theory.\footnote{Although we focus on $U(1)$ Chern-Simons theories in this work, stress tensor deformations of the boundary term for $SL(2) \times SL(2)$ Chern-Simons have been considered in \cite{Llabres:2019jtx,Ouyang:2020rpq,Ebert:2022ehb}.} In particular, we are interested in differential equations for $\mathcal{L}_{\text{bdry}}$ which are driven by conserved quantities. We will refer to any such flow equation as a ``current deformation'' regardless of whether the conserved currents driving the flow are the objects $J_\alpha^i$ and $\overbar{J}_\alpha^{\overline{i}}$ defined in equation (\ref{currents_defn}), or the energy-momentum tensor $T_{\alpha \beta}$, which is another type of conserved current in the theory.

Let us first study deformations which involve the spin-$1$ currents $J_\alpha^i$ and $\overbar{J}_\alpha^{\overline{i}}$. A general flow equation in this class takes the form
\begin{align}
    \frac{\partial \mathcal{L}_{\text{bdry}}}{\partial \lambda} = \mathcal{O} \left( J_\alpha^i , \overbar{J}_\alpha^{\overline{i}} , \lambda \right) \, ,
\end{align}
where $\mathcal{O}$ is a Lorentz scalar and $O ( N ) \times O ( \overbar{N} )$ singlet constructed from the currents. Within this class, there are fewer interesting possibilities. The most natural deformation to consider is to begin with the conventional boundary term $\mathcal{L}_{\text{bdry}} = S$ and deform by a marginal combination of the form
\begin{align}
    \mathcal{O} = k_{ij} J_\alpha^i J^{ \alpha j} \, , \; \text{ or } \quad \mathcal{O} = \overbar{k}_{\overline{i} \overline{j}} \overbar{J}_\alpha^{\overline{i}} \overbar{J}^{\alpha \overline{j}} \, .
\end{align}
However, by virtue of the chirality of the currents given in equation (\ref{holo_currents}), both of these operators vanish. One might instead construct a deforming operator which mixes the currents on the two sides, such as
\begin{align}\label{JJbar_defn}
    \mathcal{O} = C_{i \overline{j}} J_\alpha^i  \overbar{J}^{\alpha \overline{j}} \, ,
\end{align}
where $C_{i \overline{j}}$ is a constant tensor with mixed indices. For instance, in the case $N = \overbar{N}$, we do not need to distinguish between barred and unbarred indices, and can choose $C_{i \overline{j}} = \delta_{i \overline{j}} \equiv \delta_{ij}$.\footnote{Of course, when $N \neq \overbar{N}$, a deformation of this form does not preserve $O(N) \times O ( \overbar{N} )$ symmetry. For instance, a deformation by $\sum_{i=1}^{M} J^i_\alpha \overbar{J}^{i \alpha}$, where $M = \min ( N, \overbar{N} )$, treats the currents asymmetrically.} Let us consider the effect of this deformation with the simplifying assumption $k^{ij} = \overbar{k}^{ij} = \delta^{ij}$. In this case, at leading order in the deformation parameter, one finds a deformed boundary term
\begin{align}\label{JJ_deformed}
    \mathcal{L}_{\text{bdry}}^{(1)} = \frac{1}{2} \left( A_i^\alpha A^{i}_{\alpha} +  \overbar{A}_{i}^\alpha \overbar{A}^{i}_{\alpha} \right) + \lambda A_i^\alpha \overbar{A}^i_{\alpha} \, ,
\end{align}
up to the normalization of $\lambda$. That is, such an operator has introduced an off-diagonal mixing between the barred and unbarred gauge fields. Ignoring possible subtleties about quantization of the Chern-Simons levels, such a quadratic mixing can always be undone by performing a Bogoliubov-like field redefinition. Indeed, note that beginning with the undeformed boundary term
\begin{align}
    \mathcal{L}_{\text{bdry}}^{(0)} = \frac{1}{2} \left( A_i^\alpha A^{i}_{\alpha} +  \overbar{A}_{i}^\alpha \overbar{A}^{i}_{\alpha} \right)
\end{align}
and then performing a change of variables to
\begin{align}\label{CS_bogoliubov}
    A^i_\alpha = \cosh ( \mu ) B^i_\alpha + \sinh ( \mu ) \overbar{B}^i_\alpha \, , \qquad \overbar{A}^{i}_{\alpha}  = \cosh ( \mu ) \overbar{B}^i_\alpha + \sinh ( \mu ) B^i_\alpha\, ,
\end{align}
gives the transformed boundary term
\begin{align}
    \mathcal{L}_{\text{bdry}}^{(0)} = \cosh ( 2 \mu )  \left[ \frac{1}{2} \left( B_i^\alpha B^{i}_{\alpha} +  \overbar{B}_{i}^\alpha \overbar{B}^{i}_{\alpha} \right) + \tanh ( 2 \mu ) B_i^\alpha \overbar{B}^i_\alpha \right] \, .
\end{align}
Up to an overall rescaling, this is equivalent to the deformed boundary term (\ref{JJ_deformed}) if we identify $\tanh ( 2 \mu ) = \lambda$. Therefore, the marginal $J \overbar{J}$ deformation of equation (\ref{JJbar_defn}) can be viewed as inducing a rotation between the fields $A_\alpha^i$ and $\overbar{A}_\alpha^{\overline{i}}$. We will see later that the root-$\TT$ deformation, in the case $N = \overbar{N} = 1$, is qualitatively similar to this $J \overbar{J}$ deformation.

In principle, one could consider more general operators constructed from the currents $J$ and $\overbar{J}$, such as powers of the form $\mathcal{O} = \left( J_\alpha^i  \overbar{J}^{\alpha i} \right)^n$ or other structures such as $\mathcal{O} = \left( J_\alpha^i J_\beta^i  \overbar{J}^{\alpha \overline{j}}\overbar{J}^{\beta \overline{j}} \right)^m$, both of which preserve $O(N) \times O ( \overbar{N})$ symmetry. These operators are irrelevant for $n > 1$ and $m > \frac{1}{2}$, respectively. However, we will now instead turn our attention to deformations which are constructed from the energy-momentum tensor, 
\begin{align}
    \frac{\partial \mathcal{L}_{\text{bdry}}}{\partial \lambda} = \mathcal{O} \left( T_{\alpha \beta}^{(\lambda)} , \lambda \right) \, .
\end{align}
The first choice that one must make in defining such a flow is \emph{which} stress tensor to use. There are generally many definitions of the energy-momentum tensor which are all conserved but which differ by improvement transformations. One natural choice is the Hilbert stress tensor defined by varying the metric. Of course, neither the Chern-Simons action (\ref{eq:EuclideanU(1)AdS3}) nor the boundary action (\ref{boundary_term}) depend on the \emph{bulk} metric, but the term $I_{\text{bdry}}$ does depend on the boundary metric. One can therefore define a boundary stress tensor,
\begin{align}\label{hilbert_cs}
    T_{\alpha \beta} = - \frac{2}{\sqrt{g}} \frac{\delta I}{\delta g^{\alpha \beta}} = - \frac{2}{\sqrt{g}} \frac{\delta I_{\text{bdry}}}{\delta g^{\alpha \beta}} \, .
\end{align}
However, this stress tensor is qualitatively different from the one obtained in equation (\ref{general_stress_tensor}) by coupling a chiral boson theory to vielbeins. In that context, the coupling to vielbeins treated chiral and anti-chiral modes differently, and as a result the stress tensor component $T_{t \theta} = - P$ is sensitive to the difference between chiral and anti-chiral fields. Exchanging the fields $\phi$ with $\overbar{\phi}$, and vice-versa, reverses the sign of $P$ and therefore changes $T_{t \theta}$.

In contrast, since both the barred gauge fields and unbarred gauge fields couple to the boundary metric in the same way, the Hilbert stress tensor (\ref{hilbert_cs}) treats the fields $A_\alpha^i$ and $\overbar{A}_{\overline{i}}^\alpha$ on equal footing. Unlike (\ref{general_stress_tensor}), the Hilbert stress tensor associated with the standard boundary Lagrangian $\mathcal{L}_{\text{bdry}} = S$ is unchanged under the process of exchanging barred and unbarred gauge fields. To make this point explicit, let us write this boundary term as 
\begin{align}\label{Lbdry_A}
    \mathcal{L}_{\text{bdry}} = \frac{1}{2} \tensor{S}{_\alpha^\alpha} \, , \qquad S_{\alpha \beta} = k^{ij} A_{i \alpha} A_{j \beta} + \overbar{k}^{\overline{i} \overline{j}} \overbar{A}_{\overline{i} \alpha} \overbar{A}_{\overline{j} \beta} \, .
\end{align}
With this definition, one has $\tensor{S}{_\alpha^\alpha} = 2 S$. The Hilbert stress tensor computed from (\ref{Lbdry_A}), after rescaling to eliminate the overall prefactor of $- \frac{1}{16 \pi}$ in $I_{\text{bdry}}$, is
\begin{align}\label{std_lbdry_hilbert}
    T_{\alpha \beta} = - S_{\alpha \beta} + g_{\alpha \beta} S \, .
\end{align}
Deforming the standard boundary term by a generic function of the stress tensor (\ref{std_lbdry_hilbert}), which necessarily involves the single independent non-vanishing Lorentz invariant $T^{\alpha \beta} T_{\alpha \beta}$, will introduce dependence on the new variable
\begin{align}
    S_2 = S_{\alpha \beta} S^{\alpha \beta} \, .
\end{align}
Note that $S_2$ is functionally independent from the invariant $P = \frac{1}{2} \left(  k^{ij}  A_i^\alpha A^{j}_{\alpha} - \overbar{k}^{\overline{i} \overline{j}} \overbar{A}_{\overline{i}}^\alpha \overline{A}^{\overline{j}}_{\alpha} \right)$. Therefore, the class of boundary terms that can be described by functions $f ( S, P )$ is not closed under deformations by the Hilbert stress tensor. Instead, to describe flows driven by this choice of stress tensor, we should instead parameterize the boundary term as a function of different invariants:
\begin{align}\label{boson_like_class}
    \mathcal{L}_{\text{bdry}} = f ( S_1 , S_2 ) \, ,
\end{align}
where
\begin{align}
    S_1 = \Tr ( S ) = \tensor{S}{_\alpha^\alpha} = 2 S \, , \qquad S_2 = \Tr ( S^2 ) = S_{\alpha \beta} S^{\alpha \beta} \, .
\end{align}
The structure of Hilbert stress tensor deformations of the class of functions (\ref{boson_like_class}) is identical to the structure of such flows for a collection of \emph{non-chiral} bosons. Indeed, as was worked out in \cite{Ferko:2022cix}, a general Lagrangian for a collection of $N$ non-chiral bosons $\varphi^i$ with target space metric $G_{ij}$ is a function of the matrix
\begin{align}
    \tensor{X}{_\alpha^\beta} = G_{ij} \partial_\alpha \varphi^i \partial^\beta \varphi^j \, ,
\end{align}
which has two independent traces, 
\begin{align}
    x_1 = \Tr ( X ) = \tensor{X}{_\alpha^\alpha} \, , \qquad x_2 = \Tr ( X^2 ) = \tensor{X}{_\alpha^\beta} \tensor{X}{_\beta^\alpha} \, .
\end{align}
All higher traces can be expressed in terms of $x_1$ and $x_2$ using identities derived from the Cayley-Hamilton theorem for $2 \times 2$ matrices. Precisely the same results apply in the Chern-Simons context, except replacing the matrix $\tensor{X}{_\alpha^\beta} $ with $\tensor{S}{_\alpha^\beta}$ and thus replacing the invariants $x_1$, $x_2$ with $S_1$, $S_2$. For instance, the Hilbert stress tensor associated with a general boundary term (\ref{boson_like_class}) is
\begin{align}
    T_{\alpha \beta} = - 2 \frac{\partial f}{\partial S_1} S_{\alpha \beta} - 4 \frac{\partial f}{\partial S_2} S_{\alpha \gamma} \tensor{S}{^\gamma_\beta} + g_{\alpha \beta} f \, .
\end{align}
One can then construct deformations of the boundary term which depend on the two independent traces of the stress tensor, which can be written as
\begin{align}
    T^{\alpha \beta} T_{\alpha \beta} &= 2 \left( f + 2 S_1^2 \frac{\partial f}{\partial S_2} \right) \left( f - 2 S_1 \left( \frac{\partial f}{\partial S_1} + S_1 \frac{\partial f}{\partial S_2} \right) \right) + 8 S_2^2 \left( \frac{\partial f}{\partial S_2} \right)^2 \nonumber \\
    &\quad + 4 S_2 \left( \left( \frac{\partial f}{\partial S_1} \right)^2 + 6 S_1 \frac{\partial f}{\partial S_1} \frac{\partial f}{\partial S_2} - 2 \frac{\partial f}{\partial S_2} \left( f - 2 S_1^2 \frac{\partial f}{\partial S_2} \right) \right)  \, , \\
    \tensor{T}{^\alpha_\alpha} &= - 2 S_1 \frac{\partial f}{\partial S_1} - 4 S_2 \frac{\partial f}{\partial S_2} + 2 f \, .
\end{align}
All of the results concerning stress tensor flows for non-chiral bosons in two dimensions (see, for instance, \cite{Ferko:2022cix} and section 4 of \cite{Ferko:2023ruw}) therefore immediately apply to deformations of Chern-Simons boundary terms which take the form (\ref{boson_like_class}).

One way to think about this class of deformations, using the parameterization (\ref{boson_like_class}) and the Hilbert stress tensor, is the following. In the case $N = \overbar{N}$ -- when the unbarred gauge fields $A^i_\alpha$ and barred gauge fields $\overbar{A}^i_\alpha$ are dual to equal numbers of chiral bosons $\phi^i$ and anti-chiral bosons $\overbar{\phi}^i$, respectively -- one can collect these fields into a collection of non-chiral bosons $\varphi^i$ as
\begin{align}
    \varphi^i = \frac{1}{\sqrt{2}} \left( \phi^i + \overbar{\phi}^{i} \right) \, .
\end{align}
We will revisit the quantization of the boundary theory after performing this repackaging of the field content into non-chiral fields in section \ref{sec:cian}. We claim that deformations using the Hilbert stress tensor and the parameterization (\ref{boson_like_class}) are appropriate for understanding flows in which the bosons are assembled into non-chiral fields in this way. This is why such flows are naturally studied using the invariants $(S_1, S_2)$, which have the same structure as the ones appearing in $\TT$-like deformations of non-chiral bosons, rather than the invariants $(S, P)$, which we have used in section \ref{sec:classical} to understand stress tensor flows for chiral bosons. 

One might ask whether there is a different presentation of stress tensor deformations for the boundary term whose structure is more similar to that of flows in the Floreanini-Jackiw description of section \ref{sec:classical}. This brings us to the second natural choice of stress tensor, besides the Hilbert definition in equation (\ref{hilbert_cs}). Rather than coupling the boundary theory to a metric on $\partial \mathcal{M}_3$, one could instead couple to vielbeins in the same way as we did in equation (\ref{general_class_vielbeins}) for chiral boson theories. To do this, we again introduce frame fields $\tensor{E}{^a_\alpha}$, although now the flat indices will be raised or lowered with the \emph{Euclidean} tangent-space metric $\eta_{ab} = \left[ \begin{smallmatrix} 0 & 1 \\ 1 & 0 \end{smallmatrix} \right]$. In this case, the appropriate flat-space values for the vielbeins are
\begin{align}\label{euclidean_flat}
    E^+_w = E^-_{\wb} = \frac{1}{\sqrt{2}} \, , \qquad E^+_{\wb} = E^-_{w} = 0 \, ,
\end{align}
whose inverses produce the desired spacetime metric $ds^2 = dw \, d \wb$,
\begin{align}
    E^a_\alpha E^b_\beta \eta_{a b} = g_{\alpha \beta} = \begin{bmatrix} 0 & \frac{1}{2} \\ \frac{1}{2} & 0 \end{bmatrix} \, .
\end{align}
One can then couple the Chern-Simons boundary term $I_{\text{bdry}}$ to vielbeins as 
\begin{equation}\label{CS_bdry_coupled}
    I_{\text{bdry}} = - \frac{i}{16 \pi} \int_{\partial \mathcal{M}_3} d^2 x \, \left(  2 \left( E^+_w E^-_w - E^+_{\wb} E^-_{\wb} \right) P + 2 E f ( S, P ) \right) \, ,
\end{equation}
where we include factors of $2$ since, in the conventions of this section, $E = \frac{1}{2}$. Likewise, the overall factor of $i$ in (\ref{CS_bdry_coupled}) arises because $\sqrt{g} = \frac{i}{2}$ but $E = \frac{1}{2}$. To compare with equation (\ref{general_class_vielbeins}), note that in the conventions of section \ref{sec:classical}, we instead had $E = 1$. Now $S$ and $P$ are coupled to vielbeins as
\begin{align}
    S &= \frac{1}{4 \left( E^+_w E^-_{\wb} + E^-_w E^+_{\wb} + E^+_w E^-_w + E^+_{\wb} E^-_{\wb} \right) } \left( k^{ij} A_{i w} A^{j}_{j \wb} + \overbar{k}^{\overline{i} \overline{j}} \overbar{A}_{\overline{i} w} \overbar{A}^{\overline{j}}_{\wb} \right) \, , \nonumber \\
    P &= \frac{1}{4 \left( E^+_w E^-_{\wb} + E^-_w E^+_{\wb} + E^+_w E^-_w + E^+_{\wb} E^-_{\wb} \right)} \left(  k^{ij}  A_{i w} A^{j}_{\wb} - \overbar{k}^{\overline{i} \overline{j}} \overbar{A}_{\overline{i} w} \overbar{A}^{\overline{j}}_{\wb} \right) \, ,
\end{align}
in such a way that they reduce to their flat-space values when the vielbeins are given by (\ref{euclidean_flat}). Because these expressions are written with explicit $(w, \wb)$ indices, the resulting coupling to gravity is not manifestly Lorentz-invariant. However, this is to be expected since we are performing the equivalent of the procedure used in equation (\ref{general_class_vielbeins}) for coupling Floreanini-Jackiw bosons to gravity, which is also not manifestly Lorentz-invariant.

We now compute the stress tensor (\ref{stress_tensor_flat_curved}) using this coupling to the frame fields. In order to make comparison with the results of section \ref{sec:classical} easier, we will re-scale the stress tensor by an overall factor to absorb the multiplicative constant of $- \frac{i}{16 \pi}$ in the boundary term (\ref{CS_bdry_coupled}), as well as the relative factor of $2$ due to the conventions for the vielbein in this section. Therefore we instead compute
\begin{align}\label{stress_tensor_flat_curved_rescaled}
    \tensor{T}{_\beta^a} = - \frac{ 8 \pi i}{E} \frac{\delta S}{\delta \tensor{E}{^\beta_a}} \, ,
\end{align}
and convert to spacetime indices to find
\begin{align}
    T_{ww} &= - \frac{1}{4} \left( 2 S V_S + 2 P \left( 1 + V_P \right) \right) \, , \nonumber \\
    T_{w \wb} &= T_{\wb w} = \frac{1}{2} \left( V - S V_S - P V_P \right) \, , \nonumber \\
    T_{\wb \wb} &= \frac{1}{2} \left( P - P V_P - S V_S \right) \, .
\end{align}
The two Lorentz scalars that we use for constructing flows are therefore
\begin{align}
    \tensor{T}{^\alpha_\alpha} &= 2 \left( V - S V_S - P V_P \right) \, , \nonumber \\
    T^{\alpha \beta} T_{\alpha \beta} &= V^2 - 2 P^2 + \left( V - 2 \left( S V_S + P V_P \right) \right)^2 \, ,
\end{align}
which exactly matches equations (\ref{trT}) and (\ref{trT2}).

It now follows that all of our comments about stress tensor flows in section \ref{sec:classical} immediately apply to deformations of Chern-Simons boundary terms which are constructed using the stress tensor (\ref{stress_tensor_flat_curved_rescaled}) obtained from coupling to vielbeins, as opposed to the standard Hilbert stress tensor. For instance, any deformation by a function of the vielbein stress tensor (\ref{stress_tensor_flat_curved_rescaled}) necessarily preserves the condition (\ref{CS_chirality_constraint}). This means that, if one begins with a seed Chern-Simons boundary term which is invariant under the symmetry (\ref{cs_sd_trans}) that guarantees the chirality (or self-duality) of the theory, and then deforms this seed by any function of the energy-momentum tensor, the resulting deformed boundary term will also be invariant under the same symmetry. Furthermore, any one-parameter family of Chern-Simons boundary terms which are all invariant under the duality rotation (\ref{cs_sd_trans}) must satisfy a differential equation driven by a function of the vielbein stress tensor.

It also follows that the closed-form solutions to flow equations driven by functions of the stress tensor discussed in section \ref{sec:classical} -- such as the two-parameter family of solutions (\ref{scalar_modified_nambu_goto}) to the commuting $\TT$ and root-$\TT$ flow equations -- also have obvious analogs for deformations of Chern-Simons boundary terms. Besides solving these differential equations directly, a complementary way to analyze stress tensor deformations is by performing a perturbative expansion which computes the deformed action order-by-order in the flow parameter. This approach is discussed in appendix \ref{app:TTn} for deformations by various functions of the energy-momentum tensor, using the version of $T_{\alpha \beta}$ defined by coupling to vielbeins.

To conclude this section, let us summarize and mention some applications. We have seen that the boundary term of a bulk $U(1)$ Chern-Simons theory can be deformed either by functions of the Hilbert stress tensor or by functions of the vielbein stress tensor (\ref{stress_tensor_flat_curved_rescaled}). The former deformations lead to a class of modified boundary terms $\mathcal{L}_{\text{bdry}} ( S_1, S_2 )$ with the same properties as Lagrangians obtained by stress tensor deformations of non-chiral boson theories. Conversely, the latter flows generate a family of boundary terms $\mathcal{L}_{\text{bdry}} ( S, P )$ with the same structure as the Lagrangians in section \ref{sec:classical} arising from stress tensor deformations of chiral boson theories. We have thus described two complementary ways to view deformations of Chern-Simons boundary terms by functions of the energy-momentum tensor.

These results provide a general framework for studying three-dimensional $U(1)$ Chern-Simons theories subject to boundary deformations. Throughout our discussion, we have been agnostic as to the specific setting in which such Chern-Simons terms arise, but let us briefly mention two specific applications of the formalism we have developed. One context in which these results could be useful is when considering $\mathrm{AdS}_3/\mathrm{CFT}_2$ holography with $U(1)$ gauge fields. One could use our machinery to derive flow equations for various observables under stress tensor deformations, just as \cite{Ebert:2022ehb} found expressions for $\TT$-deformed Wilson lines and loops, and \cite{Ebert:2023tih} obtained formulas for the masses of BTZ black holes under a boundary root-$\TT$ deformation. For instance, one could use the results of this section to analyze the dependence of the $U(1)$ charges of charged BTZ black holes as a function of the deformation parameter for boundary $\TT$ or root-$\TT$ deformations. Another possible application of these results is to study quantum Hall systems subject to boundary deformations, which we will briefly describe in the conclusion of this paper.

\section{Quantization Along Classical Flows}\label{sec:quantization}

In this section, we will consider the quantization of a member of the general class of interacting chiral boson models. We will work purely within the Floreanini-Jackiw description, described by an action of the form (\ref{interacting_again}), rather than in the Chern-Simons formulation of section \ref{sec:cs}. We will also work in Lorentzian signature with spacetime coordinates $(t, \theta)$. Although in the preceding discussion we have been agnostic as to the spacetime topology, within this section we will assume that $\theta$ is compact and subject to the identification $\theta \sim \theta + 2 \pi$. We focus on the case of a compact spatial manifold because our primary observable of interest is the finite-volume spectrum of energy levels $E_n$, and in particular how these energies depend on a deformation parameter along a stress tensor flow.

The most well-studied example of a stress tensor deformation for which the deformed cylinder spectrum can be determined is the $\TT$ deformation. Under the $\TT$ flow, the energy levels of the deformed theory obey the inviscid Burgers' equation,
\begin{align}\label{burgers}
    \frac{\partial E_n}{\partial \lambda} = E_n \frac{\partial E_n}{\partial R} + \frac{P_n^2}{R} \, ,
\end{align}
where $R$ is the radius of the cylinder and $E_n, P_n$ are the energy and momentum of the eigenstate under consideration \cite{Zamolodchikov:2004ce, Smirnov:2016lqw,Cavaglia:2016oda}.\footnote{One can also study various generalizations of this flow for the spectrum, such as the energy levels of tensor product theories where the factors are sequentially deformed by multiple $\TT$ flows \cite{Ferko:2022dpg}.}

This example is remarkable because the flow equation (\ref{burgers}) can be proven directly at the quantum level using the properties of the local $\TT$ operator, which is defined by
\begin{align}\label{TT_point_split}
    \mathcal{O}_{\TT} ( x ) = \lim_{y \to x} \left( T^{\alpha \beta} ( x ) T_{\alpha \beta} ( y ) - \tensor{T}{^\alpha_\alpha} ( x ) \tensor{T}{^\beta_\beta} ( y ) \right) \, .
\end{align}
It was demonstrated in \cite{Zamolodchikov:2004ce} that the coincident point limit on the right side of (\ref{TT_point_split}) actually gives rise to a well-defined local operator, up to total derivative ambiguities which can be ignored. One can therefore prove results about a $\TT$-deformed quantum field theory at the quantum level using the properties of this operator; for instance, an argument involving a certain factorization property of $\mathcal{O}_{\TT}$ and the interpretation of the components of the stress tensor in terms of energy and momentum lead to the flow equation (\ref{burgers}).\footnote{These statements are special to $d = 2$, and it is not known whether there exists a higher-dimensional version of the quantum $\TT$ operator with similar properties; see \cite{Taylor:2018xcy} for a discussion of the challenges in doing so. However, in $d > 2$ one can define point-split stress tensor bilinears which factorize to leading order at large $N$ \cite{Hartman:2018tkw,Araujo-Regado:2022gvw}, and a new proposal for $\mathcal{O} \left(\frac{1}{N}\right)$ corrections was made in equation (1.58) of \cite{Ebert:2024fpc}.}

This is in contrast with a different method for attempting to learn about the quantum mechanical properties of a stress tensor deformation, which we refer to as \emph{quantization along a classical flow}. In this case, one first finds the solution to a differential equation of the form (\ref{stress_tensor_flow_defn}) for the Lagrangian of a deformed theory, and then attempts to quantize this deformed Lagrangian directly.

Assuming that a given classical deformation can be rigorously defined at the quantum level, we do not expect that quantization along the classical flow will give accurate information about \emph{all} aspects of the deformed quantum field theory. Indeed, this is already true for the $\TT$ deformation. For instance, it can be shown that the S-matrix of a $\TT$-deformed quantum field theory is equal to the S-matrix of the undeformed theory multiplied by a certain momentum-dependent phase known as a CDD factor \cite{Dubovsky:2012wk,Dubovsky:2013ira}. However, if one studies scattering using quantization along the classical $\TT$ flow, one finds that this CDD factor is not reproduced unless one adds specific counter-terms which are engineered to obtain the expected scattering behavior \cite{Rosenhaus:2019utc,Dey:2021jyl,Chakrabarti:2022lnn}. Therefore, quantization along the classical flow is not sufficient to fully characterize the properties of the $\TT$-deformed theory without additional input from the quantum definition.\footnote{Another argument for this conclusion is that quantization of fermionic fields along classical $\TT$ flows can give different Hilbert spaces depending on which definition of the stress tensor one uses \cite{Lee:2021iut,Lee:2023uxj}.}

Despite this, one may hope that quantization of a classical deformed Lagrangian will still give \emph{some} information about the corresponding deformation at the quantum level, at least in particular limiting cases. For instance, the solution to the classical $\TT$ flow equation beginning from a seed theory of free scalars is the Nambu-Goto action of string theory, and one generically expects that string theories exhibit a high-energy density of states which is Hagedorn rather than Cardy. This predicted Hagedorn scaling agrees with an analysis of the high-energy behavior of a $\TT$-deformed CFT at the quantum level, which can be seen either from the energies \cite{Cavaglia:2016oda} or the partition function \cite{Cardy:2018sdv,Datta:2018thy}. Thus certain limiting features of the quantum theory can still be inferred from the $\TT$-deformed classical Lagrangian.

For other stress tensor deformations, like the root-$\TT$ flow, it is not yet known whether one can give a rigorous definition of the deforming operator at the quantum level.\footnote{However, see \cite{Hadasz:2024pew} for a recent proposal for the quantum definition of the root-$\TT$ deformation and a computation of deformed correlation functions.} Therefore, we do not yet have any exact data about the deformed quantum theory against which to compare results obtained by other methods. However, extrapolating from the $\TT$ case, one might perform quantization along a classical root-$\TT$ flow in the hope that this procedure will still give useful information in certain limits. Our goal in this section is to carry out this quantization procedure for root-$\TT$-deformed theories of chiral bosons and examine the behavior of the deformed spectrum in such limiting cases.

One regime for which we have additional data about the root-$\TT$-deformed spectrum is the limit of a large-$c$ holographic CFT which admits a bulk $\mathrm{AdS}_3$ dual. When restricting to states for which the stress tensor is approximately constant (which are dual to BTZ black holes), one obtains the formula (\ref{zero_mode_formula}) for the root-$\TT$ deformed spectrum \cite{Ebert:2023tih}. We will see that our analysis using quantization along the classical flow agrees with this ``zero mode formula'' for states that correspond to constant stress tensor backgrounds. However, we will also be able to probe other limits of a root-$\TT$-deformed theory, such as a large-momentum limit which is \emph{not} close to a constant stress tensor configuration for which the zero mode formula is expected to apply. This result may therefore give novel information about the behavior of a putative root-$\TT$-deformed field theory in a different regime.

\label{sec:Quantization of first-order deformed}

\subsection{Generalities on quantization}\label{sec:general_quantization}

Let us now study the quantum mechanics of interacting chiral boson models such as (\ref{interacting_again}). This Floreanini-Jackiw form of the Lagrangian, although it is not manifestly Lorentz-invariant, is nonetheless convenient for quantization because it is first-order in time derivatives. This allows us to perform canonical quantization in a uniform way which does not depend on the details of the interaction function $V ( S, P )$.

We begin by reviewing some basic features of quantization of first-order systems in the simpler setting of $(0+1)$-dimensional theories, i.e. particle mechanics.

\subsubsection*{\ul{\it Quantization of first-order particle mechanics}}

We will first consider a collection of $(0 + 1)$-dimensional fields $q_i ( t )$, whose time derivatives will be denoted $\dot{q}_i ( t )$. A general first-order Lagrangian for such a system takes the form
\begin{align}\label{eq:o0iuh1fsd}
    L = \frac{1}{2} C^{ij} q_i \dot{q}_j - V ( q ) \, ,
\end{align}
where $C^{ij}$ is a non-singular constant matrix. Without loss of generality, we may assume that $C^{ij}$ is antisymmetric. Indeed, if we instead split $C^{ij} = C^{[ij]} + C^{(i j)}$ into symmetric and anti-symmetric parts, the Lagrangian would be
\begin{align}
    L = \frac{1}{2} C^{[ij]} q_i \dot{q}_j + \frac{1}{2} C^{(ij)} \frac{d}{dt} \left( q_j q_i \right) - V ( q ) \, ,
\end{align}
where the second term is a total time derivative that can be ignored. 

The canonical momentum which is conjugate to $q_j ( t )$ is
\begin{align}\label{first_order_momentum}
    p^j= \frac{\partial L}{\partial \dot{q}_j} = \frac{1}{2} C^{ij} q_i \, ,
\end{align}
and thus the Hamiltonian associated with (\ref{eq:o0iuh1fsd}) is
\begin{align}\label{first_order_hamiltonian}
    H ( q, p ) = \frac{\partial L}{\partial \dot{q}_i} \dot{q}_i - L = V ( q ) \, .
\end{align}
The Hamiltonian (\ref{first_order_hamiltonian}) appears to depend only on the position variables but not on the momenta, but this is misleading, since equation (\ref{first_order_momentum}) implies that some combinations of the $q_i$ \emph{are} momenta. The Euler-Lagrange equations arising from the Lagrangian (\ref{eq:o0iuh1fsd}) are
\begin{align}
    C^{ij} \dot{q}_i = \frac{\partial H}{\partial q_j} \, ,
\end{align}
where we now use the symbols $V$ and $H$ interchangeably. Alternatively, by defining $C_{ij}$ to be the inverse matrix $\left( C^{-1} \right)_{ij}$ of $C^{ij}$, the equations of motion can be written as
\begin{align}\label{first_order_EL}
    \dot{q}_i = C_{ij} \frac{\partial H}{\partial q_j} \, .
\end{align}
Next we consider the quantization of this model. Ordinarily, for Lagrangians which are quadratic in time derivatives, one would impose the canonical commutation relations
\begin{align}\label{ccr}
    [ x_i, p_j ] = i \delta_{ij} \, .
\end{align}
However, imposing the relations (\ref{ccr}) for a first-order system like (\ref{eq:o0iuh1fsd}) gives results that differ from the correct commutation relations by a factor of $2$. To arrive at the correct relations, we follow the prescription outlined in appendix A of \cite{Floreanini:1987as}, and further justified in \cite{Faddeev:1988qp}, which is to define commutators so that the Heisenberg-picture time evolution of operators in the quantum theory takes the same form as the classical Euler-Lagrange equations.\footnote{In conventional quantum systems with second-order Lagrangians, the fact that these two equations should take the same form is the content of the Ehrenfest theorem. We demand that the same is true here.}

In general, the Heisenberg equation of motion for an operator $\mathcal{O}$ reads $\dot{\mathcal{O}} = i [ H , \mathcal{O} ]$. In the case of the operator $\mathcal{O} = q_i$, we have
\begin{align}\label{heisenberg_eom}
    \dot{q}_i = i [ H , q_i ] = i \frac{\partial H}{\partial q_j} [ q_j, q_i ] \, .
\end{align}
Comparing (\ref{heisenberg_eom}) to (\ref{first_order_EL}), we find that the two take the same form if we identify
\begin{align}\label{correct_commutation}
    [ q_i, q_j ] = i C_{ij} \, .
\end{align}
As we mentioned, since $p^j = \frac{1}{2} C^{ij} q_i$, this differs from the canonical prescription (\ref{ccr}) which would give $[ q_i, q_j ] = 2 i C_{ij}$. The errant factor of $2$ is due to the fact that, in a first-order system, there is a constraint on the phase space.

\subsubsection*{\ul{\it Quantization of first-order field theories}}

Having reviewed the quantum mechanics of first-order $(0+1)$-dimensional systems, we now turn to the quantization of first-order $(1+1)$-dimensional field theories, and in particular the theories of chiral bosons which are the focus of this work.

As a simple example to set the stage, we will first consider a single chiral boson described by the Floreanini-Jackiw Lagrangian (\ref{fj_action}) which we repeat here:
\begin{equation}\label{FJ_later}
    \mathcal{L} = \frac{1}{2}  \left(\phi' \dot{\phi} - \phi' \phi' \right) \, .
\end{equation}
As usual, we write $\dot{\phi}$ for the time derivative of $\phi$ and $\phi'$ for the spatial derivative of $\phi$. The quantization of this system in infinite volume, i.e. with a spatial coordinate $x \in \mathbb{R}$, was first studied in \cite{Floreanini:1987as}. In short, one can view $x$ as a continuous generalization of the discrete labels $i$, $j$ in \eqref{eq:o0iuh1fsd} and rewrite the first term as
\begin{equation}
\begin{aligned}
    \frac{1}{2} \int dx\, \partial_x \phi(x,t) \dot{\phi}(x,t) &= \frac{1}{2} \int dx \int dy \, \delta(x-y) \partial_x \phi(x,t) \dot{\phi}(y,t) \\
    &= - \frac{1}{2} \int dx \int dy \, [\partial_x \delta(x-y)] \phi(x,t) \dot{\phi}(y,t) \, .
\end{aligned}
\end{equation}
The role of the constant antisymmetric matrix $C^{ij}$ in the particle mechanics example is now played by the function 
\begin{align}
    C(x-y) = -\partial_x \delta(x-y) \, ,
\end{align}
and the role of the inverse matrix $C_{ij}$ is played by the Green's function of $C ( x - y )$. This suggests that we impose the commutation relations
\begin{equation}
    [\phi(x), \phi(y)] =  - \frac{i}{2}\sgn(x-y)~,
\end{equation}
which is the field theory analog of (\ref{correct_commutation}) and which matches the result in \cite{Floreanini:1987as}. It is then straightforward to use the above equal-time commutation relations to confirm the Heisenberg equations of motion are indeed equivalent to the Euler-Lagrange equations of the Lagrangian (\ref{FJ_later}), which describe a chiral boson:
\begin{equation}
    \frac{\partial \mathcal{O}}{\partial t} = -i [\mathcal{O},H] \quad \implies \quad \dot{\phi} = \phi' \, .
\end{equation}
Next we will study this theory in finite volume. We now replace the spatial coordinate $x \in \mathbb{R}$ with an angular coordinate $\theta$ labeling a position on $S^1$, and subject to the identification $\theta \sim \theta + 2 \pi$. We will also assume that the \emph{target space} is compact, which means that $\phi$ likewise takes values in a circle so that $\phi \sim \phi + 2 \pi$. As we will see, the structure of this theory on a cylinder is closely related to the particle mechanics example considered above.

First let us write the function $\phi(t,\theta)$ using a mode expansion:
\begin{equation}\label{phi_mode_expansion}
    \phi(t,\theta) = \frac{1}{\pi} x(t) + p(t) \theta + \frac{1}{\sqrt{2\pi}} \sum_{n=1}^\infty \frac{1}{\sqrt{n}} \left( a_{n}(t) e^{in\theta}+ a^\dagger_n(t) e^{-in\theta} \right) \, .
\end{equation}
We have included a zero-mode term $x(t)$ in addition to a momentum contribution which is linear in $\theta$; the latter is permissible, despite not being periodic in $\theta$, since both $\theta \sim \theta + 2 \pi$ and $\phi \sim \phi + 2 \pi$, so such a term is compatible with our identifications if $p \in \mathbb{Z}$. The remaining sum is the standard Fourier expansion of the periodic part of $\phi$ in the $\theta$ direction.

It is now necessary to distinguish between the Lagrangian density $\mathcal{L}$ and the Lagrangian $L = \int d \theta \, \mathcal{L}$. Substituting the mode expansion (\ref{phi_mode_expansion}) into the Lagrangian density (\ref{FJ_later}) and performing the integral over the $\theta$ coordinate gives
\begin{equation}
    L = \int_{0}^{2\pi} d\theta \, \mathcal{L} =  p \dot{x} - \pi p^2 +  \frac{i}{2} \left(\sum_{n=1}^\infty  (\dot{a}^\dagger_n a_n - \dot{a}_n a_n^\dagger) \right) - \left(\frac{1}{2} \sum_{n =1} ^\infty n (a_n a^\dagger_n + a^\dagger_n a_n)  \right) ~,
\end{equation}
where we have dropped a term that is a total derivative in time. Because $p$ is integer-quantized, as we mentioned above, the first term describes the well-known quantum system which is a particle on a ring. The Hilbert space is generated by states $|p\rangle$ labeled by integer $p \in \mathbb{Z}$ with energy $E_p = \pi p^2$. The remaining terms are nothing but the familiar first-order particle mechanics system discussed previously. To make this analogy clearer, it is convenient to define
\begin{align}
    a_{-n} = a_n^\dagger \, ,
\end{align}
so that the Lagrangian can be written as
\begin{equation}\label{fj_rewritten_modes}
    L =  p \dot{x} - \pi p^2 +  \frac{i}{2} \left(\sum_{n=1}^\infty  (\dot{a}_{-n} a_n - \dot{a}_n a_{-n}) \right) - \left(\frac{1}{2} \sum_{n =1} ^\infty n (a_n a_{-n} + a_{-n} a_n)  \right) \, .
\end{equation}
The $a_n$'s now play the role of $q_i$'s, except the modes are labeled by $n \in \mathbb{Z}$ so the phase space is infinite-dimensional. Comparing the two sums in the Lagrangian (\ref{fj_rewritten_modes}) with the general form \eqref{eq:o0iuh1fsd}, we find that the two agree if we identify
\begin{equation}
    C^{n,m} = i \, \sgn(n) \delta_{n,-m} \, .
\end{equation}
Therefore, when we promote the $a_n$ from functions appearing in the expansion of the classical field $\phi$ to quantum operators, the appropriate commutation relations (\ref{correct_commutation}) are
\begin{align}
    [a_n,a_m] = \sgn(n) \delta_{n,-m} \, .
\end{align}
When expressed in terms of $a^\dagger_m$, this is the familiar commutation relation of ladder operators:
\begin{equation}\label{standard_ladders}
    [a_n, a_m^\dagger] = \delta_{n,m} \, .
\end{equation}
It is perhaps surprising that, if we had worked with the Fourier modes $a_n$ of the field $\phi$ from the beginning (rather than with the field $\phi$ itself), then imposing the standard commutation relations (\ref{standard_ladders}) gives the correct result, without the errant factor of $2$ which we mentioned around equation (\ref{correct_commutation}) that occurs due to the phase space constraint on first-order systems. The reason for this is that, after performing the mode expansion, the positive Fourier modes $a_n$ with $n > 0$ act as the position variables and the negative modes $a_n$ with $n < 0$ (or equivalently $a^\dagger_n$) act as the conjugate momentum variables. Therefore, in Fourier space, the separation between coordinates and momenta is automatic, and we need not impose phase space constraints or consider commutation relations like (\ref{correct_commutation}) which na\"ively appear to involve two position variables.\footnote{See section 6.1.3 of \cite{Tong:2016kpv} for a pedagogical review of the quantization of the chiral boson from this momentum-space perspective, and later sections of this reference for applications to quantum Hall physics.}

The Hamiltonian obtained from the Legendre transform of the Lagrangian (\ref{fj_rewritten_modes}), written in terms of $a_n^\dagger$ rather than $a_{-n}$, is
\begin{align}
    H &= \pi p^2 + \frac{1}{2} \sum_{n=1}^\infty n (a_n a_n^\dagger + a_n^\dagger a_n) \nonumber \\
    &= -\frac{1}{24} + \pi p^2 +  \sum_{n=1}^\infty n a^\dagger_n a_n \, ,
\end{align}
where we have used $a_n a_n^\dagger = a_n^\dagger a_n + 1$ and the well-known $\zeta$-function regularization
\begin{equation}
    \sum_{n=1}^\infty n = - \frac{1}{12} \, .
\end{equation}
It is straightforward to generalize the above discussion to the case of multiple chiral and anti-chiral bosons. We work with a Lagrangian density for $N$ chiral bosons $\phi^i$, $i = 1, \ldots, n$, and $\overbar{N}$ anti-chiral bosons $\overbar{\phi}^{\overline{i}}$, of the form (\ref{interaction_function}) which we have been considering in section \ref{sec:classical}. For simplicity we take trivial target-space metrics for the bosons, $G_{ij} = \delta_{ij}$ and $\overbar{G}_{\overline{i} \overline{j}} = \delta_{\overline{i} \overline{j}}$. The Lagrangian density for this system is then
\begin{equation}
    \mathcal{L} = \frac{1}{2}  \left(\phi'_i \dot{\phi}^i - \overbar{\phi}_{\overline{i}}' \dot{\overbar{\phi}} {}^{\overline{i}} \right) -V ( \phi_i', \overbar{\phi}_{\overline{i}}^{\prime} ) ~.
\end{equation}
We expand both the chiral and anit-chiral fields in modes as
\begin{equation}\label{eq:phiphibarexpan}
\begin{aligned}
    \phi_i(t,\theta) &= \frac{1}{\pi} x_i(t) + p_i(t) \theta + \frac{1}{\sqrt{2\pi}} \sum_{n = 1}^{\infty} \frac{1}{\sqrt{n}} \left(a_{i,n}(t) e^{in\theta} + a_{i,n}^\dagger(t) e^{-in\theta} \right)~, \\
    \overbar{\phi}_{\overline{i}} (t,\theta) &= - \frac{1}{\pi} \overbar{x}_{\overline{i}} (t) + \overbar{p}_{\overline{i}} (t) \theta + \frac{1}{\sqrt{2\pi}} \sum^\infty_{n=1}\frac{1}{\sqrt{n}}\left( b_{\overline{i} , n}(t) e^{-in\theta} + b_{\overline{i} ,n}^\dagger(t) e^{in\theta}\right)~.
\end{aligned}
\end{equation}
The non-zero commutation relations between the various expansion coefficients are
\begin{equation}\label{general_bosons_commutators}
    [x_i, p_j] = i \delta_{ij}, \quad [ \overbar{x}_{\overline{i}}, \overbar{p}_{\overline{j}} ] = i \delta_{\overline{i} \overline{j}}, \quad [a_{i,n}, a_{j,m}^\dagger] = \delta_{i j} \delta_{n m}, \quad [b_{\overline{i},n}, b_{\overline{j},m}^\dagger] = \delta_{\overline{i} \overline{j}} \delta_{n m} \, , 
\end{equation}
with all other commutators vanishing.

Note that here we take all $\phi_i$ and $\overbar{\phi}_{\overline{i}}$ to be compact with radius $2\pi$. Therefore, the eigenvalues of $p_j$ and $\overbar{p}_{\overline{j}}$ must be integers. The Hamiltonian is given by
\begin{equation}
    H = \int_0^{2\pi} d\theta \, V(\phi'_i, \overbar{\phi}'_{\overline{j}} ) ~.
\end{equation}
The commutation relations (\ref{general_bosons_commutators}) allow us to build the Hilbert space of the quantum theory for any potential $V$. In the next subsection, we will use this to study the spectrum of the ``Modified Scalar'' theory, that is, the theory obtained by applying a root-$\TT$ deformation to a seed theory of free chiral and anti-chiral bosons.

\subsection{Root-$\TT$-deformed spectrum}
\label{sec:quantumspectrum}

We will now use the formalism reviewed in section \ref{sec:general_quantization} to study root-$\TT$-deformed free boson theories. In principle, this can be done for any numbers $(N, \overbar{N})$ of chiral and anti-chiral bosons, respectively. However, there is a sharp distinction between the case $N = \overbar{N} = 1$, for which the deformation is comparatively simple and can be interpreted as a re-scaling of the target space radius for the boson, and all other cases with $N \geq 1$ and $\overbar{N} \geq 1$, where the deformation is more non-trivial.\footnote{Note that, if either $N = 0$ or $\overbar{N} = 0$, then the theory is a fixed point of stress tensor flows so the root-$\TT$ deformation is trivial.} We will therefore first discuss the simpler case $N = \overbar{N} = 1$ in detail, and then as an illustrative example of the latter class, we will study the example with $N = 2$ and $\overbar{N} = 1$. We expect that the qualitative features of the deformed $(N, \overbar{N}) = (2, 1)$ model will be similar to those of theories with larger $N$ and $\overbar{N}$.

\subsubsection*{\ul{\it One compact boson}}

Let us begin by studying the root-$\TT$ deformation of a single (non-chiral) $c = 1$ compact boson, or equivalently, a pair of $N = 1$ left-moving and $\overbar{N} = 1$ right-moving chiral bosons. It was already mentioned in the initial work \cite{Ferko:2022cix} that, in this case, the root-$\TT$ flow simply rescales the kinetic term for the boson, which corresponds to a change in the radius if the scalar is compact. We will revisit this claim by describing the deformed model in terms of chiral bosons and determining the quantum spectrum exactly to confirm that the root-$\TT$ deformation of a compact boson is just a change of radius. 

This formalism also provides a way to realize a compact boson at an arbitrary radius -- even at irrational points where the theory does not factorize into the chiral part and anti-chiral part -- using a Lagrangian for one chiral and one anti-chiral boson with a quadratic mixing term. Furthermore, treating this example in detail will allow us to test the ``zero mode formula'' given in equation (\ref{zero_mode_formula}) that is expected, due to evidence from holography \cite{Ebert:2023tih}, to describe the energies of states in root-$\TT$-deformed CFTs for which the energy-momentum tensor is constant in space. We will see explicitly that this zero mode formula fails to give the energies of deformed states for which this assumption is violated.

The Lagrangian for a root-$\TT$-deformed seed theory of one left-moving and one right-moving chiral boson takes the form (\ref{interaction_function}) with an interaction function $V ( S, P , \gamma )$ given by the $\lambda \to 0$ limit of equation (\ref{scalar_modified_nambu_goto}). To be pedantic, the resulting Lagrangian is technically
\begin{equation}
    \mathcal{L}^{(\gamma)} = \frac{1}{2} \left( \phi' \dot{\phi} - \overbar{\phi}' \dot{\overbar{\phi}} \right) - \frac{\cosh ( \gamma )}{2} \left( \phi^{\prime 2} + \overbar{\phi}^{\prime 2} \right) - \sinh ( \gamma ) \sqrt{ \left( \phi' \right)^2 \left( \overbar{\phi}' \right)^2 }  \, .
\end{equation}
That is, because $\phi'$ and $\dot{\phi}$ can take both positive and negative values the final term is really proportional to $| \phi' | \cdot | \overbar{\phi}' |$. However, we will ignore this subtlety and simply replace $\sqrt{ \left( \phi' \right)^2 \left( \overbar{\phi}' \right)^2 }$ with $\phi' \overbar{\phi}'$. This can be justified, for instance, by restricting attention to small fluctuations of the fields around a background for which the gradients are large and positive, so that both $\phi'$ and $\overbar{\phi}'$ have fixed positive sign. This corresponds to a solution with large positive values of $p_i$ and $\overbar{p}_i$ in the expansion of equation (\ref{eq:phiphibarexpan}). We will take a similar large-momentum limit in the analysis with several bosons below, again resolving the square root, which is more non-trivial in that setting because of an additional term under the root.

After making this simplification, the Lagrangian we wish to study becomes
\begin{equation}\label{eq:1bL}
    \mathcal{L}^{(\gamma)} = \frac{1}{2} \left( \phi' \dot{\phi} - \overbar{\phi}' \dot{\overbar{\phi}} \right) - \frac{\cosh ( \gamma )}{2} \left( \phi^{\prime 2} + \overbar{\phi}^{\prime 2} \right)  - \sinh ( \gamma )  \phi' \overbar{\phi}'  \,.
\end{equation}
As discussed previously, the Hilbert space factorizes into two parts: the particles on a ring and the infinite tower of harmonic oscillators. Due to the special form of \eqref{eq:1bL}, the Hamiltonian does not mix the two parts. Therefore, we can study them separately.

Let us first consider the sector of the Hilbert space which describes the particles on a ring. We write the states in this Hilbert space as $|p,\overline{p}\rangle$, which are labeled by two quantized momenta $p,\overline{p} \in \mathbb{Z}$. The corresponding Hamiltonian and the momentum operator are
\begin{equation}
\label{eq:0iouv1f}
    H_{\text{PR}}^{(\gamma)} = \pi (p^2 + \overline{p}^2) \cosh ( \gamma  ) + 2\pi p \overline{p} \sinh ( \gamma ) \,, \quad P^{(\gamma)}_{\text{PR}} = \pi (p^2 - \overline{p}^2 ) = P^{(0)}_{\text{PR}} \, ,
\end{equation}
where we use the subscript PR to denote \ul{p}articles on a \ul{r}ing.

Because the corresponding undeformed states at $\gamma = 0$ have energies
\begin{align}
    H_{\text{PR}}^{(0)} = \pi (p^2 + \overline{p}^2) \, ,
\end{align}
we see that the prediction for the deformed energies from the zero mode formula (\ref{zero_mode_formula}) is
\begin{align}\label{zero_mode_works_winding}
    E_{\text{PR}}^{(\gamma)} &= H_{\text{PR}}^{(0)} \cosh ( \gamma ) + \sqrt{ \left( H_{\text{PR}}^{(0)} \right)^2 - \left( P^{(0)}_{\text{PR}} \right)^2 } \sinh ( \gamma ) \nonumber \\
    &= \pi (p^2 + \overline{p}^2) \cosh ( \gamma  ) + \sqrt{ \left( \pi (p^2 + \overline{p}^2) \right)^2 - \left( \pi (p^2 - \overline{p}^2 ) \right)^2 } \sinh ( \gamma ) \nonumber \\
    &= \pi (p^2 + \overline{p}^2) \cosh ( \gamma  ) + 2\pi p \overline{p} \sinh ( \gamma ) \, ,
\end{align}
which indeed agrees with the true deformed energies $H_{\text{PR}}^{(\gamma)}$ of equation (\ref{eq:0iouv1f}), subject to the usual caveat that we have used the assumption $\sqrt{ p^2 \overbar{p}^2 } = p \overbar{p}$.

It is not too surprising that these states have deformed energies which agree with the zero-mode formula, since the corresponding saddle points have constant stress-energy tensors, and this is the assumption under which the formula (\ref{zero_mode_formula}) was derived in holography. 

To see this explicitly, we look for solutions to the equations of motion associated with the deformed Lagrangian $\mathcal{L}^{(\gamma)}$ in equation (\ref{eq:1bL}), which are
\begin{equation}
\begin{aligned}
\label{eq:0i9ou111efs@}
  \dot{\phi}' - \phi'' \cosh ( \gamma ) - \overbar{\phi}'' \sinh ( \gamma ) = 0\,, \quad 
  \dot{\overbar{\phi}} {}^{\prime} + \overbar{\phi}'' \cosh ( \gamma ) + \phi'' \sinh ( \gamma )  = 0 \, .
\end{aligned}
\end{equation}
One can integrate these equations with respect to the spatial coordinate $\theta$, up to an undetermined integration constant $h ( t )$ which is an arbitrary function of $t$. As in the discussion around equation (\ref{eom_gauge_choice}), one can always set $h ( t ) = 0$ by a gauge transformation. Specializing to this $h = 0$ gauge, the equations of motion become
\begin{equation}\label{winding_h=0_eom}
    \dot{\phi} - \phi' \cosh ( \gamma ) - \overbar{\phi}' \sinh ( \gamma ) = 0, \quad \dot{\overbar{\phi}} + \overbar{\phi}' \cosh ( \gamma ) + \phi' \sinh ( \gamma ) = 0 \, .
\end{equation}
We wish to solve the equations of motion (\ref{winding_h=0_eom}) subject to the boundary conditions
\begin{equation}
\label{eq:deformedperiodicity}
    \phi(\theta + 2\pi, t) - \phi(\theta, t) = 2\pi p\,, \quad \overbar{\phi} ( \theta + 2\pi, t ) - \overbar{\phi} (\theta, t) = 2 \pi \overbar{p}\,,
\end{equation}
where $p, \overbar{p} \in \mathbb{Z}$. The desired solutions with such periodic boundary conditions are
\begin{equation}
\begin{aligned}
 \phi_p^{(\gamma)}&= p \theta + \left(p \cosh ( \gamma ) + \overbar{p} \sinh ( \gamma )  \right) t \, , \quad
\overbar{\phi}_{\overbar{p}}^{(\gamma)} = \overbar{p} \theta -\left(\overbar{p} \cosh ( \gamma ) + p \sinh ( \gamma ) \right) t\,.
\end{aligned}
\end{equation}
Since these solutions $\phi_{p}^{(\gamma)}$ and $\overbar{\phi}_{\overline{p}}^{(\gamma)}$ depend on $t,\theta$ linearly, the corresponding stress-energy tensor is constant. Therefore, it is reasonable that the energies of these states are indeed governed by the energy formula derived via AdS$_3$/CFT$_2$ holography for constant stress tensor backgrounds, as we found around equation (\ref{zero_mode_works_winding}).\footnote{Strictly speaking, the derivation of this zero mode formula also assumes that the boundary theory is a large-$c$ holographic CFT for which we can trust semiclassical bulk gravity. However, this assumption does not seem strictly necessary for the zero mode formula to hold, since the theory we study here has $c = 1$.}

We would also like to point out that the energies of these states agree with the energies of momentum states for a compact boson with a different radius. To see this, it is convenient to change variables as
\begin{equation}
   w= \sqrt{\pi} \left( p+ \bar{p} \right)\,, \quad \bar{w} =\sqrt{\pi} \left(p- \bar{p}\right)\,, \quad R = \exp \left( - \frac{\gamma}{2} \right) \, ,
\end{equation}
so that the deformed Hamiltonian \eqref{eq:0iouv1f} can be written as
\begin{equation}
\label{eq:okjbgfty2!}
    H^{(\gamma)}_{\text{PR}}  = \frac{1}{2} \left( \frac{w^2}{R^2} + R^2 \bar{w}^2 \right) \, .
\end{equation}
This supports the claim that the root-$\TT$ deformation, in this case, corresponds to a rescaling of the target-space radius for the compact boson. However, to verify this conclusion, we should also study the effect of the deformation in the other sector of the Hilbert space, which describes an infinite tower of harmonic oscillators.

We turn to this task now. Expanding the field $\phi$ and $\overbar{\phi}$ as in \eqref{eq:phiphibarexpan}, we find the Hamiltonian operator and the momentum operator for this oscillator sector are given by
\begin{equation}
\begin{aligned}
    H^{(\gamma)}_{\text{OS}} &= \sum_{n=1}^{\infty} n  (a_n^\dagger a_n + b^\dagger_n b_n)\cosh (\gamma) + \sum_{n=1}^\infty n (a_n^\dagger b_n^\dagger + a_n b_n)\sinh(\gamma ) - \frac{1}{12} \cosh (\gamma)\,, \\
    P^{(\gamma)}_{\text{OS}} &= \sum_{n=1}^{\infty} n(a_n^\dagger a_n - b_n^\dagger b_n) \, , 
\end{aligned}
\end{equation}
where we have performed normal ordering as before and where OS stands for oscillators. This Hamiltonian has exactly the same spectrum as its undeformed counterpart, which can be made manifest by the following Bogoliubov transformation:
\begin{equation}\label{bogoliubov}
\begin{aligned}
    a_n=\tilde{a}_n \cosh \left( \frac{\gamma}{2} \right) - \tilde{b}^\dagger_n \sinh \left( \frac{\gamma}{2} \right) \, , \quad
    b_n= \tilde{b}_n \cosh \left( \frac{\gamma}{2} \right) - \tilde{a}^{\dagger}_n \sinh \left( \frac{\gamma}{2} \right) \, .
\end{aligned}
\end{equation}
We note that this has the same structure as the change of variables which diagonalized the mixing term between the two Chern-Simons gauge fields in equation (\ref{CS_bogoliubov}). This transformation preserves the commutation relation, i.e.
\begin{equation}
    [\tilde{a}_n, \tilde{a}^\dagger_n] = [\tilde{b}_n, \tilde{b}^\dagger_n] = 1~.
\end{equation}
In terms of the new oscillators, the Hamiltonian then reduces to the undeformed one,
\begin{equation}
\label{eq:plkjngtyui12}
    H^{(\gamma)}_{\text{OS}} = - \frac{1}{12} + \sum^\infty_{n=1} n \left(\tilde{a}^{\dagger}_n \tilde{a}_n + \tilde{b}^{\dagger}_n \tilde{b}_n \right)\,,
\end{equation}
while the momentum operator is unchanged,
\begin{equation}
    P_{\text{OS}}^{(\gamma)} = \sum_{n=1}^\infty n \left(\tilde{a}_n^\dagger \tilde{a}_n - \tilde{b}_n^\dagger \tilde{b}_n\right)\,.
\end{equation}
Hence, we conclude that the energies in the oscillator sector of the Hilbert space do not flow under the root-$\TT$ deformation. This agrees with the effect of changing the radius for a compact boson, which likewise does not change the energies of oscillator excitations.

Therefore, combining this result with the flow of $H_{\text{PR}}^{(\gamma)}$, we conclude that indeed the root-$\TT$ deformation corresponds to a change of radius for a single compact boson. 

We have also verified that the zero-mode energy formula (\ref{zero_mode_formula}) proposed in \cite{Ebert:2023tih} does not apply to generic states in a root-$\TT$-deformed CFT. For instance, any state with $p = \overline{p} = 0$ but with oscillator excitations will have an energy that is unchanged by the root-$\TT$ flow, whereas the formula (\ref{zero_mode_formula}) would predict that the energy flows with $\gamma$. This is because such oscillator states have non-constant stress tensors and therefore violate the assumptions under which the zero-mode formula was derived. However, we reiterate that the states which \emph{do} have constant stress tensors -- namely, states with general $p$ and $\overbar{p}$ but no oscillator excitations -- indeed have energies which flow according to the zero mode formula.

\subsubsection*{\ul{\it Multiple compact bosons}}

Next we aim to study the spectrum for the theory of root-$\TT$-deformed free bosons when there are more fields, rather than just a single left-mover and a single right-mover. All of these cases are qualitatively similar, in the sense that the argument of the square root appearing in the Lagrangian is no longer a perfect square, and thus cannot be resolved to a simple product of fields as in the $N = \overbar{N} = 1$ case above. For simplicity, we will therefore focus on the first non-trivial case, which has $N = 2$ left-movers and $\overbar{N} = 1$ right-movers (the case with $N = 1$ and $\overbar{N} = 2$ is identical, after exchanging chiral and anti-chiral fields).

The Hamiltonian for the deformed $(N, \overbar{N}) = (2, 1)$ theory is
\begin{equation}
\begin{aligned}
\label{eq:(2,1)Ham}
    H^{(\gamma)} &=  \int d\theta \left[ \frac{1}{2} \left( \phi^{\prime 2}_1+\phi^{\prime 2}_2 + \overbar{\phi}^{\prime 2}_1  \right) \cosh ( \gamma ) + \sqrt{\phi^{\prime 2}_1 + \phi^{\prime 2}_2} \overbar{\phi}'_1 \sinh ( \gamma )\right] \, .
\end{aligned}
\end{equation}
To resolve the square root, our strategy will be to expand in large positive momenta and compute the energies perturbatively. The mode expansion for the fields takes the form
\begin{equation}\hspace{-10pt}\label{expansion_(2,1)}
    \phi_j = p_j \theta + \frac{1}{\sqrt{2\pi}} \sum_{n=1}^\infty \frac{1}{\sqrt{n}} \left( a_{j, n} e^{in \theta} +a_{j, n}^{\dagger} e^{-in \theta} \right)\,, \quad \overbar{\phi}_1 = \overbar{p}_1 \theta  +\frac{1}{\sqrt{2\pi}}  \sum_{n=1}^\infty \frac{1}{\sqrt{n}} \left( b_{1, n}^{\dagger} e^{in \theta} +b_{1, n} e^{-in \theta} \right)\,,
\end{equation}
where $j = 1, 2$ and periodicity requires $\overbar{p}_1$, $p_1$, $p_2 \in \mathbb{Z}$. Substituting the expansion (\ref{expansion_(2,1)}) into our Hamiltonian (\ref{eq:(2,1)Ham}) and expanding in large $p_1$ and $\overbar{p}_1$, to leading order we find
\begin{equation}
\begin{aligned}
\label{eq:ipou1;p;}
    H^{(\gamma)} &=  \left( \pi ( p_1^2 + p_2^2 + \bar{p}_1^2) - \frac{1}{8} + \sum^\infty_{n=1} n \left( N_{1, n} + N_{2, n} + \bar{N}_{1, n} \right) \right) \cosh ( \gamma) \\ 
    &\qquad + \left(2\pi p_1 \bar{p}_1  +  \sum^\infty_{n=1} n \left( a_{1, n} b_{1, n} + a^\dagger_{1, n} b^\dagger_{1, n} \right) \right) \sinh ( \gamma )  + \cdots \, ,
\end{aligned}
\end{equation}
where $N_{i, n} = a_{i, n}^\dagger a_{i, n}$ and $\overbar{N}_{1, n}= b_{1, n}^\dagger b_{1, n}$ are number operators at level $n$ for left- and right-movers respectively.

We would now like to compare the spectrum of the true large-momentum Hamiltonian (\ref{eq:ipou1;p;}) to the zero-mode formula (\ref{zero_mode_formula}) predicted from holography for states with constant stress tensors. The undeformed Hamiltonian and momentum are
\begin{equation}\label{undeformed_H_P_(2,1)}
\begin{aligned}
    H^{(0)} &= \pi \left( p_1^2 + p_2^2 + \overbar{p}_1^2 \right) - \frac{1}{8} + \sum^\infty_{n=1} n \left( N_{1, n} + N_{2, n} + \bar{N}_{1, n} \right) \,, \\
    P^{(0)} &= \pi \left( p_1^2 + p_2^2 - \overbar{p}_1^2\right) - \frac{1}{24} + \sum^\infty_{n=1} n \left( N_{1, n} + N_{2, n} - \bar{N}_{1, n} \right) \, .
\end{aligned}
\end{equation}
Let us restrict to an eigenstate of both the momentum operators $p_1$, $p_2$, $\overbar{p}_1$ and the number operators $N_{1, n}$, $N_{2, n}$, $\overbar{N}_{1, n}$, in the undeformed theory. The energy and momentum of such a state are also given by the expressions (\ref{undeformed_H_P_(2,1)}), if we simply re-interpret each symbol representing an operator as instead representing the corresponding eigenvalue.\footnote{We have chosen not to denote operators by decorating them with hats, which would distinguish between operators $\widehat{N}_1$ and their corresponding eigenvalues $N_1$, to avoid cluttering the formulas.}

Substituting the energy and momentum eigenvalues for this state into the zero-mode formula (\ref{zero_mode_formula}) then gives a predicted value for a deformed energy:
\begin{align}
\label{eq:iopihjvg2}
    E^{(\gamma)}_{\text{zero mode}} &= \left( \pi \left(p_1^2 + p_2^2 + \bar{p}_1^2 \right)  - \frac{1}{8} + \sum^\infty_{n=1} n \left( N_{1, n} + N_{2, n} + \bar{N}_{1, n} \right) \right) \cosh ( \gamma ) \nonumber \\
    &\qquad + 2\pi p_1 \bar{p}_1 \sinh ( \gamma ) + \cdots \, .
\end{align}
We should stress that equation (\ref{zero_mode_formula}) is a prediction for the deformed \emph{spectrum} and not for the deformed \emph{eigenstates}. Therefore, even if equation (\ref{eq:iopihjvg2}) were correct, this would simply mean that there exists \emph{some} state in the deformed theory whose energy is $E^{(\gamma)}_{\text{zero mode}}$. 

However, with this caveat aside, it is now easy to see why the formula (\ref{eq:iopihjvg2}) is incorrect, and what effect it fails to take into account. Were it not for the final term in the true Hamiltonian (\ref{eq:ipou1;p;}), which involves $\sum^\infty_{n=1} n \left( a_{1, n} b_{1, n} + a^\dagger_{1, n} b^\dagger_{1, n} \right)$, then any eigenstate of the undeformed theory would remain an eigenstate of the deformed theory at this order in the momentum expansion, and its energy would indeed be given by (\ref{eq:iopihjvg2}). This is simply because the first several terms of the true Hamiltonian (\ref{eq:ipou1;p;}) agree with the zero-mode prediction (\ref{eq:iopihjvg2}), after replacing operators with their eigenvalues. However, the presence of this final term in (\ref{eq:ipou1;p;}) means that an eigenstate of the undeformed Hamiltonian will \emph{not} remain an eigenstate in the deformed theory, since terms like $a_{1, n} b_{1, n}$ will mix such a state into other states with different oscillator numbers. We conclude that the zero-mode energy formula (\ref{eq:iopihjvg2}) is not correct for the deformed spectrum, even in this large-momentum limit.

\subsubsection*{\ul{\it Possible interpretation of root-$\TT$ deformation for higher $N$, $\overbar{N}$}}

We have seen that, in the special case $N = \overbar{N} = 1$, the root-$\TT$ deformations of chiral bosons admits a simple interpretation as a rescaling of the target-space radius. This can also be understood from the observation that, for this case, the oscillator sector of the deformed theory is equivalent to that of the undeformed theory due to the Bogoliubov transformation (\ref{bogoliubov}). To conclude this section, we would like to make some speculative remarks about possible generalizations of this interpretation to cases with higher $N$ and $\overbar{N}$, which seem considerably more complicated.

First let us point out that, for the case $(N, \overbar{N}) = (1, 1)$, the Bogoliubov transformation which returns the oscillator sector of the root-$\TT$ deformed theory to its undeformed form also has an analog at the level of the Lagrangian and Hamiltonian densities. Indeed, for the quadratic theory \eqref{eq:1bL}, one can write the Lagrangian and Hamiltonian densities as
 \begin{equation}
 \begin{aligned}\label{big_phi_L_H}
 \mathcal{L}&=  \frac{1}{2} \left( \Phi' \dot{\Phi} - \overbar{\Phi}' \dot{\overbar{\Phi}} \right) - \frac{1}{2} \left( \Phi^{\prime 2} + \overbar{\Phi}^{\prime 2} \right) \,,\\
     \mathcal{H} &= \frac{1}{2} \left( \Phi^{\prime 2} + \overbar{\Phi}^{\prime 2} \right)\,,
\end{aligned}
 \end{equation}
where we have made a field redefinition
\begin{equation} \hspace{-15pt} \label{eq:rotationphis}
 \left( \begin{array}{c}
\Phi\\ \overbar{\Phi}
 \end{array} \right) = \left( \begin{array}{cc}
    \cosh \left( \frac{\gamma}{2} \right)  &  \quad \sinh \left( \frac{\gamma}{2} \right)  \\
   \sinh \frac{\gamma}{2} & \quad  \cosh \left(\frac{\gamma}{2} \right)
 \end{array} \right)  \left( \begin{array}{c}
\phi\\ \overbar{\phi}
 \end{array} \right)\,, \quad  \left( \begin{array}{c}
\phi\\ \overbar{\phi}
 \end{array} \right) = \left( \begin{array}{cc}
    \cosh \left( \frac{\gamma}{2} \right)  &  \quad - \sinh \left( \frac{\gamma}{2} \right)  \\
 -  \sinh \left( \frac{\gamma}{2} \right) & \quad  \cosh \left( \frac{\gamma}{2} \right)
 \end{array} \right)  \left( \begin{array}{c}
\Phi\\ \overbar{\Phi}
 \end{array} \right) \, .
\end{equation}
The deformed equations of motion, written in terms of the new fields $\Phi$ and $\overbar{\Phi}$, are
\begin{equation}
\label{eq:0i9uy232}
    \Phi'' = \dot{\Phi}', \quad \overbar{\Phi}'' = - \dot{\overbar{\Phi}}' \, ,
\end{equation}
which take the same form as those in the undeformed theory. Again, this is analogous to the field redefinition (\ref{CS_bogoliubov}) in the Chern-Simons setting, which undoes a similar quadratic mixing between the barred and unbarred fields induced by a $J \overbar{J}$ deformation.

Next let us consider how this observation might extend to multiple bosons. We focus on the case of $N = \overbar{N}$ for simplicity. The deformed Hamiltonian density for an equal number of left- and right-moving chiral bosons is
\begin{equation}\label{ham_density_field_redef}
\mathcal{H}^{(\mu)} = \frac{1}{2}\left( \phi^{\prime}_j \phi^{\prime}_j + \overbar{\phi}^{\prime}_{\overbar{j}} \overbar{\phi}^{\prime }_{\overbar{j}}  \right) \cosh ( \gamma ) + \sqrt{ \phi^{\prime}_j \phi^{\prime}_j  \overbar{\phi}^{\prime}_{\overbar{j}} \overbar{\phi}^{\prime}_{\overbar{j}}   } \sinh ( \gamma ) \, . 
\end{equation}
We now ask whether some more complicated field redefinition might return this Hamiltonian to a quadratic one, as in the case of (\ref{big_phi_L_H}). When $N=\overbar{N}=2$, at least formally, one can attempt to perform a change of variables that resembles a transformation to polar coordinates in a $2d$ target space:
\begin{equation}
\begin{aligned}
\label{eq:fieldred2}
\left( \begin{array}{c}
\phi^{\prime}_1(\theta, t)\\\phi^{\prime}_2 (\theta, t)
\end{array} \right) & = \left( \begin{array}{cc}
   r'(\theta, t) \cos \left( \Theta'(\theta, t) \right)   \\   r'(\theta, t)  \sin \left( \Theta'(\theta, t) \right)
\end{array} \right)\,, \quad \left( \begin{array}{c}
\overbar{\phi}^{\prime}_1(\theta, t)\\\overbar{\phi}^{\prime}_2 (\theta, t)
\end{array} \right)  = \left( \begin{array}{cc}
 \overbar{r}'(\theta, t)\cos \left( \overbar{\Theta}'(\theta, t) \right)  \\  \overbar{r}'(\theta, t) \sin \left( \overbar{\Theta}'(\theta, t) \right)
\end{array} \right)\,.
\end{aligned}
\end{equation}
Here we interpret $\Theta'(\theta, t)$ and $r'(\theta, t)$ as spatial derivatives of new fields which depend on the derivatives $\phi' ( \theta, t )$ in a nonlinear way. In terms of these quantities, the Hamiltonian density (\ref{ham_density_field_redef}) with $N = \overbar{N} = 2$ takes the form
\begin{equation}
\begin{aligned}\label{eq:H1rdef}
    \mathcal{H}^{(\mu)}&= \frac{1}{2} \left( \phi^{\prime 2}_1 + \phi^{\prime 2}_2 + \overbar{\phi}^{\prime 2}_1 + \overbar{\phi}^{\prime 2}_2 \right) \cosh \left( \gamma \right) +  \sqrt{\left(\phi^{\prime 2}_1 + \phi^{\prime 2}_2 \right) \left( \overbar{\phi}^{\prime 2}_1 + \overbar{\phi}^{\prime 2}_2 \right)} \sinh \left( \gamma \right)\\ &= \frac{1}{2} \left( r^{\prime 2} + \overbar{r}^{\prime 2} \right) \cosh \left( \gamma \right) + r^\prime \overbar{r}^\prime \sinh \left( \gamma \right)\,,
\end{aligned}
\end{equation}
where we assumed $r'\overbar{r}'>0$ in order to simplify the square root. Now we perform a second field redefinition, just as in (\ref{eq:rotationphis}), to a new field $\rho$: 
\begin{equation}\label{eq:rotationphisrho}
    \left( \begin{array}{c}
r(\theta, t)\\ \overbar{r}(\theta, t)
    \end{array} \right) = \left(\begin{array}{cc}
     \cosh \left( \frac{\gamma}{2} \right)    & \quad - \sinh \left( \frac{\gamma}{2} \right)  \\
      - \sinh \left( \frac{\gamma}{2} \right)   & \quad  \cosh \left( \frac{\gamma}{2} \right)
    \end{array}\right)  \left( \begin{array}{c}
\rho(\theta, t)\\ \overbar{\rho}(\theta, t)
    \end{array} \right)\,.
\end{equation}
Expressing the Hamiltonian density \eqref{eq:H1rdef} in terms of the $\rho$ variables rather than the $r$ variables, we conclude
\begin{equation}\label{eq:simpleH2-2}
    \mathcal{H}^{(\mu)} = \frac{1}{2} \left( \rho^{\prime 2} + \overbar{\rho}^{\prime 2} \right)\,.
\end{equation}
Therefore, again at a formal classical level, it appears that this series of field redefinitions has returned the Hamiltonian density to that of the free theory. Furthermore, the latter change of variables (\ref{eq:rotationphisrho}) can be interpreted as rescaling the overall target space radius $r$, much as in the $(N, \overbar{N}) = (1, 1)$ case. For a larger number of bosons $N = \overbar{N} > 2$, one can perform a similar series of manipulations using higher-dimensional spherical coordinates.

Several technical issues preclude us from taking this series of field redefinitions seriously, at least without further investigation. First, the change of variables (\ref{eq:fieldred2}) was at the level of \emph{derivatives} of the fields, and it is not clear that this corresponds to a sensible change of variables for the fields themselves. Second, all of these manipulations have been purely classical, and it is not guaranteed that one could make sense of these field redefinitions within a path integral (which would produce Jacobian factors from each change of variables). And third, we have not been careful about the identifications that each field is subject to. For instance, if indeed the field $\Theta$ can be interpreted as a target-space angle in polar coordinates, then it should be subject to the identification $\Theta \sim \Theta + 2 \pi$.

Nonetheless, it would be very interesting if an argument of this form could be used to endow the root-$\TT$ deformation of $N$ chiral and anti-chiral bosons with a geometrical target-space interpretation.

\section{Perturbative Quantization Using Background Field Method} \label{sec:cian}

In the preceding sections, we have considered interacting theories with arbitrary numbers $N ,\overbar{N}$ of chiral and anti-chiral bosons, respectively, and sacrificed manifest Lorentz invariance in order to use a first-order formulation which is convenient for canonical quantization. In the special case $N = \overbar{N}$, however, we also have the option of assembling the field content of our theory into $N$ \emph{non-chiral} bosons by summing the left-movers and right-movers:
\begin{align}
    \varphi^i = \frac{1}{\sqrt{2}} \left( \phi^i + \overbar{\phi}^{i} \right) \, .
\end{align}
Here we now use the same index $i = 1 , \ldots, N$ for both the chiral and anti-chiral fields, rather than distinct indices $i$ and $\overline{i}$. As this change of variables is merely a field redefinition, stress tensor deformations of such a theory of $N$ bosons must be equivalent, regardless of whether the theory is presented in terms of left-movers and right-movers $\phi^i$, $\overbar{\phi}^i$, or in terms of their non-chiral counterparts $\varphi^i$. Indeed, for the case of the $\TT$ deformation of a free seed theory, this equivalence was checked explicitly in \cite{Ouyang:2020rpq}.

In this section, we will provide a complementary analysis of the perturbative quantization of the Modified Scalar theory using this presentation in terms of non-chiral fields $\varphi^i$. For concreteness, we will focus on the case where both the fields $\varphi^i$ and the Lorentzian spacetime coordinates $(t, x)$ are non-compact, and we will use middle Greek letters like $\mu$, $\nu$ (rather than early Greek letters like $\alpha$, $\beta$, which were used in sections \ref{sec:classical} and \ref{sec:cs}) for spacetime indices in this section. We will write $g_{\mu \nu}$ for the (Minkowski) spacetime metric.

In terms of the non-chiral fields $\varphi^i$, the Lagrangian for the Modified Scalar theory can be written in the manifestly Lorentz-invariant form
\begin{align}\label{modscalar_lorentz_invariant_lagrangian}
    \mathcal{L} = \frac{1}{2} \left( \cosh \left( \gamma \right) \partial_{\mu} \varphi^{i}\partial^{\mu} \varphi^{i} + \sinh \left( \gamma \right) \sqrt{2 \left( \partial_\mu \varphi^{i} \partial^{\nu} \varphi^{i} \right) \left( \partial_\nu \varphi^{j} \partial^{\mu} \varphi^{j} \right) - \left( \partial_\mu \varphi^{i} \partial^{\mu} \varphi^{i} \right)^2 } \right) \, .
\end{align}
The advantage of this representation is that one can more easily apply standard diagrammatic techniques to compute loop corrections in the quantum theory. Of course, the second term in the Lagrangian (\ref{modscalar_lorentz_invariant_lagrangian}) is still non-analytic around the vacuum of the theory, or around any field configuration for which
\begin{align}
    \partial_\mu \varphi^i = 0 \, .
\end{align}
We will circumvent this issue by working in a background field expansion around a field configuration $\varphi^i$ for the scalars which we assume has non-zero gradients and which satisfies the classical equations of motion for the theory, but which is otherwise arbitrary. 

\subsection{Background field expansion and Feynman rules}

Throughout this section, we will use the notation
\begin{align}
    \varphi^i = C^i + Q^i \, , 
\end{align}
where $C^i$ is a classical (background) field configuration around which we perform our expansion, and $Q^i$ is a quantum field which is allowed to fluctuate within the path integral. This classical background $C^i$ is the analog of the large-momentum configuration around which we performed our expansion in section \ref{sec:quantumspectrum}. Our goal will be to investigate the terms which contribute to the quantum effective action, as a function of the background $C^i$.

To avoid cluttering the formulas, it will also be convenient to adopt the following shorthand for spacetime derivatives of the various fields:
\begin{align}
    \tensor{\varphi}{_\mu^i} = \partial_\mu \varphi^i \, , \quad \tensor{C}{_\mu^i} = \partial_\mu C^i \, , \quad \tensor{Q}{_\mu^i} = \partial_\mu Q^i \, .
\end{align}
In our analysis of chiral boson theories, we introduced two useful quantities $S$ and $P$ in equation (\ref{S_and_P_invariants}) which were independent combinations of derivatives of the scalar fields. In the present non-chiral analysis, let us similarly introduce the quantities
\begin{align}\label{non_chiral_S_P_defn}
    S = \tensor{\varphi}{_{\mu}^{i}} \tensor{\varphi}{^{\mu i}} \, , \qquad P^2 = \tensor{\varphi}{_{\mu}^{i}} \tensor{\varphi}{^{\nu i}} \tensor{\varphi}{_{\nu}^{j}} \tensor{\varphi}{^{\mu j}} \, .
\end{align}
We note that these are not the precise analogs of $S$ and $P$ in the chiral setting; for instance, the role of the combination $S^2 - P^2$ in section \ref{sec:classical} is now played by $2 P^2 - S^2$. Therefore, in terms of these quantities (\ref{non_chiral_S_P_defn}), the Modified Scalar Lagrangian (\ref{modscalar_lorentz_invariant_lagrangian}) can be written as
\begin{align}\label{modscal_S_P_form}
    \mathcal{L} = \frac{1}{2} \left( \cosh ( \gamma ) S + \sinh ( \gamma ) \sqrt{ 2 P^2 - S^2 } \right) \, .
\end{align}
We decompose $S$ into a classical piece $S_C$ and a quantum piece $S_Q$, along with a cross term:
\begin{align}
    S &= \left( \tensor{C}{_\mu^i} + \tensor{Q}{_\mu^i} \right) \left( \tensor{C}{^\mu^i} + \tensor{Q}{^\mu^i} \right)  \nonumber \\
    &= \underbrace{\tensor{C}{_{\mu}^{i}} \tensor{C}{^{\mu i}}}_{S_C} + 2 \tensor{C}{_{\mu}^{i}} \tensor{Q}{^{\mu i}} + \underbrace{\tensor{Q}{_{\mu}^{i}} \tensor{Q}{^{\mu i}}}_{S_Q} \, .
    % \tensor{P}{_\mu^\nu} &= \left( \tensor{C}{_\mu^i} + \tensor{Q}{_\mu^i} \right) \left( \tensor{C}{^\nu^i} + \tensor{Q}{^\nu^i} \right) \nonumber \\
    % &= \underbrace{\tensor{C}{_{\mu}^{i}} \tensor{C}{^{\nu i}}}_{\tensor{\left( P_C \right)}{_\mu^\nu}} + \tensor{Q}{_{\mu}^{i}} \tensor{C}{^{\nu i}} + \tensor{C}{_{\mu}^{i}} \tensor{Q}{^{\nu i}} + \underbrace{\tensor{Q}{_{\mu}^{i}} \tensor{Q}{^{\nu i}}}_{\tensor{\left( P_Q \right)}{_\mu^\nu}} \, .
\end{align}

Next we will consider the splitting of $S^2$ and $P^2$ into classical and quantum pieces. Because we assume that the field configuration $C^i$ is a solution to the classical equations of motion, by definition the action is stationary to linear order when expanding around such a solution. This means that the effective action cannot contain any terms which are linear in the fluctuation field $Q^i$, because the sum of all such contributions must conspire to form an on-shell total derivative. We will therefore label all terms linear in $Q^{\mu i}$ as ``on-shell deriv.'' and ignore them in what follows, although with the caveat that \emph{individual} terms of this form need not separately drop out; we are only guaranteed that the \emph{combined} effect of all such terms is to form an on-shell total derivative.

With this in mind, the quantity $S^2$ can be expanded as
\begin{align}
    S^2 &= S_C^2 + \underbrace{4S_C \tensor{C}{_{\mu}^{i}} \tensor{Q}{^{\mu i}}}_{\text{on-shell deriv.}} + 2 S_C S_Q  + 4\tensor{C}{_{\mu}^{i}}Q^{\mu i} \tensor{C}{_\nu^{j}} Q^{\nu j}  + \underbrace{4S_Q \tensor{C}{_{\mu}^{i}} Q^{\mu i}}_{\mathcal{O}\left( Q^3 \right) }+ \underbrace{S_Q^2}_{\mathcal{O}\left( Q^{4} \right) } \nonumber \\
    &\simeq S_C^2 + 2 S_C S_Q  + 4\tensor{C}{_{\mu}^{i}}Q^{\mu i} \tensor{C}{_\nu^{j}} Q^{\nu j} \, ,
\end{align}
where the symbol $\simeq$ means equal modulo all terms that are either linear in $Q^i$ (which will form on-shell total derivatives) or that are of cubic order or higher in $Q^i$ (which do not contribute to the one loop effective action). A similar computation for $P^2$ gives
\begin{align}
    P^2 &= \tensor{C}{_{\mu}^{i}} \tensor{C}{^{\mu j}} \tensor{C}{_{\nu}^{i}} \tensor{C}{^{\nu j}} + \underbrace{4\tensor{C}{_{\mu}^{i}} \tensor{C}{^{\mu}^{j}} \tensor{C}{^{\nu}^{i}} \tensor{Q}{_{\nu}^{j}}}_{\text{on-shell deriv.}} + 2\tensor{C}{_{\mu}^{i}} C^{\nu i} \tensor{Q}{_{\nu}^{j}} \tensor{Q}{^{\mu j}} + 2\tensor{Q}{_{\mu}^{i}} C^{\nu i} \tensor{Q}{_{\nu}^{j}} C^{\mu j} \nonumber \\
    &\qquad + 2\tensor{Q}{_{\mu}^{i}} C^{\nu i} \tensor{C}{_{\nu}^{j}} Q^{\mu j} + \mathcal{O}\left( Q^3 \right) \, \nonumber \\
    &\simeq  \tensor{C}{_{\mu}^{i}} \tensor{C}{^{\mu j}} \tensor{C}{_{\nu}^{i}} \tensor{C}{^{\nu j}}  + 2\tensor{C}{_{\mu}^{i}} C^{\nu i} \tensor{Q}{_{\nu}^{j}} \tensor{Q}{^{\mu j}} + 2\tensor{Q}{_{\mu}^{i}} C^{\nu i} \tensor{Q}{_{\nu}^{j}} C^{\mu j}  + 2\tensor{Q}{_{\mu}^{i}} C^{\nu i} \tensor{C}{_{\nu}^{j}} Q^{\mu j} \, .
\end{align}
Therefore, the combination $2 P^2 - S^2$ under the square root in (\ref{modscal_S_P_form}) has an expansion
\begin{align}\label{2psq_sq_quadratic}
    2P^2 - S^2 &\simeq 2 P_C^2 - S_C^2 - 2 S_C S_Q  - 4\tensor{C}{_{\mu}^{i}}Q^{\mu i} \tensor{C}{_\nu^{j}} Q^{\nu j} + 4\tensor{C}{_{\mu}^{i}} C^{\nu i} \tensor{Q}{_{\nu}^{j}} \tensor{Q}{^{\mu j}} + 4\tensor{Q}{_{\mu}^{i}} C^{\nu i} \tensor{Q}{_{\nu}^{j}} C^{\mu j} \nonumber \\
    &\qquad + 4\tensor{Q}{_{\mu}^{i}} C^{\nu i} \tensor{C}{_{\nu}^{j}} Q^{\mu j} \, \nonumber \\
    &\equiv 2 P_C^2 - S_C^2 + 2 \mathcal{Q}_1 \, .
\end{align}
Here we introduce the shorthand $\mathcal{Q}_1$ which is proportional to the correction to the classical part of (\ref{2psq_sq_quadratic}) up to quadratic order in fluctuations,
\begin{align}
    \mathcal{Q}_1 =  - S_C S_Q  - 2 \tensor{C}{_{\mu}^{i}}Q^{\mu i} \tensor{C}{_\nu^{j}} Q^{\nu j} + 2 \tensor{C}{_{\mu}^{i}} C^{\nu i} \tensor{Q}{_{\nu}^{j}} \tensor{Q}{^{\mu j}} + 2 \tensor{Q}{_{\mu}^{i}} C^{\nu i} \tensor{Q}{_{\nu}^{j}} C^{\mu j} + 2 \tensor{Q}{_{\mu}^{i}} C^{\nu i} \tensor{C}{_{\nu}^{j}} Q^{\mu j} \, , 
\end{align}
which is not to be confused with $Q^i$ or $\tensor{Q}{_\mu^i} = \partial_\mu Q^i$. Let us also define $\mathcal{Q}_2 \simeq \mathcal{Q}_1^2$ to be the square of this quantity, retaining terms only up to second order in $Q^i$, so that
\begin{align}
    \mathcal{Q}_2 = 4S_C^2 \tensor{C}{_{\mu}^{i}}Q^{\mu i} \tensor{C}{_{\nu}^{j}} Q^{\nu j} - 8 S_C \tensor{C}{_{\mu}^{i}}Q^{\mu i} \tensor{C}{_{\nu}^{j}} C^{\nu k} C^{\rho j} \tensor{Q}{_{\rho}^{k}} + 16 \left( \tensor{C}{_{\nu}^{j}} C^{\nu k} C^{\rho j} \tensor{Q}{_{\rho}^{k}} \right)^2 \, .
\end{align}
In terms of these combinations, we can expand the square root appearing in (\ref{modscal_S_P_form}) as
\begin{align}\label{sqrt_expansion_Q}
    \sqrt{ 2 P^2 - S^2 } \simeq \sqrt{ 2 P_C^2 - S_C^2 } + \frac{\mathcal{Q}_1}{ \sqrt{ 2 P_C^2 - S_C^2  } } - \frac{\mathcal{Q}_2}{2 \left( 2 P_C^2 - S_C^2  \right)^{3/2}} \, .
\end{align}
Finally, we can express the Modified Scalar Lagrangian expanded to quadratic order in fluctuations around a given classical solution as
\begin{align}
    \mathcal{L} \simeq \mathcal{L}_C + \frac{1}{2} \left( \cosh ( \gamma ) S_Q + \sinh ( \gamma ) \left( \frac{\mathcal{Q}_1}{ \sqrt{ 2 P_C^2 - S_C^2  } } - \frac{\mathcal{Q}_2}{2 \left( 2 P_C^2 - S_C^2  \right)^{3/2}} \right) \right) \, ,
\end{align}
where $\mathcal{L}_C$ represents the Lagrangian evaluated on the background solution $C^i$, i.e.
\begin{align}
    \mathcal{L}_C &\equiv \frac{\cosh ( \gamma ) }{2}S_C  + \frac{\sinh ( \gamma ) }{2}\sqrt{2P_C^2 - S_C^2} \, .
\end{align}
It is also convenient to write the Lagrangian for the quantum field $Q^i$ in terms of a bilinear form. Defining the tensor
\begin{align}\label{P_tensor}
    \tensor{P}{_{\mu \nu}^{ij}} &=  - \bigg( \frac{- S_C g_{\mu \nu} \delta^{ij} - 2 \tensor{C}{_{\mu}^{i}} \tensor{C}{_{\nu}^{j}} + 2 \tensor{C}{_{\mu}^{k}} \tensor{C}{_\nu^{k}} \delta^{ij} + 2 \tensor{C}{_{\mu}^{j}} \tensor{C}{_{\nu}^{i}} + 2 \tensor{C}{_{\rho}^{i}}C^{\rho j} g_{\mu \nu}}{2 \sqrt{2P_C^2 - S_C^2} }\nonumber \\
    &\qquad \qquad -\frac{ 4S_C^2 \tensor{C}{_{\mu}^{i}} \tensor{C}{_{\nu}^{j}} - 8 S_C \tensor{C}{_{\mu}^{i}} \tensor{C}{_{\rho}^{k}} \tensor{C}{^{\rho}^{j}}\tensor{C}{_{\nu}^{k}} + 16 \tensor{C}{_{\rho}^{k}} \tensor{C}{^{\rho i}}\tensor{C}{_{\mu}^{k}} \tensor{C}{_{\tau}^{m}} \tensor{C}{^{\tau j}} \tensor{C}{_{\nu}^{m}}} {4 \left( 2 P_C^2 - S_C^2 \right)^{\frac{3}{2}}} \bigg)  \, , 
\end{align}
we can write the Lagrangian $\mathcal{L}_Q$ for the fluctuating field as
\begin{align}
    \mathcal{L}_Q &= Q^{\mu i}\left( \frac{\cosh ( \gamma ) }{2}g_{\mu \nu} \delta^{ij} + \sinh ( \gamma ) \tensor{P}{_{\mu \nu}}^{ij}  \right) Q^{\nu j} \, , 
\end{align}
or after integrating by parts to move the derivative acting on $Q^{\mu i} = \partial^\mu Q^i$, as
\begin{align}
    \mathcal{L}_Q &=  - Q^{i} \left( \frac{\cosh ( \gamma ) }{2} \delta^{ij} \partial^2 +  \sinh ( \gamma ) \left( \partial^{\mu} \tensor{P}{_{\mu \nu}^{ij}} \right) \partial^\nu + \sinh ( \gamma )  \tensor{P}{_{\mu \nu}^{ij}} \partial^\mu \partial^\nu\right)  Q^{j} \label{eq:SFQuantum_Lagrangian} \, .
\end{align}
The first term in (\ref{eq:SFQuantum_Lagrangian}) is proportional to a conventional free kinetic term for the fields $Q^i$. The second and third terms, involving $\tensor{P}{_{\mu \nu}^{ij}}$ and its derivative, encode the interactions which are induced by expanding around the classical field configuration $C^i$.

\subsubsection*{\ul{\it Feynman rules}}

Now that we have obtained the Lagrangian (\ref{eq:SFQuantum_Lagrangian}), we may read off the Feynman rules which we will need for computing diagrams. The propagator for the quantum field is
\begin{align}
        \tensor{D}{^{ij}}  &= -\frac{i}{\cosh ( \gamma ) } \frac{\delta^{ij}}{k^2 } \, . \label{eq:SFpropagator}
\end{align}
Next we must work out the vertex associated to the interaction between $Q^i$ and the classical field via the combination $\tensor{P}{_{\mu \nu}^{m n}}$. We will draw quantum fields as solid lines and the cumulative effect of the background fields as a single coiled line. Consider the trivalent interaction between a field $Q^i$ with momentum $p$, a field $Q^j$ with momentum $q$, and an insertion of the background $\tensor{P}{_{\mu \nu}^{m n}}$ with momentum $r$. This vertex can be visualized as
\begin{align}\label{interaction_vertex}
    \raisebox{-0.5\height}
    {\includegraphics[width=0.38\linewidth]{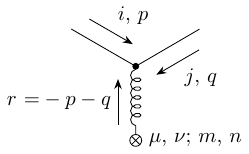}} \; .
\end{align}
Let the vertex factor for this interaction be $g_{ij}$.\footnote{The vertex factor $g_{ij}$ should not be confused with the target-space metric $G_{ij} ( \phi )$ for the bosons which appears in equation (\ref{general_class_vielbeins}). We also note that the value $r$ of the classical field momentum must be integrated over in this trivalent interaction, but we do not include this integral in the expression (\ref{vertex_value}) for $g_{ij}$.} There are four ways that we can get a contribution to this factor from the Lagrangian (\ref{eq:SFQuantum_Lagrangian}). First, there is a piece arising from the term $\sinh ( \gamma ) \left( \partial^{\mu} \tensor{P}{_{\mu \nu}^{m n}} \right) \partial^\nu$ when $m = j$ and $n = i$, which gives a term proportional to $r^\mu q^\nu$ because of the first derivative $\partial^\mu$ acting on $\tensor{P}{_{\mu \nu}^{m n}}$ and the second derivative $\partial^\nu$ acting on $Q^j$. There is another term which arises from $m = i$ and $n = j$, where the derivative acts on $Q^i$ to produce a factor of $p^\nu$. Then there are two more contributions from the term $\sinh ( \gamma )  \tensor{P}{_{\mu \nu}^{m n}} \partial^\mu \partial^\nu$, when either $m = j$ and $n = i$, or when $m = i$ and $n = j$, which come with factors of $q^\mu q^\nu$ or $p^\mu p^\nu$ from the two derivatives acting on $Q^j$ or $Q^i$, respectively. Altogether, the value of this vertex is
\begin{align}\label{vertex_value}
    g^{ij} &= i \sinh ( \gamma ) \left( \delta^{m j} \delta^{ni} P_{\mu \nu}^{mn} r^\mu q^\nu + \delta^{m i} \delta^{nj} P_{\mu \nu}^{mn} r^\mu p^\nu + \delta^{m j} \delta^{ni} P_{\mu \nu}^{mn} q^\mu q^\nu + \delta^{m i} \delta^{nj} P_{\mu \nu}^{mn} p^\mu p^\nu \right) \nonumber \\
    &= i \sinh ( \gamma ) P_{\mu \nu}^{mn}  \left( - \delta^{m j} \delta^{ni} \left( q^\mu + p^\mu \right) q^\nu - \delta^{m i} \delta^{nj} \left( q^\mu + p^\mu \right) p^\nu + \delta^{m j} \delta^{ni} q^\mu q^\nu + \delta^{m i} \delta^{nj}  p^\mu p^\nu \right) \nonumber \\
    &= - 2 i \sinh ( \gamma ) P_{\mu \nu}^{mn} \left( \delta^{mj} \delta^{ni} p^\mu q^\nu + \delta^{mi} \delta^{nj} q^\mu p^\nu \right) \nonumber \\
    &= - 2 i \sinh ( \gamma ) P_{\mu \nu}^{ij} p^\mu q^\nu \, ,
\end{align}
where in the second step we have used $r^\mu = - q^\mu - p^\mu$, and in the last step we have used $P_{\mu \nu}^{ij} = P_{\nu \mu}^{ji}$. This gives the desired value of the trivalent vertex $g_{ij}$ between $Q^i$, $Q^j$, and the classical background. However, in the calculations that follow, it will be convenient to factor out the dependence on $P_{\mu \nu}^{mn}$ and use an ``uncontracted'' vertex factor $\widetilde{g}$ defined by
\begin{align}\label{uncontracted_vertex_factor}
    g_{ij} &= P_{\mu \nu}^{mn} \left( \widetilde{g}^{mn}_{ij} \right)^{\mu \nu} \, , \nonumber \\
    \left( \tilde{g}^{mn}_{ij} \right)^{\mu \nu} &= - 2 i \sinh ( \gamma ) \tensor{\delta}{^m_i} \tensor{\delta}{^n_j} p^\mu q^\nu \, .
\end{align}
Let us emphasize that $\left( \tilde{g}^{mn}_{ij} \right)^{\mu \nu}$ is \emph{not} the full value of the interaction vertex, but rather a useful intermediate quantity which has removed all factors of $P_{\mu \nu}^{mn}$. After computing Feynman diagrams using ``uncontracted'' vertices $\tilde{g}$, we must contract the final result with one factor of $P_{\mu \nu}^{mn}$ for each vertex in order to recover the true value of the diagram.

\subsection{Quantum effective action}

We are now ready to compute the leading quantum corrections to the Modified Scalar Lagrangian. Most of our discussion will focus on the one-loop effective action, defined by the first term beyond the classical contribution in the expansion
\begin{align}\label{effective_action}
    \Gamma [ C^i ] = S [ C^i ] + \frac{i}{2} \Tr \left[ \log \left( \frac{\delta^2 S}{\delta \varphi^i \, \delta \varphi^j } \right) \Bigg\vert_{\varphi^k = C^k} \right] + \cdots \, . 
\end{align}
Although we are primarily interested in the one-loop contribution to $\Gamma$, we will also present some partial results concerning corrections at higher loop order.

There are several techniques for computing the one-loop effective action $\Gamma$. One way is to use heat kernel methods; we will not pursue this strategy here, but we refer the reader to the thesis \cite{pinelliThesis} for a discussion of this approach in the related context of the $4d$ ModMax theory. Rather, we will compute contributions to the effective action perturbatively, using the Feynman rules derived in the preceding subsection. This amount to a diagrammatic evaluation of the one-loop determinant of the operator $\frac{\delta^2 S}{\delta \varphi^i \, \delta \varphi^j }$, which is the operator appearing in $\mathcal{L}_Q$ that we have computed in equation (\ref{eq:SFQuantum_Lagrangian}).

In particular, our goal is to evaluate divergent Feynman diagrams in the Modified Scalar theory using dimensional regularization, as a function of the background configuration $C^i$. Each such divergent contribution necessitates the addition of an appropriate counterterm to cancel the divergence. The collection of all such counterterms which must be added to the classical Lagrangian therefore reproduces the additional terms that appear in the quantum effective action, giving a characterization of the corrections in the expansion (\ref{effective_action}).

\subsubsection*{\ul{\it Constant background, one-loop diagrams}}

Let us begin by considering the simpler case in which the background field configuration $C^i$ is linear in the spacetime coordinates, which means that the classical field has constant gradients. That is, we assume that $\tensor{C}{_\mu^i} = \partial_\mu C^i$ is constant for such backgrounds, so that $\partial_\mu \tensor{C}{_{\nu}^i} = 0$ for all $\mu, \nu, i$. In this case, no momentum can flow through the classical fields in the interaction vertex (\ref{interaction_vertex}), which implies that $r = 0$ and thus $p = - q$.

To obtain the one-loop effective action $\Gamma$, we must evaluate all Feynman diagrams built from the quantum field propagator and interaction vertex (\ref{interaction_vertex}) which contain at most one loop. This corresponds to an infinite series of diagrams given by
\begin{align}\label{infinite_series_constant_bg}
    \Gamma = \raisebox{-0.5\height}{\includegraphics[width=0.2\linewidth]{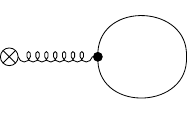}} \; + \; \raisebox{-0.5\height}{\includegraphics[width=0.25\linewidth]{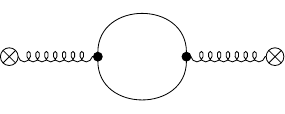}} \; + \; \raisebox{-0.5\height}{\includegraphics[width=0.2\linewidth]{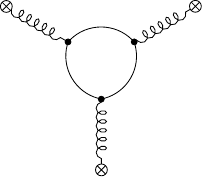}} \; + \; \cdots \, .
\end{align}
Let $\mathcal{D}_n$ represent the value of the diagram in the series (\ref{infinite_series_constant_bg}) which has $n$ insertions of the classical background. The first diagram in this infinite series is
\begin{align}
    \mathcal{D}_1 = \raisebox{-0.5\height}
 {\includegraphics[width=0.25\linewidth]{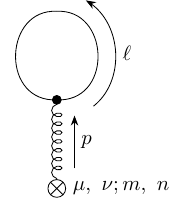}} \, .
\end{align}
Following the comments around equation (\ref{uncontracted_vertex_factor}), we will evaluate this diagram -- and the others in this section -- by using the Feynman rule associated with the uncontracted vertex factor $\tilde{g}$, and then contracting with $P_{\mu \nu}^{m n}$. Doing this and simplifying the resulting sum of Kronecker delta functions using symmetry, one finds
\begin{align}
    \mathcal{D}_1 &= P_{\mu \nu}^{m n} \sinh ( \gamma ) \delta^{im} \delta^{jn} \int \frac{\dd{^{d}\ell}}{\left( 2\pi \right)^{d}} \left( 2i \ell^{\mu} \ell^{\nu} \right) D^{ij} \, .
\end{align}
A term in the integrand which is proportional to $\ell^\mu \ell^\nu$ will produce a result which scales like $\ell^2$ and which is a symmetric tensor in $\mu$ and $\nu$. The only constant symmetric $2$-tensor in the problem is the spacetime metric $g^{\mu \nu}$, so the integral of such a term must be proportional to $\ell^2 g^{\mu \nu}$. By taking the trace, one can fix the dimensionless constant to be $\frac{1}{d}$. Thus, within the integral, we can make the replacement
\begin{align}\label{sym_replacements_1}
    \ell^{\mu} \ell^{\nu} &\to \frac{1}{d} \ell^2 g^{\mu \nu} \, .
\end{align}
Using this replacement and the propagator (\ref{eq:SFpropagator}), we find
\begin{align}
    \mathcal{D}_1 &= - \frac{2\tanh ( \gamma )}{d}  P_{\mu \nu}^{m n} \delta^{m n} \int \frac{\dd{^{d}\ell}}{\left( 2\pi \right)^{d}} g^{\mu \nu} \, .
\end{align}
The integrand is now independent of $\ell$. Although this integral diverges as $\Lambda^{d}$ with a na\"ive cutoff at momentum $\Lambda$, within dimensional regularization it is exactly zero \cite{Anselmi:2019pdm}. 

This result relies only on the momentum dependence of the integral. However, note that the insertions of additional vertices appearing in the higher one-loop diagrams $\mathcal{D}_n$ will not change the momentum dependence of the integral. In general, we will have $n$ propagators $D_{ij}$ of the form (\ref{eq:SFpropagator}), each of which is proportional to $\frac{1}{\ell^2}$, and $n$ copies of the vertex factor (\ref{uncontracted_vertex_factor}). Because the vertex factor contains products of momenta like $\ell^\mu \ell^\nu$, the integrand of $\mathcal{D}_n$ will involve a product of $2n$ momenta. We can replace such factors using a generalization of the argument which led to the replacement rule (\ref{sym_replacements_1}). That is, any integral involving a totally symmetric product of $2n$ momenta must yield a result which is proportional to $\ell^{2n}$ multiplied by a totally symmetrized combination of $n$ metric tensors, since the metric is the only symmetric tensor in the problem. This leads to the replacement
\begin{align}
    \prod_{i=1}^{n} \ell^{\mu_{2i- 1 }} \ell^{\mu_{2i}} \to \frac{\ell^{2n} \left( d -2 \right)!! \left( 2n - 1 \right)!!}{\left( d-2 + 2n \right)!!} g^{(\mu_{1}\mu_2} \cdots g^{\mu_{2n - 1} \mu_{2n})} \, ,\label{eq:symmetrization}
\end{align}
where we have used the double factorial $n!! = n \cdot (n-2) \ldots 4 \cdot 2$.
We thus find an overall factor of $\ell^{2n}$ from the vertex factors, in addition to a compensating factor of $\frac{1}{\ell^{2n}}$ from the $n$ copies of the propagator, each of which scales like $\frac{1}{\ell^2}$. Note that all of these momenta are equal due to momentum conservation around the loop, as we assumed that no momentum can be carried by the classical fields, so the powers of loop momentum precisely cancel. Therefore, every diagram $\mathcal{D}_n$ involves an integrand which is independent of momentum, and thus vanishes in dimensional regularization just as $\mathcal{D}_1$ does.

We conclude that the perturbative one-loop effective action $\Gamma [ C^i ]$, with constant background field strength $\tensor{C}{_\mu^i}$, vanishes in dimensional regularization. This implies that under these assumptions, there are no 1-loop corrections to the classical theory.

\subsubsection*{\ul{\it Constant background, multi-loop diagrams}}

Proceeding to higher loops, more vertices in the perturbative expansion become accessible, beginning at two loops with a vertex cubic in the quantum field. The first of such diagrams that is not a tadpole, shown in equation (\ref{not_tadpole}), emerges at order $\mathcal{O}(\gamma^2)$, and one can show that it nontrivially vanishes within dimensional regularization.
\begin{align}\label{not_tadpole}
    \raisebox{-0.5\height}{\includegraphics[width=0.35\linewidth]{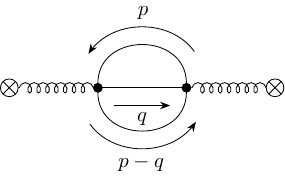}} \, .
\end{align}
The introduction of multiple loop momenta prevents the simple argument in the one-loop case from generalizing immediately. However, since with constant backgrounds there cannot be any external momenta and there is no characteristic scale present in these integrals, it will always be possible to iteratively symmetrize using (\ref{eq:symmetrization}) and integrate over each loop momentum, leaving a symmetrizable integral that will vanish in dimensional regularization. Therefore we expect that the argument presented above generalizes to all loops, implying that the full effective action $\Gamma [ C^i ]$ admits no corrections for constant background field strengths $\tensor{C}{_\mu^i}$.

\subsubsection*{\ul{\it Background-varying, one-loop diagrams}}

We now study the more general case in which we do not assume that $\partial_\mu \tensor{C}{_{\nu}^{i}}= 0$, instead allowing the background field to vary. Besides requiring that the field configuration $C^i$ is a solution to the classical equations of motion, we make no further assumptions.

For this general background analysis, let us use the same notation $\mathcal{D}_n$ for the diagrams appearing in the infinite sum (\ref{infinite_series_constant_bg}). The first diagram in this series, $\mathcal{D}_1$, is unchanged from the constant background case, and thus it identically vanishes in dimensional regularization. 

The first nontrivial diagram is
\begin{align}\label{D2_diagram}
    \mathcal{D}_2 = \raisebox{-0.5\height}{\includegraphics[width=0.35\linewidth]{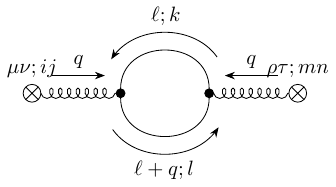}} \, .
\end{align}
As usual, it will be convenient to strip off factors of $\tensor{P}{_{\mu \nu}}^{ij}$ when computing the value of this diagram. This corresponds to evaluating the diagram using the ``uncontracted'' vertex $\tilde{g}$ of (\ref{uncontracted_vertex_factor}) and contracting the result with factors of $\tensor{P}{_{\mu \nu}}^{ij}$. To this end, let us write the value of the diagram as
\begin{align}\label{D2_stripped}
    \mathcal{D}_2 &= \frac{\tanh^2 ( \gamma )}{2} \int \frac{\dd{^{d}q}}{\left( 2\pi \right)^{d}} \tensor{P}{_{\mu \nu}}^{ij} \left( -q \right) \Bigg(  \int \frac{\dd{^{d}\ell}}{\left( 2\pi \right)^{d}} \frac{1}{\ell^2 \left( \ell + q \right)^2} \left( \delta^{ik} \delta^{jl} \ell^{\mu} \left( \ell + q \right)^{\nu} + \delta^{il} \delta^{jk} \left( \ell + q \right)^{\mu} \ell^{\nu} \right) \nonumber \\
    &\hspace{180pt} \cdot \left( \delta^{km} \delta^{ln} \ell^{\rho} \left( \ell + q \right)^{\tau} + \delta^{kn} \delta^{lm} \left( \ell + q \right)^{\rho} \ell^{\tau} \right) \Bigg) \tensor{P}{_{\rho \tau}^{m n}}\left( q \right)  \, .
\end{align}
Using the symmetry property $\tensor{P}{_\mu_\nu^i^j} = \tensor{P}{_\nu_\mu^j^i}$, this can also be expressed as
\begin{align}\label{I2_to_D2}
    \mathcal{D}_2 &= 2 \tanh^2 ( \gamma ) \int \frac{\dd{^{d}q}}{\left( 2\pi \right)^{d}} \tensor{P}{_{\mu \nu}}^{ij} \left( -q \right) {\mathcal{I}}^{\mu \nu \rho \tau}_2 \tensor{P}{_{\rho \tau}^{i j}}\left( q \right) \, ,
\end{align}
where we have defined the simpler integral
\begin{align}\label{simpler_integral_body}
    {\mathcal{I}}^{\mu \nu \rho \tau}_2 = \int \frac{\dd{^{d}\ell}}{\left( 2\pi \right)^{d}} \frac{( \ell + q )^{\nu}\ell^{\mu} ( \ell + q )^{\tau} \ell^{\rho} }{\ell^2 \left( \ell + q \right)^2} \, ,
\end{align}
and where our conventions for symmetrization are $T^{( \mu \nu ) } = \frac{1}{2} \left( T^{\mu \nu} + T^{\nu \mu} \right)$.

To study the divergence structure of the diagram $\mathcal{D}_2$, it suffices to evaluate the quantity ${\mathcal{I}}^{\mu \nu \rho \tau}_2$ in dimensional regularization, which is performed in appendix \ref{sec:scalar_field_1_loop_2_vertex_calc}. The resulting divergent contribution is
\begin{align}\hspace{-10pt}\label{divergent_final_body}
    {\mathcal{I}}^{\mu \nu \rho \tau}_2 &= \left( \frac{1}{\epsilon} \right) \frac{-i}{24\left( 4\pi \right)} \bigg[ q^{2}  \left( g^{\mu \nu} g^{\rho \tau}  + g^{\mu \rho} g^{\nu \tau} + g^{\mu \tau} g^{\nu \rho} \right) +2 \left( q^{\nu} q^{\mu} g^{\tau \rho} + g^{\mu \tau} q^{\nu} q^{\rho}+ g^{\nu \mu} q^{\tau} q^{\rho} + g^{\nu \rho} q^{\mu} q^{\tau} \right) \nonumber \\
    &\quad +4\left( g^{\mu \rho} q^{\nu} q^{\tau} + g^{\nu \tau} q^{\mu} q^{\rho} \right) \bigg] \, .
\end{align}
In order to cancel this $\frac{1}{\epsilon}$ divergence, one would introduce a counter-term which involves two factors of $\tensor{P}{_{\mu \nu}^{ij}}$ in the Lagrangian. Therefore, in the background-varying case, there is a non-trivial contribution to the quantum effective action at one loop. Because the higher diagrams $\mathcal{D}_n$ will involve higher powers of $\gamma$, the result (\ref{divergent_final_body}) represents the complete one-loop effective action at $\mathcal{O} ( \gamma^2 )$.

With the two-vertex diagram evaluated, to complete the computation of the one-loop effective action, we seek to evaluate all remaining diagrams containing one loop. Fortunately, there is only one diagram $\mathcal{D}_n$ for each number of vertices $n$. The details of the evaluation of this diagram are presented in appendix \ref{app:one_loop_n_vertex}. Here we merely summarize the results. The value of $\mathcal{I}_n$ can be written as 
\begin{align}
    \left( {\mathcal{I}}_{n} \right)^{\mu_1 \ldots \mu_{2n}} &= \left( n - 1 \right)!\int_0^{1} \left( \prod_{i=0}^{n-1} \dd{x}_{i} \right) \delta \left( \sum_{i=0}^{n-1} x_{i} - 1  \right)  \left( \vb{C}_{2n}^{\mu_1 \ldots \mu_{2n}} + \vb{D}_{2n}^{\mu_1 \ldots \mu_{2n}} \right) \, ,
\end{align}
where we have defined

\begin{align}
    \vb{C}_{2n}^{\mu_1 \ldots \mu_{2n}} &= \frac{i\left( d - 2 \right)!! \left( 2n - 1 \right)!!}{\left( d - 2 + 2n \right)!!} g^{\mu_1 \cdots \mu_{2n}}  \frac{\Gamma \left( n + \frac{d}{2} \right)  }{\left( 4\pi \right)^{\frac{d}{2}} \Gamma \left( n \right) \Gamma \left( \frac{d}{2} \right)} \Gamma \left( -\frac{d}{2} \right)\Delta^{d} \, \, \nonumber \\
    \vb{D}_{2n}^{\mu_1 \ldots \mu_{2n}} &= \frac{i\left( d - 2 \right)!! \left( 2n - 3 \right)!!}{\left( d - 4 + 2n \right)!!}\sum_{a=1}^{2n} \sum_{b>a}^{2n} g^{\{ \mu \neq \mu_a, \mu_b \}}   f^{\mu_a} \left( x,q,a \right)  f^{\mu_{b}} \left( x,q,b \right) \nonumber \\
    &\qquad \cdot \frac{\Gamma \left( n-1+\frac{d}{2} \right) }{\left( 4\pi \right)^{\frac{d}{2}} \Gamma \left( n \right) \Gamma \left( \frac{d}{2} \right)  } \Gamma \left( 1 - \frac{d}{2} \right) \Delta^{d - 2} \, .
\end{align}
The notation $g^{\mu_1 \cdots \mu_{2n}}$ refers to a symmetrized product of metric tensor factors, which is defined in equation (\ref{shorthand_metric}). Similarly, $g^{\{ \mu \neq \mu_a, \mu_b \}}$ is shorthand for such a symmetrized product of metrics which which omits the two indices $\mu_a$ and $\mu_b$, which is explained in more detail around equation (\ref{vertex_intermediate}). Finally, the function $f^\mu ( x, q, a )$ is defined in equation (\ref{cursed_defn}).

In dimensional regularization, with $d = 2 ( 1 + \epsilon )$ and as $\epsilon \to 0$, the overall momentum dependence and divergence structure of these terms is
\begin{align}\label{one_loop_n_vertex_final}
    \vb{C}_{2n}^{\mu_1 \ldots \mu_{2n}} &\sim \frac{1}{\epsilon} q^2 g^{\mu_1 \cdots \mu_{2n}}  \, , \nonumber \\
    \vb{D}_{2n}^{\mu_1 \ldots \mu_{2n}} &\sim \frac{1}{\epsilon} \sum_{a=1}^{2n} \sum_{b>a}^{2n} q^{\mu_a} q^{\mu_b} g^{\{ \mu \neq \mu_a, \mu_b \}}  \, ,
\end{align}
which is of the same qualitative form as the one-loop, two-vertex contribution (\ref{divergent_final_body}).

Therefore, the full one-loop effective action for the Modified Scalar theory is obtained by introducing counterterms that cancel the divergent contributions which we have described in equations (\ref{divergent_final_body}) and (\ref{one_loop_n_vertex_final}). Because, after Fourier transforms, only two derivatives arise acting on the external background vertices, the counterterms are invariant under classical conformal transformations.

\subsubsection*{\ul{\it Background varying, two vertex, $m$-loop diagrams}}

One could imagine computing the quantum effective action (\ref{effective_action}) using a double expansion in both the number $n$ of vertices and the number $m$ of loops. The preceding subsections have discussed the contributions at one loop but for any number of vertices. We have also argued that higher loop corrections vanish when expanding around constant backgrounds.

It is then natural to ask what one can say about the higher-loop contributions in the general case of varying backgrounds. Although the structure of the problem quickly becomes quite complicated, we can make some general remarks by restricting to two vertices but any number of loops. For instance, we can consider a diagram with $m + 1$ internal quantum field lines, each of which runs between two interaction vertices with a classical background field, thus forming $m$ loops:
\begin{align} \label{two_vertex_n_loop} % a loop with 'n' in the center
    \mathcal{D}_{m, 2} \; = \; \vcenter{\hbox{
                \begin{tikzpicture}
                    \begin{feynman}
                        \vertex (a);
                        \vertex [below right=of a] (b);
                        \vertex [below left=of b] (c);
                        \vertex [right=of b] (z);
                        \vertex [above left=of c] (d);
                        \vertex [above right=of d] (e);
                        \vertex [left=of d] (x);
                        \diagram* {
                            (b) -- [quarter left] (d),
                            (b) -- [quarter right] (d),
                            (a) [dot] -- [quarter left] (b) -- [quarter left] (c) [dot] -- [quarter left] (d) -- [quarter left] (e),
                            (z) [crossed dot] -- [rmomentum=$q$, gluon] (b), (d) -- [rmomentum=$q$, gluon] (x) [crossed dot],
                        };
                        % make line invisible but node not
                        \draw [draw = none, ] (c.south west) -- (a.north west) node [pos=0.5] {$m + 1$} node [pos=0.2] {$\vdots$} node [pos=0.9] {$\vdots$};
                    \end{feynman}
                \end{tikzpicture}
        }} \, .
\end{align}
We use the notation $\mathcal{D}_{m, n}$ for a diagram which has $m$ loops and $n$ vertices. In this notation, the one-loop diagrams which we called $\mathcal{D}_n$ in the preceding subsections would be denoted $\mathcal{D}_{1, n}$. For example, the diagram $\mathcal{D}_2$ of equation (\ref{D2_diagram}) would be written as $\mathcal{D}_{1, 2}$, since it is of the form in equation (\ref{two_vertex_n_loop}) with $m = 1$ because it has $2 = 1 + 1$ internal lines between two vertices and thus one loop. Similarly, a diagram with $4$ internal lines between two vertices would have three loops and be denoted $\mathcal{D}_{3, 2}$.

In order to study the diagrams $\mathcal{D}_{m, 2}$, we will need to derive a new Feynman rule for the $(m + 2)$-valent vertex involving $( m + 1)$ quantum fields lines and one insertion of the classical background. These higher vertex factors will come from further terms in the expansion of the square root in equation (\ref{sqrt_expansion_Q}),
\begin{align}\label{implicit_m_vertex}
    \sqrt{ 2 P^2 - S^2 } &= \sqrt{ 2 P_C^2 - S_C^2 } + \sum_{N=1}^{\infty} \binom{\frac{1}{2}}{N} \frac{ 2^N \mathcal{Q}_N }{\left( 2P_C^2 - S_C^2 \right)^{N - \frac{1}{2}}} \nonumber \\
    &= \sqrt{ 2 P_C^2 - S_C^2 } + \sum_{M=2}^{\infty} \tensor{P}{^{i_{1} \cdots i_{M}}_{\mu_1 \cdots \mu_{M}}} \prod_{k=1}^{M} \partial^{\mu_{k}} Q_{i_k} \, .
\end{align}
In the first line, the factor of $2^N$ is a choice of normalization which is needed to match our conventions for $\mathcal{Q}_1$ and $\mathcal{Q}_2$ above. We will not compute the higher terms $\mathcal{Q}_N$ explicitly, but we instead schematically denote the collection of all contributions from these terms which involve a product of $M$ derivatives of the quantum fields by writing the tensor $\tensor{P}{^{i_{1} \cdots i_{M}}_{\mu_1 \cdots \mu_{M}}}$. When $M = 2$, this is precisely the tensor $\tensor{P}{_\mu_\nu^i^j}$ of equation (\ref{P_tensor}). We have changed the summation variable to $M$ in the second line to emphasize that one must collect contributions from several $\mathcal{Q}_N$ at each fixed order in $M$. There are no linear vertices in $Q^i$, so the $M = 1$ term is absent, but both the $N = 1$ term $\mathcal{Q}_1$ and the $N = 2$ term $\mathcal{Q}_2$ of the first sum contributes to the quadratic $M = 2$ interaction of the second sum, and so on.

In terms of the tensors $\tensor{P}{^{i_{1} \cdots i_{M}}_{\mu_1 \cdots \mu_{M}}}$ which are defined implicitly through the expansion in equation (\ref{implicit_m_vertex}), the Feynman rule for an $(M+1)$-valent interaction with one classical field insertion is
\begin{align} % 3-arm diagram with ... between 2 arms
% \vcenter{\hbox{
%                             \begin{tikzpicture}
%                             \begin{feynman}
%     \diagram [small,horizontal=a to d] {
%         e [dot] -- [gluon,rmomentum'=$q\,\,\mathrm{=}-\Sigma_i p^{\mu_i}$] c [crossed dot,label=1:$\beta_1\text{,}\cdots\text{,}\beta_n\text{;~}\nu_1\text{,}\cdots\text{,}\nu_n$];
%         a -- [momentum'=$\alpha_1\text{, }p^{\mu_1}$] e,  
%         b -- [momentum=$\alpha_n\text{, }p^{\mu_n}$] e, 
%     }; \end{feynman}
%     % \ldots in arc between a and b using tikz
%     \draw [decorate,decoration={brace,amplitude=5pt},xshift=-4pt,yshift=0pt]
%     ($(a.north west) + (0,0.2)$) -- ($(b.south west) + (0,0.2)$) node [black, midway, above=5pt] {\footnotesize $n$}
%     node [black, midway, below=4pt] {\footnotesize $\cdots$};
% \end{tikzpicture} }}
    \raisebox{-0.5\height}{\includegraphics[width=0.53\linewidth]{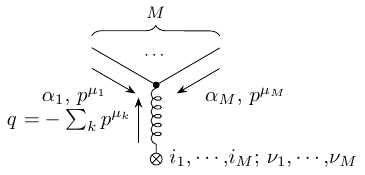}}
    &= \int \dd{^{d}q}\tensor{P}{^{i_{1}}^{\cdots}^{i_M}_{\nu_1 \cdots \nu_{M}}} \left( q \right) \prod_{k=1}^{M}  p^{\nu_k}_{i_k}  \, . \label{n_vertex_rule}
\end{align}
Using this Feynman rule, we can compute the value of the diagram $\mathcal{D}_{m, 2}$ in equation (\ref{two_vertex_n_loop}). Such a diagram has two vertices of the form (\ref{n_vertex_rule}), each with $M = m + 1$, along with $m$ loop momenta $\ell_i$. The contribution from this diagram is given by the integral
\begin{align}\hspace{-25pt}
    \mathcal{D}_{m, 2} &= \frac{\sinh^2 ( \gamma )}{\cosh^{m + 1} ( \gamma ) }\int \frac{\dd{^{d}q}}{( 2 \pi )^d} \left( \int \prod_{i=1}^{m}  \frac{\dd{^d \ell_{i} }}{ ( 2 \pi  )^d } \right) \left( \frac{1}{\prod_{j=1}^{m + 1} p^\mu_j p_{\mu j} } \right) \nonumber \\
    &\qquad \qquad \qquad \cdot \tensor{P}{^{ i_1 }^{ \cdots }^{ i_{m + 1} }_{ \nu_1 }_{ \cdots }_{\nu_{m + 1}}} \left( q \right) \left( \prod_{k=1}^{m + 1}  p_{i_k}^{\nu_k} p_{j_k}^{\mu_k} \right)  \tensor{P}{^{j_1}^{\cdots}^{j_{m + 1}}_{\mu_1}_{\cdots}_{\mu_{m + 1}}} \left( -q \right) \, .
\end{align}
Here the momenta of the internal lines are chosen to be $p_1 = q - \ell_1$, $p_{i} = \ell_{i-1} - \ell_{i}$ for $1 < i < m + 1$, and $p_{m+1} = -\ell_{m}$, so that the total momentum satisfies
\begin{align}
    \sum_{i= 1}^{m+1} p_{i} &= q \, .
\end{align}
Besides the diagrams $\mathcal{D}_{m, 2}$ drawn in equation (\ref{two_vertex_n_loop}), one might ask whether we should account for additional diagrams where a loop begins and ends on the same vertex. However, diagrams of this form do not contribute, as they vanish in dimensional regularization. We can see this by noting that the momentum $\ell$ running in such a loop will appear in the vertex factor only in the combination $\ell^{\mu} \ell^{\nu}$, and in the propagator in the form $\frac{1}{\ell^2}$. Therefore, the value of any diagram will be proportional to
\begin{align}
    \int \frac{\dd{^{d}\ell}}{\left( 2\pi \right)^{d}} \frac{\ell^{\mu} \ell^{\nu}}{\ell^2} &= \frac{g^{\mu \nu}}{d}\int \frac{\dd{^{d}\ell}}{\left( 2\pi \right)^{d}} 1 \ , 
\end{align}
which we have seen vanishes in dimensional regularization in the limit $d \to 2$, as desired.

Next let us consider the divergence structure of the diagram $\mathcal{D}_{m, 2}$. It is convenient to isolate the part of the integrand which depends on the loop momenta and evaluate it separately. To do this, let us define
\begin{align}\label{frak_L_defn}
    \left( \mathfrak{L}_{m, 2} \right)_{\{ i j \}}^{\{ \mu \nu \}} &= \int \, \left(  \prod_{i=1}^{m}  \frac{\dd{^d \ell_{i} }}{ ( 2 \pi  )^d } \right) \left( \frac{1}{\prod_{j=1}^{m + 1} p^\mu_j p_{\mu j} } \right) \left( \prod_{k=1}^{m + 1}  p_{i_k}^{\nu_k} p_{j_k}^{\mu_k} \right) \, .
\end{align}
Here we use $\{ i j \}$ as a shorthand for the multi-index $\{ i_1 \ldots i_{m+1} j_1 \ldots j_{m+1} \}$ and $\{ \mu \nu \}$ for $\{ \mu_1 \ldots \mu_{m+1} \nu_1 \ldots \nu_{m+1} \}$. We will sometimes suppress these multi-indices in writing $\mathfrak{L}_{m, 2}$ for convenience. The quantity $\mathfrak{L}_{m, 2}$ determines the value of the diagram $\mathcal{D}_{m, 2}$ as
\begin{align}
    \mathcal{D}_{m, 2} &= \frac{\sinh^2 ( \gamma )}{\cosh^{m + 1} ( \gamma ) }\int \frac{\dd{^{d}q}}{( 2 \pi )^d} \tensor{P}{^{ i_1 }^{ \cdots }^{ i_{m + 1} }_{ \nu_1 }_{ \cdots }_{\nu_{m + 1}}}(q) \left( \mathfrak{L}_{m, 2} \right)_{\{ i j \}}^{\{ \mu \nu \}} \tensor{P}{^{j_1}^{\cdots}^{j_{m + 1}}_{\mu_1}_{\cdots}_{\mu_{m + 1}}} \left( -q \right) \, ,
\end{align}
so to understand the divergences in $\mathfrak{D}_{m, 2}$, it suffices to understand those in $\mathfrak{L}_{m, 2}$.

One can evaluate $\mathfrak{L}_{m, 2}$ by performing the integral over each loop momentum in succession. The details of one such integration, namely the integral over the final variable $\ell_m$, are presented in appendix \ref{sec:scalar_field_n_loop_2_vertex_calc}. After evaluating this single integral over $\ell_m$, one obtains a result proportional to $\Gamma \left( -\frac{d}{2} \right) \ell^{d}_{m - 1}$. One can then apply the same argument recursively to conclude that performing all $m$ of the integrals generates $m$ factors of this form. After evaluating all $m$ integrals, the final dependence on the momentum $q$ takes the form
\begin{align}\label{final_n_loop_2_vertex_dimreg}
    \left( \mathfrak{L}_{m, 2} \right)_{\{ i j \}}^{\{ \mu \nu \}} \sim \Gamma\left( -\frac{d}{2} \right)^{m} q^{d m} \, ,
\end{align}
where we show only the dependence on $q$ and $d$ but suppress the tensor structure in the $i, j, \mu, \nu$ indices.\footnote{Each integral yields $6$ different symmetrizations of the external indices. Thus the exact form of an $m$-loop diagram contains many different index structures and is challenging to write explicitly in general.}

It is also useful to translate the divergence structure of equation (\ref{final_n_loop_2_vertex_dimreg}) in dimensional regularization to an equivalent dependence on a momentum cutoff $\Lambda$. For $d = 2 ( 1 + \epsilon )$ we have the limiting behavior $\Gamma \left( - \frac{d}{2} \right) \sim \frac{1}{\epsilon}$, and a divergence proportional to $\frac{1}{\epsilon}$ in dimensional regularization corresponds to a logarithmic divergence of the form $\log ( \Lambda )$. Therefore, the $m$-loop, $2$-vertex contributions from (\ref{final_n_loop_2_vertex_dimreg}) yield divergences of the form
\begin{align}
    \left( \mathfrak{L}_{m, 2} \right)_{\{ i j \}}^{\{ \mu \nu \}} \sim \left( \frac{1}{\epsilon} \right)^{m} q^{2m} \sim \left( \log \Lambda \right)^m \, .
\end{align}
This is a different divergence structure than the one which we have seen in our study of the $1$-loop effective action, which would necessitate the addition of different counterterms. It is interesting to note that each of these counterterms is classically conformal and has a different higher derivative dependence on the external classical field momenta.

We conclude this section with some further comments. Even though our analysis for non-constant backgrounds is very preliminary, no clear organizational principle seems to emerge in this hierarchy of divergences and necessary counterterms. Though this might be a feature of our perturbative approach, it begins to suggest that this non-analytic model is non-renormalizable, which might also spoil the quantum conformal invariance of the model. Ultimately, the theory might retain a sensible interpretation only as an effective field theory.
Yet, it remains a very interesting fact that there are no quantum corrections for constant background fields $C_\mu{}^i$. We leave other open questions for further future investigations.

\section{Conclusion} \label{sec:Conclusion&Outlook1}

In this work, we have explored the space of interacting chiral boson theories from several perspectives. We showed that, when written in a Floreanini-Jackiw representation, the property of non-manifest Lorentz invariance is closely related to stress tensor deformations: indeed, every parameterized family of Lorentz-invariant chiral boson theories can be interpreted as a deformation by some function of the energy-momentum tensor. In the dual description using $U(1)$ gauge fields with a Chern-Simons action, Lorentz invariance is manifest but chirality (or self-duality) is not, and in this setting we find that every family of \emph{self-dual} Chern-Simons boundary terms likewise obeys a flow equation driven by a function of the stress tensor. We have also explained how a general boundary term for such a bulk $U(1)$ Chern-Simons theory imposes modified boundary conditions on the gauge fields which lead to a non-linear self-duality condition for the currents; this mirrors the analogous non-linear self-duality constraints obeyed by interacting Floreanini-Jackiw bosons.

We then studied the quantization of interacting chiral boson models, focusing on a root-$\TT$-deformed system of free bosons. We characterized the finite-volume spectrum both for one left-moving and one right-moving boson, where the root-$\TT$ deformation acts as a rescaling of the target space radius, and also for two left-moving bosons and one right-moving boson, where the deformation is more complicated but can be analyzed perturbatively in a large-momentum expansion.  In doing so, we confirmed that the zero-mode formula (\ref{zero_mode_formula}) derived via holography does not apply to generic states, but does apply in certain states with constant stress tensors. 
We also gave a classical/heuristic argument on how a set of field redefinitions might turn all these models into free ones.
Finally, we have studied the quantum effective action for the theory of root-$\TT$-deformed bosons with equal numbers of left- and right-movers. Intriguingly, we find that the one-loop effective action vanishes around classical backgrounds which are linear in the spacetime coordinates.

There are several interesting directions for future research, some of which we summarize in what follows. Understanding more about these issues, and in particular developing a clearer picture of field theories with non-analytic interaction terms such as the Modified Scalar theory, may teach us new lessons about previously unexplored models within the space of quantum field theories.

\subsubsection*{\ul{\it Supersymmetry}}

There has been a great deal of work on supersymmetric extensions of deformations constructed from the energy-momentum tensor \cite{Baggio:2018rpv,Chang:2018dge,Jiang:2019hux,Chang:2019kiu, Coleman:2019dvf,Ferko:2019oyv,He:2019ahx,Ebert:2020tuy,Ferko:2021loo} and other conserved currents \cite{Jiang:2019trm}, including analogous deformations of $1d$ theories by conserved charges \cite{Gross:2019ach,Gross:2019uxi,Ebert:2022xfh,Ebert:2022ehb,Ferko:2023ozb}.

A natural direction for further investigation is to seek such a supersymmetric generalization of the results in this work. This would involve coupling a supersymmetric theory of interacting chiral bosons and their fermionic superpartners to supergravity, which would give expressions for the fields in the stress tensor supermultiplet.

In the case of a single free chiral boson and its fermionic partner, the procedure for performing this coupling to supergravity was explained in \cite{Bastianelli:1989cu}, building on earlier results for the supergravity couplings of non-chiral fields \cite{Brink:1976sc}. The bosonic truncation of this supergravity coupling reproduces the coupling to vielbeins which we have used in this work. It would be interesting to generalize this technique and couple an arbitrary number of chiral and anti-chiral bosons, and their fermionic counterparts, to supergravity, and then consider flows in the space of such supersymmetric interacting theories, much as we have done here. In principle, one could perform this analysis either using component fields -- which was the strategy adopted in \cite{Bastianelli:1989cu} -- or using a superspace formulation, such as the one employed in \cite{Gates:1987sy,Bellucci:1987mj}. One might also hope to interpret these theories using a bulk description involving a supersymmetric Chern-Simons theory, which would give a supersymmetric generalization of the results in section \ref{sec:cs}.

\subsubsection*{\ul{\it Quantum Hall physics}}

A famous application of $U(1)$ Chern-Simons theories, and the chiral bosons which describe their edge modes, occurs in the study of the quantum Hall effect. The essential reason for this, as we mentioned in section \ref{sec:cs}, is that the Chern-Simons term is more relevant at low energies than the Maxwell term. Therefore, in an effectively $(2 + 1)$-dimensional system -- such as a flat slab of material subject to a background magnetic field -- one expects that the low-energy effective action $S_{\text{eff}} [ A ]$ will be controlled by the Chern-Simons term $S_{\text{CS}} [ A ]$. Computing the associated current which we defined in equation (\ref{currents_defn}), 
\begin{align}\label{currents_conclusion}
    J_i \sim \frac{\delta S_{\text{CS}}}{\delta A_i} \, ,
\end{align}
therefore gives predictions for the behavior of the system. For instance, in the integer quantum Hall effect, this current $J_i$ agrees with the Hall conductivity of an integer number of filled Landau levels, if this integer $\nu \in \mathbb{Z}$ is related to the Chern-Simons level appropriately.

We have seen that a Chern-Simons theory on a manifold with boundary supports chiral bosons on the edge. In the quantum Hall setting, these chiral edge modes describe propagating fluctuations in the charge density at the edge of the physical sample. Remarkably, the quantum mechanics of this chiral boson theory contains a great deal of information about the interior of the sample. For instance, by carrying out the quantization of a single Floreanini-Jackiw boson as we described in section \ref{sec:general_quantization}, one finds a Hamiltonian which correctly predicts the spectrum (including degeneracies) of excited modes for the Laughlin wavefunction which describes the fractional quantum Hall effect.\footnote{See the reviews \cite{Tong:2016kpv,RevModPhys.75.1449}, or the incomplete sampling of some of the original works \cite{PhysRevB.41.12838,PhysRevB.43.11025,1992IJMPB...6.1711W}, for further discussion on this subject.}

One might ask whether the modified Chern-Simons boundary terms which we have considered in this work could be used to model some variant of a conventional quantum Hall system. For instance, it would be very interesting if an experimentally realizable modification of a quantum Hall droplet would subject the system to a boundary term like the one which is generated by the root-$\TT$ deformation. If so, this could offer a way to study the effective dynamics of the Modified Scalar theory -- and other theories obtained via stress tensor deformations -- in the laboratory.

\subsubsection*{\ul{\it Non-perturbative analysis}}

All of the results concerning the quantum theory of root-$\TT$-deformed bosons presented in this work have been obtained in perturbation theory, by expanding around a classical background. For instance, we have attempted a perturbative analysis of the effective action and noticed that a hierarchy of counterterms emerged in the Modified Scalar theory. However, it seems likely that the most interesting features of root-$\TT$-deformed theories at the quantum level -- assuming that they exist -- will only be visible non-perturbatively. It is therefore important to find a way to study the quantization of such root-$\TT$ deformed theories beyond perturbation theory, which will likely require a new perspective.

One way to re-frame these deformed theories, which may be useful for a non-perturbative analysis, is via geometry. In the case of the related $\TT$ deformation, many insights have resulted from presentations of the flow in terms of coupling to gravity \cite{Dubovsky:2017cnj,Dubovsky:2018bmo} or random geometry \cite{Cardy:2018sdv}, or realizing the deformation via a field-dependent change of variables \cite{Conti:2018tca,Conti:2022egv,Morone:2024ffm,Ferko:2024yhc}. The root-$\TT$ deformation appears to admit a similar geometrical interpretation \cite{Babaei-Aghbolagh:2024hti,Tsolakidis:2024wut}. Perhaps relatedly, the Modified Scalar Lagrangian (\ref{modscalar_lorentz_invariant_lagrangian}) can be rewritten as
\begin{align}\label{rtt_background_metric}
    \mathcal{L} &= \frac{1}{2} g^{\mu \nu} \partial_\mu \varphi^i \partial_\nu \varphi^i \, , \nonumber \\
    g^{\mu \nu} &= \cosh ( \gamma ) \eta^{\mu \nu} + \sinh ( \gamma ) \left( \frac{2 \partial^\mu \tensor{\varphi}{^{j}} \partial^\nu \varphi^{j} - \eta^{\mu \nu} \partial_\rho \tensor{\varphi}{^{j} } \partial^\rho \varphi^{j} }{\sqrt{ 2 \partial_\sigma \varphi^i \partial^\tau \varphi^i \partial_\tau \varphi^k \partial^\sigma \varphi^k  - \left( \partial_\sigma \varphi^i \partial^\sigma \varphi^i \right)^2 } } \right) \, ,
\end{align}
which is equivalent to a theory of \emph{free} scalar fields coupled to a field-dependent metric. Even at the perturbative level, such a rewriting of the deformation may be useful -- for instance, it may be possible to adapt existing heat kernel techniques\footnote{See \cite{Vassilevich:2003xt} and references therein for a review.} which compute the quantum effective actions for theories on background metrics to handle field-dependent metrics such as (\ref{rtt_background_metric}), which could reproduce results like those in section \ref{sec:cian} from a different point of view. However, it would be even more useful if such a geometrical presentation of the root-$\TT$ flow could furnish us with a non-perturbative definition of the quantum theory.

Another potential way to approach the study of renormalisation of the Modified Scalar theory, and analyse its quantum conformal symmetry, is by using non-perturbative functional renormalisation group approaches. An attempt to use such techniques for $\TT$ deformed scalar theories has been made in \cite{Liu:2023omp}. It would be intriguing to reattempt this analysis for non-analytic models and generic $\TT$-like deformations, including root-$\TT$.

A third strategy is to bypass the classical Lagrangian (\ref{modscalar_lorentz_invariant_lagrangian}) and attempt to define the quantum Modified Scalar theory directly by characterizing the set of local operators in the theory along with their correlation functions. For instance, one could proceed under the assumption that the theory in question is a CFT, and see whether this leads to a contradiction.\footnote{An example of such a contradiction would be finding an operator which can be neither a primary nor a descendant, which is used to demonstrate that the Maxwell theory is not conformal except in four dimensions \cite{El-Showk:2011xbs}. Alternatively, one could use the more formal machinery of algebraic/axiomatic QFT.} Here there appears to be an interesting tension. Standard lore suggests that, in any $\mathrm{CFT}_2$ with a conserved vector current $J$, its Hodge dual $\ast J$ must also be conserved. For a putative theory of root-$\TT$-deformed $\varphi^i$, it appears that the operators $J^i_\mu = \partial_\mu \varphi^i$ should not be conserved at finite $\gamma$ due to the source terms in the equations of motion, although their duals $\widetilde{J}^i_\mu = \epsilon_{\mu \nu} \partial^\nu \varphi^i$ \emph{are} conserved (at least for non-compact scalars).\footnote{The analogous tension for the ModMax theory can be phrased in terms of generalized global symmetries: if a $4d$ CFT has a $U(1)_1$ magnetic one-form global symmetry, then it must also have the corresponding $U(1)_1$ electric one-form global symmetry, and vice-versa \cite{Hofman:2018lfz}. A $4d$ ModMax CFT would appear to have the magnetic $1$-form symmetry of the Maxwell theory but not the electric one, since $\partial_\mu \widetilde{F}^{\mu \nu} = 0$ but $\partial_\mu F^{\mu \nu} \neq 0$.} If the quantum Modified Scalar theory does exist, it would be very interesting to see how this tension is resolved. Perhaps the quantum theory is not a CFT, or perhaps it is not even a local quantum field theory, much like a $\TT$-deformed CFT is believed to become non-local due to its Hagedorn density of states at high energies.

\section*{Acknowledgements}
We thank Peter Bouwknegt, Per Kraus, Sergei Kuzenko, Savdeep Sethi, Alessandro Sfondrini, Dmitri Sorokin, and Roberto Tateo for useful discussions and collaboration on related projects, and we are especially grateful to Zhengdi Sun for collaboration in the early stages of this project. 
C. F., C. L. M., and G. T.-M. acknowledge kind hospitality and financial support at the MATRIX Program ``New Deformations of Quantum Field and Gravity Theories,'' where part of this work was performed, and thank the participants of this meeting for productive conversations on related topics. This research was supported in part by grant NSF PHY-2309135 to the Kavli Institute for Theoretical Physics (KITP). 
S.E. is supported by the Bhaumik Institute and by the Dissertation Year Fellowship from the UCLA Graduate Division. C. F. is supported by U.S. Department of Energy grant DE-SC0009999 and by funds from the University of California.
 C. L. M. is supported by a
postgraduate scholarship at the University of Queensland.
 G. T.-M. has been supported by the Australian Research Council (ARC)
Future Fellowship FT180100353, ARC Discovery Project DP240101409, and the Capacity
Building Package of the University of Queensland.

\appendix

\section{Perturbative \texorpdfstring{$ f( \tensor{T}{^\alpha_\alpha}, T^{\alpha \beta} T_{\alpha \beta})$}{TT}-Deformed Actions}\label{app:TTn}

Throughout this paper, we have considered various deformations which are constructed from the energy-momentum tensor. Although the most important examples within this class are the $\TT$ and root-$\TT$ flows, it appears that \emph{general} stress tensor deformations nonetheless have interesting properties -- for instance, we have shown that every parameterized family of interacting $2d$ chiral boson theories which enjoys non-manifest Lorentz invariance admits an interpretation as a stress tensor deformation. This is a $2d$ analog of similar theorems about $4d$ theories of duality-invariant electrodynamics \cite{Ferko:2023wyi} or $6d$ chiral tensor theories \cite{Ferko:2024zth}.

Motivated by these observations, one may wish to study $2d$ deformations by other functions of the energy-momentum tensor, besides the ones considered in the body of this manuscript. One way to do this is to solve the resulting flow equations perturbatively, i.e. order-by-order in the deformation parameter. In this appendix we will use the symbol $g$ for the parameter of a general stress tensor flow, which is not to be confused with the metric $g_{\alpha \beta}$ or its determinant.

Let us therefore consider the following general class of operators in $2d$ which can be expressed in terms of the two independent Lorentz scalars that can be built from the stress tensor, namely $\Tr ( T ) = \tensor{T}{^\alpha_\alpha}$ and $\Tr ( T^2 ) = T^{\alpha \beta} T_{\alpha \beta}$:
\begin{equation}\label{eq:TTn}
 f( \tensor{T}{^\alpha_\alpha}, T^{\alpha \beta} T_{\alpha \beta})\, .
\end{equation}
We note that all higher traces of the stress tensor, $\Tr ( T^n )$ for $n > 2$, can be expressed in terms of these two lower traces. Given such an operator, we wish to study the flow equation\footnote{One can also consider flows driven by a function $f$ which has explicit dependence on the deformation parameter $g$. For instance, the so-called $\TT + \Lambda_2$ deformation is defined by performing a $\TT$ deformation and then activating a cosmological constant proportional to $\frac{1}{\lambda}$. See \cite{Gorbenko:2018oov, Lewkowycz:2019xse, Coleman:2021nor, Shyam:2021ciy, Torroba:2022jrk, Batra:2024kjl} for further details.}
\begin{equation}
\label{eq:TTn op}
    \frac{\partial S^{(g)}}{\partial g} = \int d^2x \, E \,  f( \tensor{T}{^\alpha_\alpha}, T^{\alpha \beta} T_{\alpha \beta}) \, .
\end{equation}
The solution to \eqref{eq:TTn op} can be written as a series expansion,
\begin{equation}\label{perturbative_S}
      S^{(g)} = S^{(0)} +\sum^\infty_{m=1} \frac{g^m}{m}\int d^2x \, E \,   f( \tensor{T}{^\alpha_\alpha}, T^{\alpha \beta} T_{\alpha \beta})_{m-1} \, ,
\end{equation}
where we write $ f( \tensor{T}{^\alpha_\alpha}, T^{\alpha \beta} T_{\alpha \beta})_{m}$ for the term of order $g^{m}$ in the expression for the $ f( \tensor{T}{^\alpha_\alpha}, T^{\alpha \beta} T_{\alpha \beta})$ operator computed from the action at order $g^{m-1}$. Because each term in this expansion only depends on the data of lower-order terms, one can build up the solution iteratively in powers of $g$.

As in section \ref{sec:classical} of the main text, we will work in the tetrad formalism with a Lorentzian tangent-space metric and with spacetime coordinates $x^\alpha = ( t, \theta )$. A general spacetime metric can therefore be expanded in terms of vielbeins as
\begin{equation}
    g_{\alpha \beta} =E^a{}_\alpha E^b{}_\beta \eta_{ab} = -\left( \begin{array}{cc}
     2 E^+{}_t E^-{}_t    & \quad   E^+{}_t  E^-{}_\theta+  E^-{}_t  E^+{}_\theta \\
       E^+{}_t  E^-{}_\theta+  E^-{}_t  E^+{}_\theta  & \quad  2 E^+{}_\theta E^-{}_\theta  
    \end{array} \right) \, .
\end{equation}
The stress tensor associated with a general action $S$, which has been coupled to gravity using the vielbeins $\tensor{E}{^a_\alpha}$, can be written as
\begin{equation}\label{perturbative_stress_tensor}
    T^\alpha{}_\beta= - \frac{1}{E} \frac{\partial S}{\partial E^a{}_\alpha} E^a{}_\beta = - \frac{1}{E}\left( \begin{array}{cc}
      \frac{\partial S}{\partial E^+{}_t} E^+{}_t + \frac{\partial S}{\partial E^-{}_t} E^-{}_t   & \quad \frac{\partial S}{\partial E^+{}_\theta} E^+{}_t + \frac{\partial S}{\partial E^-{}_\theta} E^-{}_t  \\ \frac{\partial S}{\partial E^+{}_t} E^+{}_\theta + \frac{\partial S}{\partial E^-{}_t} E^-{}_\theta
         & \quad     \frac{\partial S}{\partial E^+{}_\theta} E^+{}_\theta + \frac{\partial S}{\partial E^-{}_\theta} E^-{}_\theta 
    \end{array} \right)\,.
\end{equation}
We will use the general expression (\ref{perturbative_stress_tensor}) for the stress tensor, along with the expansion (\ref{perturbative_S}), to perturbatively solve the flow equation (\ref{eq:TTn op}) for various choices of the $ f( \tensor{T}{^\alpha_\alpha}, T^{\alpha \beta} T_{\alpha \beta})$ operator. 

We begin by finding perturbative solutions for some of the flow equations considered in the main text, before generalizing to other deformations which were not considered in the body. In our examples, we compute the stress tensor (\ref{perturbative_stress_tensor}) using the vielbein formalism due to computational speed in \texttt{Mathematica}, but we note that the metric formalism gives identical results.

\subsubsection*{\ul{\it Root-$\TT$ Perturbative Flow for Multiple Bosons}}

For our first example, we will consider the perturbative solution to the root-$\TT$ flow equation for an arbitrary number of non-chiral bosons $\varphi^i$, $i = 1, \ldots, N$. This flow equation was first solved in closed-form in \cite{Ferko:2022cix}.

We take a seed action which describes $N$ free massless bosons in Lorentzian signature,
\begin{equation}
\begin{aligned}\label{eq:seedapp}
        S^{(0)} &=  \frac{1}{2} \int d^2x \sqrt{-g} g^{\alpha \beta} \partial_\alpha \varphi^i \partial_\beta \varphi^i
    \\&= \int d^2x \frac{ E^-{}_\theta E^+{}_\theta \dot{\varphi}^i \dot{\varphi}^i + E^-{}_t E^+{}_t \varphi^{\prime i} \varphi^{\prime i} - \left( E^-{}_\theta E^+{}_t + E^-{}_t E^+{}_\theta \right) \dot{\varphi}^i \varphi^{\prime i}  }{E} \, ,
\end{aligned}
\end{equation}
which have been coupled to gravity using the tetrad formalism. We then deform using the root-$\TT$ operator, which corresponds to the general $ f( \tensor{T}{^\alpha_\alpha}, T^{\alpha \beta} T_{\alpha \beta})$ operator of equation (\ref{eq:TTn}) being
\begin{equation}
    f( \tensor{T}{^\alpha_\alpha}, T^{\alpha \beta} T_{\alpha \beta}) = \mathcal{R}^{(\gamma)} = \frac{1}{\sqrt{2}} \sqrt{ T^{\alpha \beta} T_{\alpha \beta} - \frac{1}{2} \left( \tensor{T}{^\alpha_\alpha} \right)^2 } \, .
\end{equation}
In this case, the perturbative solution (\ref{perturbative_S}) to the flow equation takes the form
\begin{equation}\label{eq:ansatzLorentz}
    S^{(\gamma)} = S^{(0)} +\sum^\infty_{m=1} \frac{\gamma^m}{m}\int d^2x E~ \mathcal{R}^{(\gamma)}_{m-1} \, .
\end{equation}
Following the conventions in the main text, we use the symbol $\gamma$ for the flow parameter of a root-$\TT$ deformation, rather than the variable $g$ which stood for the parameter in a general deformation above.

The first few terms in this perturbative expansion are
\begin{equation}
\begin{aligned}\label{eq:coeffsLorentz}
    \mathcal{R}^{(\gamma)}_0|_{\text{flat}} &= \frac{1}{2} \sqrt{\left( \dot{\varphi}^i - \varphi^{\prime i} \right) \left( \dot{\varphi}^i - \varphi^{\prime i} \right) \left( \dot{\varphi}^j + \varphi^{\prime j} \right) \left( \dot{\varphi}^j + \varphi^{\prime j} \right)  }\,, \quad  \mathcal{R}^{(\gamma)}_1|_{\text{flat}} = \frac{1}{2} \left( -\dot{\varphi}^i \dot{\varphi}^i +\varphi^{\prime i} \varphi^{\prime i} \right) \,, \\
      \mathcal{R}^{(\gamma)}_2|_{\text{flat}} &= \frac{1}{2}    \mathcal{R}^{(\gamma)}_0|_{\text{flat}}\,, \quad
      \mathcal{R}^{(\gamma)}_3|_{\text{flat}} = \frac{1}{6}     \mathcal{R}^{(\gamma)}_1|_{\text{flat}}\,, \quad  
          \mathcal{R}^{(\gamma)}_4|_{\text{flat}} =  \frac{1}{24}     \mathcal{R}^{(\gamma)}_0|_{\text{flat}}\,, \quad \mathcal{R}^{(\gamma)}_5|_{\text{flat}} = \frac{1}{120} \mathcal{R}^{(\gamma)}_1|_{\text{flat}}\,,
\end{aligned}
\end{equation}
where ``flat'' means that we have set the vielbeins to their flat-space values \eqref{flat_vielbeins}.

We note that the quantities appearing in (\ref{eq:coeffsLorentz}) can be written in terms of the manifestly Lorentz-invariant combinations
\begin{equation}
\begin{aligned}
\left( \partial_\mu \varphi^{i} \partial^{\nu} \varphi^{i} \right) \left( \partial_\nu \varphi^{j} \partial^{\mu} \varphi^{j} \right)   &= \left( -\dot{\varphi}^i \dot{\varphi}^j + \varphi^{\prime i} \varphi^{\prime j}\right) \left( -\dot{\varphi}^i \dot{\varphi}^j + \varphi^{\prime i} \varphi^{\prime j}\right)
\\&= (\dot{\varphi}^i \dot{\varphi}^i)^2 + (\varphi^{\prime i} \varphi^{\prime i})^2 -2 (\dot{\varphi}^{i} \varphi^{\prime i})^2  \, , 
\end{aligned}
\end{equation}
and
\begin{equation}
\begin{aligned}
    2 \left( \partial_\mu \varphi^{i} \partial^{\nu} \varphi^{i} \right) \left( \partial_\nu \varphi^{j} \partial^{\mu} \varphi^{j} \right) - \left( \partial_\mu \varphi^{i} \partial^{\mu} \varphi^{i} \right)^2 &= \left( \dot{\varphi}^i \dot{\varphi}^i \right)^2 + \left( \varphi^{\prime i} \varphi^{\prime i} \right)^2 - 4 \left( \dot{\varphi}^i \varphi^{\prime i} \right)^2 + 2 \dot{\varphi}^i \dot{\varphi}^i \varphi^{\prime j} \varphi^{\prime j}
    \\&=\left( \dot{\varphi}^i - \varphi^{\prime i} \right) \left( \dot{\varphi}^i - \varphi^{\prime i} \right) \left( \dot{\varphi}^j + \varphi^{\prime j} \right) \left( \dot{\varphi}^j + \varphi^{\prime j} \right) \,.
\end{aligned}
\end{equation}
In terms of these quantities, one finds that the perturbative expansion to the flow equation converges to the solution \eqref{modscalar_lorentz_invariant_lagrangian},

\begin{equation}
\begin{aligned}
     S^{(\gamma)} &= S^{(0)} \\&+ \int dt \, d\theta \, \left( \gamma   \mathcal{R}^{(\gamma)}_0|_{\text{flat}} + \frac{\gamma^2}{2}   \mathcal{R}^{(\gamma)}_1|_{\text{flat}} + \frac{\gamma^3}{6}   \mathcal{R}^{(\gamma)}_0|_{\text{flat}} + \frac{\gamma^4}{24}   \mathcal{R}^{(\gamma)}_1|_{\text{flat}} +\frac{\gamma^5}{120}   \mathcal{R}^{(\gamma)}_0|_{\text{flat}}+ \frac{\gamma^6}{720}   \mathcal{R}^{(\gamma)}_1|_{\text{flat}}  + \cdots  \right)
    \\&= \frac{1}{2} \int dt \, d\theta \, \bigg [ \partial_\alpha \varphi^{i} \partial^{\alpha} \varphi^{i}  \left(1 + \frac{\gamma^2}{2}+ \frac{\gamma^4}{24} + \frac{\gamma^6}{720} + \mathcal{O}(\gamma^8) \right) \\&+ \sqrt{   2 \left( \partial_\mu \varphi^{i} \partial^{\nu} \varphi^{i} \right) \left( \partial_\nu \varphi^{j} \partial^{\mu} \varphi^{j} \right) - \left( \partial_\mu \varphi^{i} \partial^{\mu} \varphi^{i} \right)^2 } \left(  \gamma + \frac{\gamma^3}{6} + \frac{\gamma^5}{120} + \mathcal{O}(\gamma^7) \right)  \bigg] 
    \\&= \frac{1}{2} \int dt \, d\theta \, \bigg[  \cosh (\gamma) \partial_\alpha \varphi^{i} \partial^{\alpha} \varphi^i  + \sinh (\gamma)\sqrt{    2 \left( \partial_\mu \varphi^{i} \partial^{\nu} \varphi^{i} \right) \left( \partial_\nu \varphi^{j} \partial^{\mu} \varphi^{j} \right) - \left( \partial_\mu \varphi^{i} \partial^{\mu} \varphi^{i} \right)^2} \bigg] \, .
\end{aligned}
\end{equation}

\subsubsection*{\ul{\it Root-$\TT$ Perturbative Flow for Chern-Simons}}

An almost identical calculation can be performed to study the perturbative root-$\TT$ deformation of the Chern-Simons boundary action given in \eqref{free_CS_bdry}. The first few terms in the expansion are
\begin{equation}
\begin{aligned}
    \mathcal{R}^{(\gamma)}_0|_{\text{flat}} &= \frac{1}{4\pi} \sqrt{ \left( k^{ij} A_{iw} A_{jw} + \overbar{k}^{\overline{i} \overline{j}} \overbar{A}_{\overline{i}w} \overbar{A}_{\overline{j}w} \right) \left( k^{mn} A_{m\overbar{w}} A_{n\overbar{w}} + \overbar{k}^{\overline{m} \overline{n}} \overbar{A}_{\overline{m}\overbar{w}} \overbar{A}_{\overline{n}\overbar{w}} \right) }\,, \\
      \mathcal{R}^{(\gamma)}_1|_{\text{flat}} &=  -\frac{1}{4\pi}  \left(k^{ij} A_{i w} A_{j \overbar{w}}+\overbar{k}^{\overline{i} \overline{j}} \overbar{A}_{\overline{i} w} \overbar{A}_{\overline{j} \overbar{w}}\right)\,, \quad 
      \mathcal{R}^{(\gamma)}_2|_{\text{flat}} = \frac{1}{2}    \mathcal{R}^{(\gamma)}_0|_{\text{flat}}\,, \quad    \mathcal{R}^{(\gamma)}_3|_{\text{flat}} = \frac{1}{6}    \mathcal{R}^{(\gamma)}_1|_{\text{flat}}\,,
\end{aligned}
\end{equation}
where now ``flat'' means that we have set the vielbeins equal to the values \eqref{euclidean_flat} appropriate for a flat \emph{Euclidean} tangent space metric, following the conventions of section \ref{sec:cs}. 

Therefore, the root-$\TT$-deformed Chern-Simons boundary action is
\begin{equation}
\begin{aligned}
    I^{(\gamma)}_{\partial \mathcal{M}_3} &=  \frac{i}{8 \pi} \int_{\partial \mathcal{M}_3}  dw d\overbar{w} \left( -\left(k^{ij} A_{i w} A_{j \overbar{w}}+\overbar{k}^{\overline{i} \overline{j}} \overbar{A}_{\overline{i} w} \overbar{A}_{\overline{j} \overbar{w}} \right) \left(1 + \frac{\gamma^2}{2} + \mathcal{O}(\gamma^4) \right)\right)
    \\&+ \frac{i}{8 \pi}\int_{\partial \mathcal{M}_3}  dw d\overbar{w} \sqrt{ \left( k^{ij} A_{iw} A_{jw} + \overbar{k}^{\overline{i} \overline{j}} \overbar{A}_{\overline{i}w} \overbar{A}_{\overline{j}w} \right) \left( k^{mn} A_{m\overbar{w}} A_{n\overbar{w}} + \overbar{k}^{\overbar{m} \overbar{n}} \overbar{A}_{\overbar{m}\overbar{w}} \overbar{A}_{\overbar{n}\overbar{w}} \right) } \left( \gamma + \frac{\gamma^3}{6} + \mathcal{O}(\gamma^5) \right)
    \\&= \frac{i}{8 \pi } \int_{\partial \mathcal{M}_3}  dw d\overbar{w} \bigg[- \cosh (\gamma) \left(k^{ij} A_{i w} A_{j \overbar{w}}+\overbar{k}^{\overline{i} \overline{j}} \overbar{A}_{\overline{i} w} \overbar{A}_{\overline{j} \overbar{w}} \right)  \\&+ \sinh (\gamma)  \sqrt{ \left( k^{ij} A_{iw} A_{jw} + \overbar{k}^{\overline{i} \overline{j}} \overbar{A}_{\overline{i}w} \overbar{A}_{\overline{j}w} \right) \left( k^{mn} A_{m\overbar{w}} A_{n\overbar{w}} + \overbar{k}^{\overbar{m} \overbar{n}} \overbar{A}_{\overbar{m}\overbar{w}} \overbar{A}_{\overbar{n}\overbar{w}} \right) }   \bigg] \, .
\end{aligned}
\end{equation}

\subsubsection*{\ul{\it $\TT$ Perturbative Flow for a Single Boson}}

For our next example, we will consider the irrelevant $\TT$ flow rather than the marginal root-$\TT$ flow. For simplicity, we will restrict to a deformation of a single bosonic field $\varphi$ whose seed action is that of a free massless field. From the general $ f( \tensor{T}{^\alpha_\alpha}, T^{\alpha \beta} T_{\alpha \beta})$ deformation of (\ref{eq:TTn}), we recover the usual $\TT$ deformation by taking
\begin{equation}
    f( \tensor{T}{^\alpha_\alpha}, T^{\alpha \beta} T_{\alpha \beta}) = -\frac{1}{2} \left(  T^\alpha{}_\beta T^\beta{}_\alpha-  (T^{\alpha}{}_\alpha)^2 \right) \, .
\end{equation}
Evaluating a few of the terms in the perturbative expansion, we find
\begin{equation}
\begin{aligned}
 \TT_{0}|_{\text{flat}} &=- \frac{1}{4} \left( - \dot{\varphi}^2 + \varphi^{\prime 2 } \right)^2\,, \quad  \TT_{1}|_{\text{flat}} = \frac{1}{2}\left( - \dot{\varphi}^2 + \varphi^{\prime 2 } \right)^3\,, \quad   \TT_{2}|_{\text{flat}} &= -\frac{15}{16}\left( - \dot{\varphi}^2 + \varphi^{\prime 2 } \right)^4 \, .
\end{aligned}
\end{equation}
This series expansion then converges to the well-known $\TT$-deformed action,
\begin{equation}
\begin{aligned}
    S^{(\lambda)} &= \int dt d\theta  \bigg[ \frac{1}{2} \left( - \dot{\varphi}^2 + \varphi^{\prime 2 } \right) - \frac{\lambda}{4} \left( - \dot{\varphi}^2 + \varphi^{\prime 2 } \right)^2+\frac{\lambda^2}{4}\left( - \dot{\varphi}^2 + \varphi^{\prime 2 } \right)^3 -\frac{5 \lambda^3}{16}\left( - \dot{\varphi}^2 + \varphi^{\prime 2 }\right)^4 +\cdots \bigg]
    \\&= \int dt d\theta \frac{1}{2\lambda}\bigg[\sqrt{1+2\lambda \left(  - \dot{\varphi}^2 + \varphi^{\prime 2 } \right)} -1\bigg]\,.
\end{aligned}
\end{equation}

\subsubsection*{\ul{\it $\TT^{\frac{1}{3}}$ Perturbative Flow for Multiple Bosons}}

Next we turn our attention to a deformation which was not considered in the body of this manuscript. Consider a deformation by the relevant $\TT^{\frac{1}{3}}$ operator, which we define by
\begin{equation}
    \TT^{\frac{1}{3}} =  \frac{1}{2} \left(\frac{1}{2} \left(T^\alpha{}_\beta T^\beta{}_\alpha-  (T^{\alpha}{}_\alpha)^2 \right)  \right)^{\frac{1}{3}}\,. 
\end{equation}

We will again consider a seed action for $N$ massless free bosons, given in equation \eqref{eq:seedapp}. The perturbative expansion for the $\TT^{\frac{1}{3}}$-deformed action takes the form
\begin{equation}
    S^{(\lambda)} = S^{(0)} +\sum^\infty_{m=1} \frac{\lambda^m}{m}\int d^2x E~ \TT^{\frac{1}{3}}_{m-1} \, ,
\end{equation}
and a few of the coefficients are
\begin{equation}
\begin{aligned}
     \TT^{\frac{1}{3}}_0|_{\text{flat}} &= \frac{1}{2^{\frac{5}{3}}} \bigg[\left( \dot{\varphi}^i - \varphi^{\prime i} \right) \left( \dot{\varphi}^i - \varphi^{\prime i} \right) \left( \dot{\varphi}^j + \varphi^{\prime j} \right) \left( \dot{\varphi}^j + \varphi^{\prime j} \right)\bigg]^{\frac{1}{3}}\,, \\    \TT^{ \frac{1}{3}}_1|_{\text{flat}} &= \frac{-\dot{\varphi}^{i} \dot{\varphi}^{i} + \varphi^{\prime i}  \varphi^{\prime i} }{9 \cdot 2^{\frac{1}{3}} \bigg[ \left( \dot{\varphi}^i - \varphi^{\prime i} \right) \left( \dot{\varphi}^i - \varphi^{\prime i} \right) \left( \dot{\varphi}^j + \varphi^{\prime j} \right) \left( \dot{\varphi}^j + \varphi^{\prime j} \right) \bigg]^{\frac{1}{3}}}\,, \\
          \TT^{\frac{1}{3}}_2|_{\text{flat}} &= - \frac{\left( \dot{\varphi}^i \dot{\varphi}^i \right)^2 + (\varphi^{\prime i} \varphi^{\prime i} )^2 + 12 (\dot{\varphi}^i \varphi^{\prime i} )^2 - 14 \dot{\varphi}^i \dot{\varphi}^i \varphi^{\prime j} \varphi^{\prime j} }{216 \bigg[  \left( \dot{\varphi}^i - \varphi^{\prime i} \right) \left( \dot{\varphi}^i - \varphi^{\prime i} \right) \left( \dot{\varphi}^j + \varphi^{\prime j} \right) \left( \dot{\varphi}^j + \varphi^{\prime j} \right)  \bigg]}\,.
\end{aligned}
\end{equation}
For this deformation, it does not seem possible to find an all-orders closed-form solution to the flow equation, but the perturbative $\TT^{\frac{1}{3}}$-deformed action to $\mathcal{O} \left( \lambda^{3} \right)$ is
\begin{equation}
\begin{aligned}
    S^{(\lambda)} &= \int dt \, d\theta \, \bigg( \frac{1}{2} \left( - \dot{\varphi}^i  \dot{\varphi}^i +  \varphi^{\prime i} \varphi^{\prime i}  \right)  + \frac{\lambda}{2^{\frac{5}{3}}} \bigg[\left( \dot{\varphi}^i - \varphi^{\prime i} \right) \left( \dot{\varphi}^i - \varphi^{\prime i} \right) \left( \dot{\varphi}^j + \varphi^{\prime j} \right) \left( \dot{\varphi}^j + \varphi^{\prime j} \right)\bigg]^{\frac{1}{3}}\\&+ \frac{\lambda^2}{9 \cdot 2^{\frac{4}{3}}} \frac{-\dot{\varphi}^{i} \dot{\varphi}^{i} + \varphi^{\prime i}  \varphi^{\prime i} }{\bigg[ \left( \dot{\varphi}^i - \varphi^{\prime i} \right) \left( \dot{\varphi}^i - \varphi^{\prime i} \right) \left( \dot{\varphi}^j + \varphi^{\prime j} \right) \left( \dot{\varphi}^j + \varphi^{\prime j} \right) \bigg]^{\frac{1}{3}}} \\& - \frac{\lambda^3}{648} \frac{\left( \dot{\varphi}^i \dot{\varphi}^i \right)^2 + (\varphi^{\prime i} \varphi^{\prime i} )^2 + 12 (\dot{\varphi}^i \varphi^{\prime i} )^2 - 14 \dot{\varphi}^i \dot{\varphi}^i \varphi^{\prime j} \varphi^{\prime j} }{  \left( \dot{\varphi}^i - \varphi^{\prime i} \right) \left( \dot{\varphi}^i - \varphi^{\prime i} \right) \left( \dot{\varphi}^j + \varphi^{\prime j} \right) \left( \dot{\varphi}^j + \varphi^{\prime j} \right)  } + \cdots \bigg)\,. 
\end{aligned}
\end{equation}

\subsubsection*{\ul{\it $f( \tensor{T}{^\alpha_\alpha}, T^{\alpha \beta} T_{\alpha \beta})$ Perturbative Flow for Multiple Bosons}}

To conclude this appendix, we note that one can also study the perturbative solution to the flow driven by the $ f( \tensor{T}{^\alpha_\alpha}, T^{\alpha \beta} T_{\alpha \beta}) = f(z, x)$ operator of equation \eqref{eq:TTn} for an \emph{arbitrary} function $f$. We again take the initial condition for the flow to be the action \eqref{eq:seedapp} for $N$ free massless bosons. The first few terms in the perturbative expansion are
\begin{equation}
\begin{aligned}\label{eq:coeffsttgeneral}
      f( \tensor{T}{^\alpha_\alpha}, T^{\alpha \beta} T_{\alpha \beta})_0|_{\text{flat}} &= f\left(x \right)\,, \quad 
       f( \tensor{T}{^\alpha_\alpha}, T^{\alpha \beta} T_{\alpha \beta})_1|_{\text{flat}} = 4x y \left( \frac{\partial f(x)}{\partial x} \right)^2 \,, 
\end{aligned}
\end{equation}
where
\begin{equation}
    \begin{aligned}
x&= \frac{1}{2} \left( \dot{\varphi}^i - \varphi^{\prime i} \right) \left( \dot{\varphi}^i - \varphi^{\prime i} \right) \left( \dot{\varphi}^j + \varphi^{\prime j} \right) \left( \dot{\varphi}^j + \varphi^{\prime j} \right) = \frac{1}{2} \left(  2 \left( \partial_\mu \varphi^{i} \partial^{\nu} \varphi^{i} \right) \left( \partial_\nu \varphi^{j} \partial^{\mu} \varphi^{j} \right) - \left( \partial_\mu \varphi^{i} \partial^{\mu} \varphi^{i} \right)^2 \right)\,, \\ 
y&=   -\dot{\varphi}^i \dot{\varphi}^i +\varphi^{\prime i} \varphi^{\prime i} = \partial_\mu \varphi^i \partial^\mu \varphi^i\,.
    \end{aligned}
\end{equation}
The perturbative action at $\mathcal{O}(g^2)$ is
\begin{equation}
\begin{aligned}\label{eq:generalperturbative action}
    S^{(g)} &= \int dt d\theta \bigg[ \frac{y}{2} + g f\left(x \right) + 2g^2 xy \left( \frac{\partial f (x)}{\partial x} \right)^2  + \cdots \bigg]\,.
\end{aligned}
\end{equation}
Furthermore, to summarize in the table below, one can check equation \eqref{eq:generalperturbative action} recovers the correct coefficients at $\mathcal{O}(g^2)$ for the perturbative actions describing $N$ free massless bosons considered in this appendix.
\begin{figure*}[htbp]
	\centering
{\includegraphics[width=.95\linewidth]{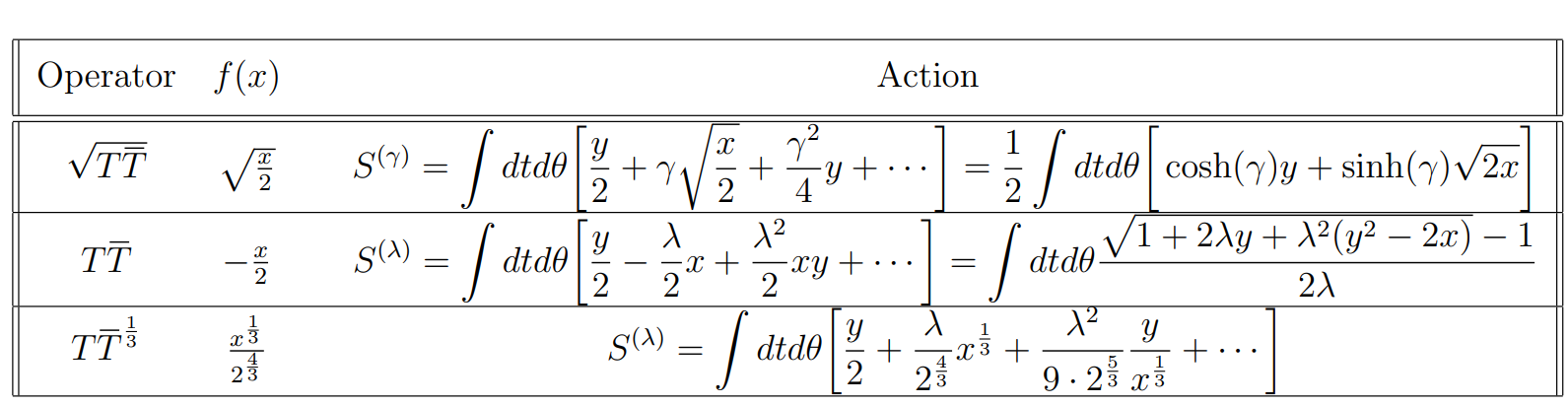}}
\end{figure*}

\begin{comment}
\begin{center}
\begin{tabular}{||c c c ||} 
 \hline
Operator & $f(x)$ & Action  \\ [0.5ex] 
 \hline\hline
 $\sqrt{\TT}$ & $\sqrt{\frac{x}{2}}$ &  \quad    \(\displaystyle S^{(\gamma)} = \int dt d\theta \bigg[ \frac{y}{2} + \gamma \sqrt{\frac{x}{2}} + \frac{\gamma^2}{4} y + \cdots \bigg] = \frac{1}{2} \int dtd\theta \bigg[ \cosh (\gamma) y + \sinh(\gamma) \sqrt{2x}\bigg] \)  \\ 
 \hline
$\TT$ & $-\frac{x}{2}$ & \quad \(\displaystyle S^{(\lambda)} = \int dt d\theta \bigg[ \frac{y}{2} - \frac{\lambda}{2} x + \frac{\lambda^2}{2} xy + \cdots \bigg] =  \int dtd\theta \frac{\sqrt{1+2\lambda y + \lambda^2 (y^2 - 2x)}-1}{2\lambda}  \)   \\
 \hline
 $\TT^{\frac{1}{3}}$ & $\frac{x^{\frac{1}{3}}}{2^{\frac{4}{3}}}$ & \quad \(\displaystyle S^{(\lambda)} = \int dt d\theta \bigg[ \frac{y}{2} + \frac{\lambda}{2^{\frac{4}{3}}} x^{\frac{1}{3}}+ \frac{\lambda^2}{9 \cdot 2^{\frac{5}{3}}} \frac{y}{x^{\frac{1}{3}}} + \cdots \bigg]  \)     \\ [1ex] 
 \hline
\end{tabular}
\end{center}
\end{comment}
In principle, one could also study the perturbative quantization of these more general $ f( \tensor{T}{^\alpha_\alpha}, T^{\alpha \beta} T_{\alpha \beta})$-deformed scalar models. For instance, one could use the background field expansion and determine their Feynman rules as done in section \ref{sec:cian} for the Modified Scalar theory, or study canonical quantization following section \ref{sec:Quantization of first-order deformed}.

\section{Details of Feynman Diagram Calculations}\label{app:feynman}

In this appendix, we collect the technical details of certain evaluations of Feynman diagrams which occur in the analysis of section \ref{sec:cian}.

\subsection{One-loop, $2$-vertex calculation} \label{sec:scalar_field_1_loop_2_vertex_calc}

Let us first focus on the divergence structure of the diagram $\mathcal{D}_2$ of equation (\ref{D2_diagram}), which we repeat here for convenience:
\begin{align}\label{first_diagram}
    \mathcal{D}_2 \; = \; \raisebox{-0.5\height}{\includegraphics[width=0.38\linewidth]{divergent_contribution.pdf}} \; .
\end{align}
As we mentioned around equation (\ref{simpler_integral_body}), the value of this diagram can be expressed in terms of the simpler quantity
\begin{align}
    {\mathcal{I}}^{\mu \nu \rho \tau}_2 = \int \frac{\dd{^{d}\ell}}{\left( 2\pi \right)^{d}} \frac{( \ell + q )^{\nu}\ell^{\mu} ( \ell + q )^{\tau} \ell^{\rho} }{\ell^2 \left( \ell + q \right)^2} \, .
\end{align}
All of the dependence on loop momenta is encoded within ${\mathcal{I}}^{\mu \nu \rho \tau}_2$, which we will also write as ${\mathcal{I}}_2$ with indices suppressed for simplicity. From the value of $\mathcal{I}_2$, the original diagram $\mathcal{D}_2$ is recovered from the expression (\ref{I2_to_D2}), which only involves additional dependence on the classical background via the tensor $\tensor{P}{_\mu_\nu^i^j}$ and an additional integration over the momentum $q$. Therefore, in order to study the divergences arising from the loop, it suffices to perform dimensional regularization of the quantity $\mathcal{I}_2$.

Expanding out the products and introducing a Feynman parameter $x$ in order to resolve the denominator, we find
\begin{align}
    \mathcal{I}_2 &= \int \frac{\dd{^{d}\ell}}{\left( 2\pi \right)^{d}} \frac{\ell^{\nu}\ell^{\mu} \ell^{\tau} \ell^{\rho} + \ell^{\nu} \ell^{\mu} q^{\tau} \ell^{\rho} + q^{\nu} \ell^{\mu} \ell^{\tau} \ell^{\rho}  + q^{\nu} \ell^{\mu} q^{\tau} \ell^{\rho}}{\ell^2 \left( \ell + q \right)^2} \nonumber \\
    &=\int \frac{\dd{^{d}\ell}}{\left( 2\pi \right)^{d}} \int_0^{1} \dd{x}\frac{\ell^{\nu}\ell^{\mu} \ell^{\tau} \ell^{\rho} + \ell^{\nu} \ell^{\mu} q^{\tau} \ell^{\rho} + q^{\nu} \ell^{\mu} \ell^{\tau} \ell^{\rho}  + q^{\nu} \ell^{\mu} q^{\tau} \ell^{\rho}}{\left[ \ell^2 \left( 1 - x \right)  +  x\left( \ell + q \right)^2 \right]^2} \nonumber \\
    &= \int \frac{\dd{^{d}\ell}}{\left( 2\pi \right)^{d}} \int_0^{1} \dd{x}\frac{\ell^{\nu}\ell^{\mu} \ell^{\tau} \ell^{\rho} + \ell^{\nu} \ell^{\mu} q^{\tau} \ell^{\rho} + q^{\nu} \ell^{\mu} \ell^{\tau} \ell^{\rho}  + q^{\nu} \ell^{\mu} q^{\tau} \ell^{\rho}}{\left[ \ell^2  + x \left( 2\ell_{\mu} q^{\mu} + q^2 \right) \right]^2} \nonumber \\
    &= \int \frac{\dd{^{d}\ell}}{\left( 2\pi \right)^{d}} \int_0^{1} \dd{x}\frac{\ell^{\nu}\ell^{\mu} \ell^{\tau} \ell^{\rho} + \ell^{\nu} \ell^{\mu} q^{\tau} \ell^{\rho} + q^{\nu} \ell^{\mu} \ell^{\tau} \ell^{\rho}  + q^{\nu} \ell^{\mu} q^{\tau} \ell^{\rho}}{\left[ \left( \ell^{\mu} + xq^{\mu}  \right) ^2  +x \left( 1 - x \right)q^2  \right]^2} \, .
\end{align}
In the final step, we have completed the square in the denominator by adding and subtracting $q^2 x^2$. We now shift the integration variable from $\ell^\mu$ to
\begin{align}
    \ell^{\prime \mu} = \ell^\mu - x q^\mu \, ,
\end{align}
which causes the denominator to become even in $\ell^\prime$, and thus terms in the numerator which are odd in $\ell^{\prime \mu}$ will vanish by symmetry. We immediately drop the primes on $\ell^{\prime \mu}$ and write the surviving terms as
\begin{align}
    {\mathcal{I}}_2 &= \int \frac{\dd{^{d}\ell}}{\left( 2\pi \right)^{d}} \int_0^{1} \dd{x} \Bigg[ \frac{\ell^{\nu} \ell^{\mu} \ell^{\tau} \ell^{\rho}}{\left( \ell^2  + q^2x\left( 1 - x \right)  \right)^2} + \frac{x^2  \ell^{\nu} q^{\mu} \ell^{\tau} q^{\rho} }{\left( \ell^2 + q^2 x \left( 1 - x \right)  \right)^2} + \frac{\left( x^2 - 2x + 1 \right) q^{\nu} \ell^{\mu} q^{\tau} \ell^{\rho}}{\left( \ell^2  + q^2x\left( 1 - x \right)  \right)^2} \nonumber \\
    &\qquad +  \frac{\left( x^2 - x  \right) \left( q^{\nu} q^{\mu} \ell^{\tau} \ell^{\rho} + q^{\nu} \ell^{\mu} \ell^{\tau} q^{\rho} + \ell^{\nu} \ell^{\mu} q^{\tau} q^{\rho} + \ell^{\nu} q^{\mu} q^{\tau} \ell^{\rho} \right) }{\left[ \ell^2  + q^2x \left( 1 - x \right)  \right]^2}  + \frac{\left( x^{4} -2x^3 + x^2 \right)  q^{\nu} q^{\mu} q^{\tau} q^{\rho}}{\left( \ell^2  + q^2x\left( 1 - x \right)  \right)^2} \Bigg] \, \nonumber \\
    &= \int \frac{\dd{^{d}\ell}}{\left( 2\pi \right)^{d}} \int_0^{1} \dd{x} \Bigg[ \frac{\ell^{\nu} \ell^{\mu} \ell^{\tau} \ell^{\rho}}{\left( \ell^2  + q^2x\left( 1 - x \right)  \right)^2} + \frac{ x^2  \ell^{\nu} q^{\mu} \ell^{\tau} q^{\rho} }{\left( \ell^2 + q^2 x \left( 1 - x \right)  \right)^2} + \frac{\left( 1 - x \right)^2 q^{\nu} \ell^{\mu} q^{\tau} \ell^{\rho}}{\left[ \ell^2  + q^2x\left( 1 - x \right)  \right]^2}  \nonumber\\
    &\quad + \frac{ x\left( 1 -x \right) \left( q^{\nu} q^{\mu} \ell^{\tau} \ell^{\rho} + q^{\nu} \ell^{\mu} \ell^{\tau} q^{\rho} +  \ell^{\nu} \ell^{\mu} q^{\tau} q^{\rho} + \ell^{\nu} q^{\mu} q^{\tau} \ell^{\rho} \right) }{\left[ \ell^2  + q^2x \left( 1 - x \right)  \right]^2}   + \frac{\left[ x^2 \left( 1 - x \right)^2 \right]  q^{\nu} q^{\mu} q^{\tau} q^{\rho}}{\left[ \ell^2  + q^2x\left( 1 - x \right)  \right]^2} \Bigg] \, ,
\end{align}
where in the last expression we have factored various polynomials.

By a symmetry argument similar to the one discussed around equations (\ref{sym_replacements_1}) and (\ref{eq:symmetrization}), within the integral we can replace products of loop momenta with symmetrized combinations of metric tensors:
\begin{align}\label{sym_replacements_app}
    \ell^{\mu} \ell^{\nu} &\to \frac{1}{d} \ell^2 g^{\mu \nu} \, , \nonumber \\
    \ell^{\mu} \ell^{\nu} \ell^{\rho} \ell^{\tau} &\to \frac{1}{d \left( d + 2 \right) } \ell^4 \left( g^{\mu \nu} g^{\rho \tau} + g^{\mu \rho} g^{\nu \tau} + g^{\mu \tau} g^{\nu \rho} \right) \, .
\end{align}
Applying the replacements (\ref{sym_replacements_app}), the integral ${\mathcal{I}}_2$ becomes
\begin{align}\label{app_A1_intermediate}
    {\mathcal{I}}_2 &= \int \frac{\dd{^{d}\ell}}{\left( 2\pi \right)^{d}} \int_0^{1} \dd{x} \Bigg( \frac{\ell^{4}}{d \left( d + 2 \right) }\frac{g^{\mu \nu} g^{\rho \tau} + g^{\mu\tau} g^{\nu \rho} + g^{\mu \rho} g^{\nu \tau}  }{\left[ \ell^2  + q^2x\left( 1 - x \right)  \right]^2} + \frac{\ell^2}{d}\frac{x^2 g^{\nu \tau}q^{\mu} q^{\rho} }{\left[ \ell^2  + q^2x \left( 1 - x \right)  \right]^2} \nonumber\\
    &\qquad + \frac{\ell^2}{d}\frac{ \left( 1 - x \right)^2  g^{\mu \rho} q^{\nu} q^{\tau} }{\left[ \ell^2  + q^2x\left( 1 - x \right)  \right]^2} + \frac{\ell^2}{d}\frac{x\left( 1 -x \right) \left( q^{\nu} q^{\mu} g^{\tau\rho} + g^{\mu \tau} q^{\nu} q^{\rho} + g^{\nu \mu} q^{\tau} q^{\rho} + g^{\nu \rho}q^{\mu} q^{\tau} \right) }{\left[ \ell^2  + q^2x \left( 1 - x \right)  \right]^2} \nonumber\\
    &\qquad + \frac{ x^2 \left( 1 - x \right)^2  q^{\nu} q^{\mu} q^{\tau} q^{\rho}}{\left[ \ell^2  + q^2x\left( 1 - x \right)  \right]^2} \Bigg) \, .
\end{align}
It will be convenient to make use of the standard result
\begin{align}\label{first_anselmi}
    \int \frac{\dd{^{d}\ell}}{\left( 2\pi \right)^{d}} \frac{\ell^{2\beta}}{\left( \ell^2 - \Delta^2 \right)^{\alpha}} = i \left( -1 \right)^{\alpha + \beta} \frac{\Gamma \left( \beta + \frac{d}{2} \right) \Gamma \left( \alpha - \beta - \frac{d}{2} \right) }{\left( 4\pi \right)^{\frac{d}{2}} \Gamma \left( \alpha \right) \Gamma \left( \frac{d}{2} \right)  } \Delta^{2 \left( \frac{d}{2} - \alpha + \beta \right) } \, ,
\end{align}
which can be found, for instance, in equation (A.4) in \cite{Anselmi:2019pdm}. Using (\ref{first_anselmi}) with
\begin{align}\label{delta_sq_defn}
    \Delta^2 = -q^2 x\left( 1 - x \right) 
\end{align}
in equation (\ref{app_A1_intermediate}), we find
\begin{align}
    {\mathcal{I}}_2 &= \frac{i}{\left( 4\pi \right)^{\frac{d}{2}} \Gamma \left( 2 \right) \Gamma \left( \frac{d}{2} \right)  } \int_0^{1} \dd{x} \Bigg( \frac{\Delta^{d}}{d \left( d + 2 \right) }\Gamma \left( 2 + \frac{d}{2} \right) \Gamma \left( -\frac{d}{2} \right) \left( g^{\mu \nu} g^{\rho \tau} + g^{\mu \rho} g^{\nu \tau} + g^{\mu \tau} g^{\nu \rho} \right)  \nonumber\\
    &\quad +\frac{\Delta^{d - 2}}{d} \Gamma \left( 1 + \frac{d}{2} \right) \Gamma \left( 1 - \frac{d}{2} \right)  x \left( 1 - x \right)  \left( q^{\nu} q^{\mu} g^{\tau \rho} + g^{\mu \tau} q^{\nu} q^{\rho} +  g^{\nu \mu} q^{\tau} q^{\rho} + g^{\nu \rho} q^{\mu} q^{\tau} \right) \nonumber\\
    &\quad + \frac{\Delta^{d-2}}{d} \Gamma \left( 1 + \frac{d}{2} \right) \Gamma \left( 1 - \frac{d}{2} \right) x^2 g^{\nu \tau} q^{\mu} q^{\rho} +\frac{\Delta^{d-2}}{d} \Gamma \left( 1 + \frac{d}{2} \right) \Gamma \left( 1 - \frac{d}{2} \right) \left( 1 - x \right)^2  g^{\mu \rho} q^{\nu} q^{\tau}\nonumber \\
    &\quad +\Delta^{d-4}\Gamma \left( \frac{d}{2} \right) \Gamma \left( 2 - \frac{d}{2} \right)  \left[ x^2 \left( 1 - x \right)^2  \right] q^{\nu} q^{\mu} q^{\tau} q^{\rho} \Bigg) \, .
\end{align}
Using gamma function identities and some algebra, one can simplify this to
\begin{align}
    {\mathcal{I}}_2 &= \frac{i\Gamma \left( -\frac{d}{2} \right)}{\left( 4\pi \right)^{\frac{d}{2}}} \int_0^{1} \dd{x} \Bigg( \frac{\Delta^{d}}{4} \left( g^{\mu \nu} g^{\rho \tau} +   g^{\mu \rho} g^{\nu \tau} + g^{\mu \tau} g^{\nu \rho} \right) -\frac{d\Delta^{d-2}}{4} x^2 g^{\nu \tau} q^{\mu} q^{\rho}  \nonumber \\
    &\quad -\frac{d\Delta^{d-2}}{4} \left( 1 - x \right)^2 g^{\mu \rho} q^{\nu} q^{\tau} - \frac{d\Delta^{d - 2}}{4} x \left( 1 - x \right) \left( q^{\nu} q^{\mu} g^{\tau \rho} + g^{\mu\tau} q^{\nu} q^{\rho} + g^{\nu \mu} q^{\tau} q^{\rho} + g^{\nu \rho} q^{\mu} q^{\tau} \right) \nonumber\\
    &\quad  + \frac{d\left( d - 2 \right) \Delta^{d-4}}{4} x^2 \left( 1 - x \right)^2 q^{\nu} q^{\mu} q^{\tau} q^{\rho} \Bigg) \, .
\end{align}
After substituting in for $\Delta^2$ using the definition (\ref{delta_sq_defn}), we can now evaluate the resulting integrals using the formula
\begin{align}
    \int_0^{1} \dd{x} x^{\alpha - 1} \left( 1 - x \right)^{\beta - 1} &= \frac{\Gamma \left( \alpha \right) \Gamma \left( \beta \right) }{\Gamma \left( \alpha + \beta \right) } = \mathrm{B} ( \alpha, \beta ) \, , 
\end{align}
which we recognize as the definition of the beta function $\mathrm{B} ( \alpha , \beta )$. By doing this, we find
\begin{align}\label{appendix_intermediate_two}
    {\mathcal{I}}_2 &= \frac{i\Gamma \left( -\frac{d}{2} \right)}{4\left( 4\pi \right)^{\frac{d}{2}}} \int_0^{1} \dd{x} \Bigg( d \left( d - 2 \right) q^{d-4}\left[ -x\left( 1 - x \right)   \right]^{\frac{d}{2}} q^{\nu} q^{\mu} q^{\tau} q^{\rho} - dq^{d-2}  \left( -x \right)^{\frac{d}{2}-1}\left( 1 - x \right)^{\frac{d}{2}+1} g^{\mu \rho} q^{\nu} q^{\tau} \nonumber\\
    &\quad + q^{d} \left[ -x\left( 1 - x \right) \right]^{\frac{d}{2}} \left( g^{\mu \nu} g^{\rho \tau} + g^{\mu \rho} g^{\nu \tau} + g^{\mu \tau} g^{\nu \rho} \right)    -dq^{d-2} \left[ \left( -x \right)^{\frac{d}{2}+1} \left( 1 - x \right)^{\frac{d}{2}-1}\right] g^{\nu \tau} q^{\mu} q^{\rho}  \nonumber \\
    &\quad +dq^{d - 2} \left[-x \left( 1 - x \right)  \right]^{\frac{d}{2}} \left( q^{\nu} q^{\mu} g^{\tau \rho} + g^{\mu \tau} q^{\nu} q^{\rho} + g^{\nu \mu} q^{\tau} q^{\rho} + g^{\nu \rho} q^{\mu} q^{\tau} \right)   \Bigg) \nonumber \\
    &= \frac{i \left( -1 \right)^{\frac{d}{2}}\Gamma \left( -\frac{d}{2} \right) }{4\left( 4\pi \right)^{\frac{d}{2}}} \Bigg( q^{d} \frac{\Gamma \left( \frac{d}{2} + 1 \right)^2}{\Gamma \left( d + 2 \right) } \left( g^{\mu \nu} g^{\rho \tau} + g^{\mu \rho} g^{\nu \tau} + g^{\mu \tau} g^{\nu \rho} \right)   \nonumber \\
    &\quad +dq^{d-2} \frac{\Gamma \left( \frac{d}{2} \right) \Gamma \left( \frac{d}{2} + 2 \right) }{\Gamma \left( d + 2 \right) } \left[ g^{\mu \rho} q^{\nu} q^{\tau} + g^{\nu \tau} q^{\mu} q^{\rho} \right] + d\left( d - 2 \right) q^{d - 4} \frac{\Gamma\left( \frac{d}{2} + 1 \right)^2}{\Gamma \left( d + 2 \right)} q^{\nu} q^{\mu} q^{\tau} q^{\rho} \nonumber \\
    &\quad +dq^{d-2} \frac{\Gamma \left( \frac{d}{2} + 1 \right)^2}{\Gamma \left( d + 2 \right) } \left( q^{\nu} q^{\mu} g^{\tau \rho} + g^{\mu \tau} q^{\nu}q^{\rho} +  g^{\nu \mu} q^{\tau} q^{\rho} + g^{\nu \rho} q^{\mu} q^{\tau} \right)  \Bigg) \, .
\end{align}
Note that each term in (\ref{appendix_intermediate_two}) scales as $q^d$, as expected. Factoring out the gamma functions, we have found
\begin{align}
    {\mathcal{I}}_2 &= \frac{i \left( -1 \right)^{\frac{d}{2}}\Gamma \left( -\frac{d}{2} \right) }{4\left( 4\pi \right)^{\frac{d}{2}}} \frac{\Gamma \left( \frac{d}{2} + 1 \right)^2}{\Gamma \left( d + 2 \right) } \Bigg[ q^{d}  \left( g^{\mu \nu} g^{\rho \tau} + g^{\mu \rho} g^{\nu \tau} + g^{\mu \tau} g^{\nu \rho} \right) + d \left( d - 2 \right) q^{d - 4}q^{\nu} q^{\mu} q^{\tau} q^{\rho} \nonumber \\
    &\quad + dq^{d-2} \left( q^{\nu} q^{\mu} g^{\tau \rho} + g^{\mu \tau} q^{\nu} q^{\rho} + g^{\nu \mu} q^{\tau} q^{\rho} + g^{\nu \rho} q^{\mu} q^{\tau} \right)  +\left( d + 2 \right) q^{d-2} \left( g^{\mu \rho} q^{\nu} q^{\tau} + g^{\nu \tau} q^{\mu} q^{\rho} \right)  \Bigg] \, .
\end{align}
Finally, to perform dimensional regularization, we set the spacetime dimension to $d = 2 + 2 \epsilon$ and take $\epsilon \to 0$ using the limiting behavior
\begin{align}
    \Gamma \left( -1-\epsilon \right) = \frac{1}{\epsilon} - \gamma + 1 +  \mathcal{O}\left( \epsilon \right)
\end{align}
for the gamma functions. Keeping only divergent terms, we arrive at the final expression
\begin{align}\label{two_vertex}
    {\mathcal{I}}_2 &= \left( \frac{1}{\epsilon} \right) \frac{-i}{24\left( 4\pi \right)} \bigg[ q^{2}  \left( g^{\mu \nu} g^{\rho \tau}  + g^{\mu \rho} g^{\nu \tau} + g^{\mu \tau} g^{\nu \rho} \right) +2 \left( q^{\nu} q^{\mu} g^{\tau \rho} + g^{\mu \tau} q^{\nu} q^{\rho}+ g^{\nu \mu} q^{\tau} q^{\rho} + g^{\nu \rho} q^{\mu} q^{\tau} \right) \nonumber \\
    &\qquad \qquad \qquad \qquad + 4\left( g^{\mu \rho} q^{\nu} q^{\tau} + g^{\nu \tau} q^{\mu} q^{\rho} \right) \bigg] \, .
\end{align}
This completes the evaluation of the divergent contribution from ${\mathcal{I}}_2$, which justifies the result (\ref{divergent_final_body}) which was quoted in the body of the paper.

\subsection{One-loop, $n$-vertex calculation}\label{app:one_loop_n_vertex}

The $n$-vertex diagram $\mathcal{D}_n$ can be computed via a generalization of the method used in appendix \ref{sec:scalar_field_1_loop_2_vertex_calc}. We again write $\ell$ for the loop momentum and we label the external momenta as $q_{i}$, for $i = 0, \ldots, n-1$, with momentum conservation implying that
\begin{align}
    q_{n-1} = - \sum_{i=0}^{n-2} q_{i} \, .
\end{align}
As we did with $\mathcal{D}_2$ in equation (\ref{D2_stripped}), let us strip off various factors of $\tensor{P}{_\mu_\nu^i^j}$ to write
\begin{align}\label{Dn_defn}
    \mathcal{D}_n &= \frac{\left( - 2 \tanh ( \gamma ) \right)^n}{n} \int \left(\frac{d^d q_0}{ ( 2 \pi )^d } P_{\mu_1 \mu_2}^{i_1 i_2} ( q_0 ) \right) \left(\frac{d^d q_0}{ ( 2 \pi )^d } P_{\mu_3 \mu_4}^{i_2 i_3} ( q_1 ) \right) \cdots \left(\frac{d^d q_{n-2}}{(2 \pi)^d} P_{\mu_{2n-3} \mu_{2n-2}}^{i_{n-1} i_{n}} ( q_{n-2} ) \right) \nonumber \\
    &\qquad \qquad \qquad \qquad \qquad \cdot P_{\mu_{2n-1} \mu_{2n}}^{i_{n} i_{1}} ( q_{n-1} ) \left( {\mathcal{I}}_n \right)^{\mu_1 \mu_2 \ldots \mu_{2n-1} \mu_{2n}} \, ,
\end{align}
where ${\mathcal{I}}_n$ is the simpler integral
\begin{align}\label{In_defn}\hspace{-10pt}
    \left( {\mathcal{I}}_n \right)^{\mu_1 \mu_2 \ldots \mu_{2n-1} \mu_{2n} } &= \int \frac{d^d \ell}{( 2\pi )^d} \, \left( \prod_{i = 0}^{n-1} \left( \ell + \sum_{j=0}^{i} q_{j}  \right)^{-2} \right) \left( \prod_{k=1}^{n} \left( \ell + \sum_{j=0}^{k-1} q_{j} \right)^{\mu_{2k-1}} \left( \ell + \sum_{j=0}^{k-1} q_{j}  \right)^{\mu_{2k}} \right) \, .
\end{align}
We will further break up $\mathcal{I}_n$ into pieces and evaluate each piece in turn. Let us write the integrand of (\ref{In_defn}) as a product
\begin{align}\label{In_PV}
    \left( {\mathcal{I}}_n \right)^{\mu_1 \mu_2 \ldots \mu_{2n-1} \mu_{2n} } = \int \frac{d^d \ell}{( 2\pi )^d} \, \mathcal{P}_n \mathcal{V}^{\mu_1 \mu_2 \ldots \mu_{2n-1} \mu_{2n} } \, , 
\end{align}
where the symbol
\begin{align}
    \mathcal{P}_n &= \prod_{i = 0}^{n-1} \left( \ell + \sum_{j=0}^{i} q_{j}  \right)^{-2}
\end{align}
refers to the collection of all factors in $\mathcal{I}_n$ which come from propagators, and the symbol
\begin{align}\label{Vn_defn}
    \mathcal{V}^{\mu_1 \mu_2 \ldots \mu_{2n-1} \mu_{2n} } &= \prod_{k=1}^{n} \left( \ell + \sum_{j=0}^{k-1} q_{j} \right)^{\mu_{2k-1}} \left( \ell + \sum_{j=0}^{k-1} q_{j}  \right)^{\mu_{2k}} \, ,
\end{align}
which we will sometimes abbreviate as $\mathcal{V}$, refers to the pieces coming from vertex factors.

In order to highlight the divergence structure of the diagram $\mathcal{D}_n$, we will focus on performing the loop momentum integral of various terms appearing in the product $\mathcal{P} \mathcal{V}$ of equation (\ref{In_PV}), and neglect the additional structure arising from the contraction with the various tensors $P_{\mu \nu}^{ij}$ to obtain $\mathcal{D}_n$ in (\ref{Dn_defn}).

Let us begin by simplifying the product $\mathcal{P}$ of $n$ propagator factors. In general, we can write the product of $n$ propagators using a Feynman parameterization:
\begin{align}
    \prod_{i=0}^{n-1} A_{i}^{-1} &= \int_0^{1} \left( \prod_{i=0}^{n-1} \dd{x}_{i} \right) \delta \left( \sum_{i=0}^{n-1} x_{i} - 1  \right) \frac{\left( n - 1 \right)!}{\left[ \sum_{i}^{}x_{i} A_{i}  \right]^{n} } \, .
\end{align}
The product of propagators inside the loop can thus be expressed as 
\begin{align}
    \mathcal{P} = \int_0^{1} \left( \prod_{i=0}^{n-1} \dd{x}_{i} \right) \delta \left( \sum_{i=0}^{n-1} x_{i} - 1  \right) \left[ \sum_{i=0}^{n-1}x_{i} \left( \ell + \sum_{j=0}^{i} q_{j} \right)^2   \right]^{-n} \, .
\end{align}
As $\sum_{i}^{} x_{i} = 1$, we can expand and reduce the square bracketed term to
\begin{align}\hspace{-10pt}
    \left[ \sum_{i}^{}x_{i} \left( \ell + \sum_{j=0}^{i} q_{j} \right)^2   \right]^{-n} &= \left[ \ell^2  + \sum_{i}^{}x_{i} \left( 2\ell_{\mu} \sum_{j=0}^{i} q_{j}^{\mu}+ \left( \sum_{j=0}^{i} q_{j} \right)^2  \right)   \right]^{-n} \, \nonumber \\
    &=  \left[ \left( \ell + \sum_{i}^{} \sum_{j=0}^{i} x_{i} q_{j}  \right)^2 - \left( \sum_{i}^{} \sum_{j=0}^{i} x_{i} q_{j} \right)^2  + \sum_{i}^{}x_{i}   \left( \sum_{j=0}^{i} q_{j} \right)^2  \right]^{-n} \, .
\end{align}
We change variables in the loop momentum by shifting
\begin{align}\label{loop_momentum_shift}
    \ell^\mu \to \ell^\mu - \sum_{i=0}^{n-1} \sum_{j=0}^{i} x_{i} q^\mu_{j} \, ,
\end{align}
so that the bracketed expression becomes
\begin{align}
    \left[ \sum_{i}^{}x_{i} \left( \ell + \sum_{j=0}^{i} q_{j} \right)^2   \right]^{-n} &=  \left[ \ell^2 - \left( \sum_{i}^{} \sum_{j=0}^{i} x_{i} q_{j} \right)^2  + \sum_{i}^{}x_{i}   \left( \sum_{j=0}^{i} q_{j} \right)^2  \right]^{-n} \nonumber \\
    &=  \left[ \ell^2 - \Delta^2  \right]^{-n} \, ,
\end{align}
where we have defined
\begin{align}
    \Delta^2 &= \left( \sum_{i}^{} \sum_{j=0}^{i} x_{i} q_{j} \right)^2  - \sum_{i}^{}x_{i}   \left( \sum_{j=0}^{i} q_{j} \right)^2 \, .
\end{align}
Overall this allows us to write the propagators as
\begin{align}
    \mathcal{P} = (n - 1)! \int_0^{1} \left( \prod_{i=0}^{n-1} \dd{x}_{i} \right) \delta \left( \sum_{i=0}^{n-1} x_{i} - 1  \right) \left( \ell^2 - \Delta^2  \right)^{-n} \, .
\end{align}
Next let us turn to the contributions from the vertices in equation (\ref{Vn_defn}), which yield factors of momenta in the numerator of the integrand. Under the change of variables (\ref{loop_momentum_shift}) which renders the denominator of $\mathcal{P}$ quadratic in $\ell$, the vertex factor contribution becomes
\begin{align}
    \mathcal{V}^{\mu_1 \mu_2 \ldots \mu_{2n-1} \mu_{2n} } =  \prod_{k=1}^{n} \left( \ell - \sum_{i=0}^{n-1} \sum_{j=0}^{i} x_{i} q_{j}  + \sum_{j=0}^{k-1} q_{j} \right)^{\mu_{2k-1}} \left( \ell - \sum_{i=0}^{n-1} \sum_{j=0}^{i} x_{i} q_{j}+ \sum_{j=0}^{k-1} q_{j}  \right)^{\mu_{2k}} \, .
\end{align}
We expand this product in descending powers of $\ell$ as only powers $\ell^{2n}$ and $\ell^{2n-2}$ will lead to divergent terms. We have that
\begin{align}\label{V_formula}
    \mathcal{V}^{\mu_1 \mu_2 \ldots \mu_{2n-1} \mu_{2n} } &= \prod_{k=1}^{n} \ell^{\mu_{2k-1}} \ell^{\mu_{2k}} + \sum_{a=1}^{2n} \sum_{b>a}^{2n} \left( \prod_{c \neq a,b}^{2n} \ell^{\mu_c} \right)  f^{\mu_{a}} \left( x,q,a \right) f^{\mu_{b}} \left( x,q,b  \right) + \mathcal{O}\left( \ell^{2n-4} \right) \, ,
\end{align}
where we have defined for brevity
\begin{align}\label{cursed_defn}
    f^{\mu} \left( x,q,a \right) = \sum_{i=0}^{n-1} \sum_{j=0}^{i} x_{i} q_{j}^{\mu}  + \sum_{j=0}^{\lfloor \frac{a-1}{2} \rfloor} q_{j}^{\mu} \, .
\end{align}
% Given a=2k or 2k-1 we need to get k-1
% floor (a-1)/2 achieves this

Next we will replace products of loop momenta using the generalized symmetrization rule of equation (\ref{eq:symmetrization}). To ease notation, let us write
\begin{align}\label{shorthand_metric}
    g^{\mu_1 \ldots \mu_n} = g^{(\mu_1 \mu_2 } \cdots g^{\mu_{n-1} \mu_{n})} \, 
\end{align}
for the symmetrized combination of derivatives appearing in this expression. When no confusion is possible, we will also write $g^{\{ \mu \}}$ for (\ref{shorthand_metric}), where $\{ \mu \}$ is understood to refer to a multi-index $\{ \mu \} = \mu_1 \ldots \mu_n$. With this notation, the replacement rule becomes
\begin{align}
    \prod_{i=1}^{n} \ell^{\mu_{2i- 1 }} \ell^{\mu_{2i}} \to \frac{ \ell^{2n} \left( d -2 \right)!! \left( 2n - 1 \right)!!}{\left( d-2 + 2n \right)!!} g^{\mu_{1}\ldots \mu_{2n}} \, .
\end{align}
This transforms the vertex factor contribution to
\begin{align}\label{vertex_intermediate}
    \mathcal{V}^{\mu_1 \mu_2 \ldots \mu_{2n-1} \mu_{2n} } &=  \frac{ \ell^{2n} \left( d - 2 \right)!! \left( 2n - 1 \right)!!}{ \left( d  - 2 + 2n \right)!!} g^{\mu_1 \cdots \mu_{2n}} \nonumber\\
    &\quad + \frac{  \ell^{2n-2} \left( d - 2 \right)!! \left( 2n - 3 \right)!!}{\left( d - 4 + 2n \right)!!}\sum_{a=1}^{2n} \sum_{b>a}^{2n} g^{\{ \mu \neq \mu_a, \mu_b \}}   f^{\mu_a} \left( x,q,a \right)  f^{\mu_{b}} \left( x,q,b \right) + \mathcal{O}\left( \ell^{2n-4} \right) \, .
\end{align}
In equation (\ref{vertex_intermediate}), we have written $g^{\{ \mu \neq \mu_a, \mu_b \}}$ to refer to a product of the form (\ref{shorthand_metric}) in which the multi-index $\{ \mu \}$ runs over all possible values except for the two indices $\mu_a$ and $\mu_b$, which are excluded.

Let us now combine the pieces and identify the divergent terms in $\mathcal{D}_n$. We can evaluate
\begin{align}\hspace{-10pt}\label{Dn_Itilde_intermediate}
    &\left( {\mathcal{I}}_{n} \right)^{\mu_1 \mu_2 \ldots \mu_{2n}} \nonumber \\
    &\quad = \int \frac{d^d \ell}{( 2\pi )^d} \mathcal{P} \mathcal{V}^{\mu_1 \mu_2 \ldots \mu_{2n}} \, \nonumber \\
    &\quad = \int \frac{d^d \ell}{( 2\pi )^d} \left( n - 1 \right)!\int_0^{1} \left( \prod_{i=0}^{n-1} \dd{x}_{i} \right) \delta \left( \sum_{i=0}^{n-1} x_{i} - 1  \right) \left( \ell^2 - \Delta^2  \right)^{-n}  \cdot \Bigg( \frac{ \ell^{2n} \left( d - 2 \right)!! \left( 2n - 1 \right)!!}{\left( d  - 2 + 2n \right)!!} g^{\mu_1 \ldots \mu_{2n}} \nonumber \\
    &\quad \qquad + \frac{ \ell^{2n-2} \left( d - 2 \right)!! \left( 2n - 3 \right)!!}{\left( d - 4+ 2n  \right)!!}\sum_{a=1}^{2n} \sum_{b>a}^{2n} g^{\{ \mu \neq \mu_a, \mu_b \}}   f^{\mu_a} \left( x,q,a \right)  f^{\mu_{b}} \left( x,q,b \right)  + \mathcal{O}\left( \ell^{2n-4} \right) \Bigg) \, ,
\end{align}
in terms of the known integral
\begin{align}\label{known_integral}
    \int \frac{\dd{^{d}\ell}}{\left( 2\pi \right)^{d}} \frac{\ell^{2\beta}}{\left( \ell^2 - \Delta^2 \right)^{\alpha}} &= i \left( -1 \right)^{\alpha + \beta} \frac{\Gamma \left( \beta + \frac{d}{2} \right) \Gamma \left( \alpha - \beta - \frac{d}{2} \right) }{\left( 4\pi \right)^{\frac{d}{2}} \Gamma \left( \alpha \right) \Gamma \left( \frac{d}{2} \right)  } \Delta^{2 \left( \frac{d}{2} - \alpha + \beta \right) } \, .
\end{align}
First let us justify why the terms of order $\ell^{2n-4}$ and lower in equation (\ref{Dn_Itilde_intermediate}) will not give divergent contributions. A term proportional to $\ell^{2n-4}$ in the parentheses of (\ref{Dn_Itilde_intermediate}), after multiplying the propagator factor $( \ell^2 - \Delta^2 )^{-n}$, gives a term in the integrand of the form (\ref{known_integral}) with $\alpha = n$ and $\beta = n - 2$. Such a term gives a contribution
\begin{align}\label{known_integral_finite}
    \int \frac{\dd{^{d}\ell}}{\left( 2\pi \right)^{d}} \frac{\ell^{2 ( n - 2 ) }}{\left( \ell^2 - \Delta^2 \right)^{n}} &= i \left( -1 \right)^{n +  ( n - 2 )} \frac{\Gamma \left( n - 2 + \frac{d}{2} \right) \Gamma \left( 2 - \frac{d}{2} \right) }{\left( 4\pi \right)^{\frac{d}{2}} \Gamma \left( n \right) \Gamma \left( \frac{d}{2} \right)  } \Delta^{2 \left( \frac{d}{2} - n + ( n - 2 ) \right) } \, .
\end{align}
In the limit as $d \to 2$, the two factors of gamma functions in the numerator of (\ref{known_integral_finite}) tend to $\Gamma ( n - 1 )$ and $\Gamma ( 1 )$, which are both finite since $n > 2$. Since we are only interested in computing the divergent contributions arising from these diagrams, we ignore these terms. Similarly, any terms of lower order in $\ell$ can be evaluated in the same way but with even smaller values of $\beta$, which also lead to finite contributions from the gamma functions.

Let us therefore focus on the divergent terms. The term in the integrand proportional to $\ell^{2n}$ in equation (\ref{Dn_Itilde_intermediate}) takes the form (\ref{known_integral}) with $\alpha = \beta = n$. Similarly, the term in the integrand that scales as $\ell^{2n-2}$ is of the form (\ref{known_integral}) with $\alpha = n$ and $\beta = n - 1$. Evaluating the loop momentum integrals then gives 
\begin{align}\label{In_final}
    \left( {\mathcal{I}}_{n} \right)^{\mu_1 \mu_2 \ldots \mu_{2n}} &= \left( n - 1 \right)!\int_0^{1} \left( \prod_{i=0}^{n-1} \dd{x}_{i} \right) \delta \left( \sum_{i=0}^{n-1} x_{i} - 1  \right)  \nonumber \\
    &\quad \cdot \Bigg( \frac{i \left( d - 2 \right)!! \left( 2n - 1 \right)!!}{ \left( d  - 2 + 2n\right)!!} g^{\mu_1 \cdots \mu_{2n}} \frac{\Gamma \left( n + \frac{d}{2} \right)  }{\left( 4\pi \right)^{\frac{d}{2}} \Gamma \left( n \right) \Gamma \left( \frac{d}{2} \right)} \Gamma \left( -\frac{d}{2} \right)\Delta^{ d }  \nonumber \\
    &\quad + \frac{i\left( d - 2 \right)!! \left( 2n - 3 \right)!!}{\left( d  - 4+ 2n \right)!!}\sum_{a=1}^{2n} \sum_{b>a}^{2n} g^{\{ \mu \neq \mu_a, \mu_b \}}   f^{\mu_a} \left( x,q,a \right)  f^{\mu_{b}} \left( x,q,b \right) \nonumber \\
    &\quad \cdot \frac{\Gamma \left( n-1+\frac{d}{2} \right) }{\left( 4\pi \right)^{\frac{d}{2}} \Gamma \left( n \right) \Gamma \left( \frac{d}{2} \right)  } \Gamma \left( 1 - \frac{d}{2} \right) \Delta^{d - 2} \Bigg) \, .
\end{align}
To better analyze the divergence structure, it is useful to define shorthand notation for the two terms appearing in (\ref{In_final}), which we call $\vb{C}_{2n}^{\mu_1 \ldots \mu_{2n}}$ and $\vb{D}_{2n}^{\mu_1 \ldots \mu_{2n}}$:
\begin{align}
    \vb{C}_{2n}^{\mu_1 \ldots \mu_{2n}} &\equiv \frac{i \left( d - 2 \right)!! \left( 2n - 1 \right)!!}{ \left( d - 2 + 2n \right)!!} g^{\mu_1 \cdots \mu_{2n}}  \frac{\Gamma \left( n + \frac{d}{2} \right)  }{\left( 4\pi \right)^{\frac{d}{2}} \Gamma \left( n \right) \Gamma \left( \frac{d}{2} \right)} \Gamma \left( -\frac{d}{2} \right)\Delta^{d} \, \, \nonumber \\
     \vb{D}_{2n}^{\mu_1 \ldots \mu_{2n}} &\equiv \frac{i \left( d - 2 \right)!! \left( 2n - 3 \right)!!}{\left( d - 4 + 2n \right)!!}\sum_{a=1}^{2n} \sum_{b>a}^{2n} g^{\{ \mu \neq \mu_a, \mu_b \}}   f^{\mu_a} \left( x,q,a \right)  f^{\mu_{b}} \left( x,q,b \right)\nonumber \\
    &\qquad \cdot \frac{\Gamma \left( n-1+\frac{d}{2} \right) }{\left( 4\pi \right)^{\frac{d}{2}} \Gamma \left( n \right) \Gamma \left( \frac{d}{2} \right)  } \Gamma \left( 1 - \frac{d}{2} \right) \Delta^{d - 2} \, .
\end{align}
Both contributions $\vb{C}_{2n}^{\mu_1 \ldots \mu_{2n}}$ and $\vb{D}_{2n}^{\mu_1 \ldots \mu_{2n}}$ scale as $q^d$ and contain a polynomial in $x_{i}$ of degree $d$. In $\vb{C}_{2n}^{\mu_1 \ldots \mu_{2n}}$, this $q^{d}$ dependence is contained within $\Delta^{d}$, and for $\vb{D}_{2n}^{\mu_1 \ldots \mu_{2n}}$, the power of $q^{d-2}$ from $\Delta^{d-2}$ is compensated by two factors of $q$, one of which sits in each function $f^\mu$.

Therefore, we conclude that in the limit $d \to 2$, all such $n$ vertex diagrams have the same general structure as the $2$-vertex diagram which we saw in (\ref{two_vertex}). In particular, both $\vb{C}_{2n}^{\mu_1 \ldots \mu_{2n}}$ and $\vb{D}_{2n}^{\mu_1 \ldots \mu_{2n}}$ generate divergences of the form $\frac{1}{\epsilon}$ because they are proportional to $\Gamma \left( - \frac{d}{2} \right)$ and $\Gamma \left( 1 - \frac{d}{2} \right)$, respectively.

We conclude that
\begin{align}
    \vb{C}_{2n}^{\mu_1 \ldots \mu_{2n}}&\sim \frac{1}{\epsilon} q^2 g^{\mu_1 \cdots \mu_{2n}}  \, , \nonumber \\
    \vb{D}_{2n}^{\mu_1 \ldots \mu_{2n}} &\sim \frac{1}{\epsilon} \sum_{a=1}^{2n} \sum_{b>a}^{2n} q^{\mu_a} q^{\mu_b} g^{\{ \mu \neq \mu_a, \mu_b \}}  \, .
\end{align}

\subsection{$m$-loop, two-vertex calculation} \label{sec:scalar_field_n_loop_2_vertex_calc}

In this appendix, we will show how to evaluate the integral over one of the $m$ loop momenta $\ell_i$ which appear in the expression for the two-vertex, $m$-loop diagram of equation (\ref{two_vertex_n_loop}). It suffices to integrate over the final momentum $\ell_{m}$, since the result may then be iterated to evaluate the other $m-1$ integrals.

Therefore, let us focus on performing the integration over $\ell_m$ in the quantity $\mathfrak{L}_{m, 2}$ of equation (\ref{frak_L_defn}). Specifically, we will compute the quantity
\begin{align}\label{one_integral_Lm2_defn}
    L_{m, 2} &= \int \dd{^{d}\ell_{m}} \frac{\ell_{m}^{\nu_{m+1}} \ell_{m}^{\tau_{m+1}} \left( \ell_{m} - \ell_{m-1} \right)^{\nu_{m}} \left( \ell_{m} - \ell_{m-1} \right)^{\tau_{m}}}{\ell_{m}^2 \left( \ell_{m} - \ell_{m-1} \right)^2} \, .
\end{align}
This object $L_{m, 2}$ is proportional to the remaining integrand that one finds by performing the integral over $\ell_m$ in the definition of $\mathfrak{L}_{m, 2}$. As we will see, after obtaining an expression for $L_{m,2}$, this result can be used recursively to evaluate $\mathfrak{L}_{m, 2}$ itself.

We notice the integral in equation (\ref{one_integral_Lm2_defn}) is exactly of the form of the one appearing in the $1$-loop, $2$-vertex diagram which we evaluated in appendix \ref{sec:scalar_field_1_loop_2_vertex_calc}. Proceeding in the same way, we introduce a Feynman parameter $x$ to write
\begin{align}
    L_{m, 2} &= \int \dd{^{d}\ell_{m}} \dd{x} \frac{\ell_{m}^{\nu_{m+1}} \ell_{m}^{\tau_{m+1}} \left( \ell_{m} - \ell_{m-1} \right)^{\nu_{m}} \left( \ell_{m} - \ell_{m-1} \right)^{\tau_{m}}}{\left[ \left( 1 - x \right) \ell_{m}^2  + x \left( \ell_{m} - \ell_{m-1} \right)^2 \right]^2} \nonumber \\
    &= \int \dd{^{d}\ell_{m}} \dd{x} \frac{\ell_{m}^{\nu_{m+1}} \ell_{m}^{\tau_{m+1}} \left( \ell_{m} - \ell_{m-1} \right)^{\nu_{m}} \left( \ell_{m} - \ell_{m-1} \right)^{\tau_{m}}}{\left[ \ell_{m}^2  + x \left(  2\ell_{m} \cdot \ell_{m-1}- \ell_{m-1}^2 \right) \right]^2} \nonumber \\
    &= \int \dd{^{d}\ell_{m}} \dd{x} \frac{\ell_{m}^{\nu_{m+1}} \ell_{m}^{\tau_{m+1}} \left( \ell_{m} - \ell_{m-1} \right)^{\nu_{m}} \left( \ell_{m} - \ell_{m-1} \right)^{\tau_{m}}}{\left[ \left( \ell_{m} + x \ell_{m-1} \right)^2 - x^2 \ell_{m-1}^2  + x \ell_{m-1}^2 \right]^2} \, ,
\end{align}
or after shifting the integration variable as $\ell_{m} \to \ell_{m} - x \ell_{m-1}$,
\begin{align}\hspace{-20pt}
    L_{m, 2} &= \int \dd{^{d}\ell_{m}} \dd{x} \frac{\left( \ell_{m} - x\ell_{m-1} \right)^{\nu_{m+1}} \left( \ell_{m} - x\ell_{m-1} \right) ^{\tau_{m+1}} \left( \ell_{m} - \left( 1 - x \right) \ell_{m-1} \right)^{\nu_{m}} \left( \ell_{m} - \left( 1 - x \right) \ell_{m-1} \right)^{\tau_{m}}}{\left[ \ell_{m}^2 + x \left( 1 - x \right)  \ell_{m-1}^2  \right]^2} \, .
\end{align}
We may keep only even powers of $\ell_{m}$ in the integrand, as odd powers vanish by symmetry:
\begin{align}
    &L_{m, 2} = \int \dd{^{d}\ell_{m}} \dd{x} \Bigg( \frac{  \ell_{m}^{\nu_{m+1}} \ell_{m}^{\tau_{m+1}} \ell_{m}^{\nu_{m}} \ell_{m}^{\tau_{m}}  + x^2 \ell_{m-1}^{\nu_{m+1}} \ell_{m-1}^{\tau_{m+1}} \ell_{m}^{\nu_{m}} \ell_{m}^{\tau_{m}} + x \left( 1 - x \right)  \ell_{m}^{\nu_{m+1}} \ell_{m-1}^{\tau_{m+1}} \ell_{m}^{\nu_{m}} \ell_{m-1}^{\tau_{m}} }{\left[ \ell_{m}^2 + x \left( 1 - x \right)  \ell_{m-1}^2  \right]^2} \nonumber \\
    &+ \frac{x \left( 1 - x \right)  \ell_{m-1}^{\nu_{m+1}} \ell_{m}^{\tau_{m+1}} \ell_{m-1}^{\nu_{m}} \ell_{m}^{\tau_{m}} + \left( 1 - x \right)^2 \ell_{m}^{\nu_{m+1}} \ell_{m}^{\tau_{m+1}} \ell_{m-1}^{\nu_{m}} \ell_{m-1}^{\tau_{m}} + x^2 \left( 1 - x \right)^2 \ell_{m-1}^{\nu_{m+1}} \ell_{m-1}^{\tau_{m+1}} \ell_{m-1}^{\nu_{m}} \ell_{m-1}^{\tau_{m}} }{\left[ \ell^2_{m} + x\left( 1 - x \right)\ell_{m-1}^2 \right]^2} \Bigg) \, .
\end{align}
We now replace products of $\ell_m^\mu$ with powers of $\ell_m$ and symmetrized metric factors, following the generalized symmetrization rule (\ref{eq:symmetrization}), which yields
\begin{align}
    &\hspace{-10pt} L_{m, 2} = \int \dd{^{d}\ell_{m}} \dd{x} \Bigg( \frac{ \frac{3\ell_{m}^{4}}{d \left( d + 2 \right) }  g^{(\nu_{m+1} \tau_{m+1}} g^{\nu_{m} \tau_{m} )}  + x^2 \frac{\ell_{m}^2}{d} g^{\mu_{m} \nu_{m}}\ell_{m-1}^{\nu_{m+1}} \ell_{m-1}^{\tau_{m+1}}  + x \left( 1 - x \right)  \frac{\ell_{m}^2}{d} g^{\nu_{m+1} \nu_{m}} \ell_{m-1}^{\tau_{m+1}}  \ell_{m-1}^{\tau_{m}} }{\left[ \ell_{m}^2 + x \left( 1 - x \right)  \ell_{m-1}^2  \right]^2} \nonumber \\
    &\hspace{-10pt}  + \frac{x \left( 1 - x \right)  \frac{\ell_{m}^2}{d} g^{\tau_{m+1} \nu_{m}}\ell_{m-1}^{\nu_{m+1}} \ell_{m-1}^{\tau_{m}}  + \left( 1 - x \right)^2 \frac{\ell_{m}^2}{d}  g^{{\nu_{m+1}}{\tau_{m+1}}} \ell_{m-1}^{\nu_{m}} \ell_{m-1}^{\tau_{m}} + x^2 \left( 1 - x \right)^2 \ell_{m-1}^{\nu_{m+1}} \ell_{m-1}^{\tau_{m+1}} \ell_{m-1}^{\nu_{m}} \ell_{m-1}^{\tau_{m}} }{\left[ \ell^2_{m} + x\left( 1 - x \right)\ell_{m-1}^2 \right]^2} \Bigg) \, .
\end{align}
Splitting the numerator up, we can once again apply the standard formula (\ref{known_integral}) with 
\begin{align}\label{deltasq_app_defn}
    \Delta^2 = - x \left(  1 - x \right) \ell_{m-1}^2 \, ,
\end{align}
which gives
\begin{align}
    L_{m, 2} &= \frac{i}{\left( 4\pi \right)^{\frac{d}{2}} \Gamma \left( \frac{d}{2} \right) }\int_{0}^{1} \dd{x} \Bigg( \frac{3}{d \left( d + 2 \right) } g^{(\nu_{m+1} \tau_{m+1}} g^{\nu_{m} \tau_{m})}  \Gamma \left( 2 + \frac{d}{2} \right) \Gamma \left( -\frac{d}{2} \right) \Delta^{d}  \nonumber \\
    &\qquad \qquad \qquad + \frac{x^2}{d} g^{\mu_{m} \nu_{m}} \ell_{m-1}^{\nu_{m+1}} \ell_{m-1}^{\tau_{m+1}} \Gamma \left( 1 + \frac{d}{2} \right) \Gamma \left( 1 - \frac{d}{2} \right) \Delta^{d-2}  \nonumber\\
    &\qquad \qquad \qquad + \frac{x\left( 1 - x \right) }{d} g^{\nu_{m+1} \nu_{m}} \ell_{m-1}^{\tau_{m+1}} \ell_{m-1}^{\tau_{m}} \Gamma \left( 1 + \frac{d}{2} \right) \Gamma \left( 1 - \frac{d}{2} \right) \Delta^{d-2}  \nonumber\\
    &\qquad \qquad \qquad + \frac{x\left( 1 - x \right) }{d} g^{\tau_{m+1} \nu_{m}} \ell_{m-1}^{\nu_{m+1}} \ell_{m-1}^{\tau_{m}} \Gamma \left( 1 + \frac{d}{2} \right) \Gamma \left( 1 - \frac{d}{2} \right) \Delta^{d-2}  \nonumber\\
    &\qquad \qquad \qquad + \frac{\left( 1 - x \right)^2 }{d} g^{\nu_{m+1} \tau_{m+1}} \ell_{m-1}^{\nu_{m}} \ell_{m-1}^{\tau_{m}} \Gamma \left( 1 + \frac{d}{2} \right) \Gamma \left( 1 - \frac{d}{2} \right) \Delta^{d-2}  \nonumber\\
    &\qquad \qquad \qquad + x^2 \left( 1 - x \right)^2  \ell_{m-1}^{\nu_{m+1}} \ell_{m-1}^{\tau_{m+1}} \ell_{m-1}^{\nu_{m}} \ell_{m-1}^{\tau_{m}} \Gamma \left( \frac{d}{2} \right) \Gamma \left( 2 - \frac{d}{2} \right) \Delta^{d-4} \Bigg) \, .
\end{align}
Replacing $\Delta$ using its definition in equation (\ref{deltasq_app_defn}), we see that each term contains $d$ overall factors of loop momenta:
\begin{align}
    L_{m, 2} &= \frac{i}{\left( 4\pi \right)^{\frac{d}{2}} \Gamma \left( \frac{d}{2} \right) }\int_{0}^{1} \dd{x} \Bigg( \frac{3}{d \left( d + 2 \right) }  g^{(\nu_{m+1} \tau_{m+1}} g^{\nu_{m} \tau_{m})}  \Gamma \left( 2 + \frac{d}{2} \right) \Gamma \left( -\frac{d}{2} \right) \left( x\left( 1 - x \right) \right)^{\frac{d}{2}} \ell_{m-1}^{d}   \nonumber \\
    &\qquad \qquad \qquad + \frac{1}{d} g^{\mu_{m} \nu_{m}} \ell_{m-1}^{\nu_{m+1}} \ell_{m-1}^{\tau_{m+1}} \Gamma \left( 1 + \frac{d}{2} \right) \Gamma \left( 1 - \frac{d}{2} \right) x^{\frac{d}{2}+1}\left( 1 - x \right)^{\frac{d}{2} - 1} \ell_{m-1}^{d-2} \nonumber\\
    &\qquad \qquad \qquad + \frac{1}{d} g^{\nu_{m+1} \nu_{m}} \ell_{m-1}^{\tau_{m+1}} \ell_{m-1}^{\tau_{m}} \Gamma \left( 1 + \frac{d}{2} \right) \Gamma \left( 1 - \frac{d}{2} \right) \left( x\left( 1- x \right)  \right)^{\frac{d}{2}} \ell_{m-1}^{d-2} \nonumber\\
    &\qquad \qquad \qquad + \frac{1}{d} g^{\tau_{m+1} \nu_{m}}  \Gamma \left( 1 + \frac{d}{2} \right) \Gamma \left( 1 - \frac{d}{2} \right) \left( x\left( 1- x \right)  \right)^{\frac{d}{2}} \ell_{m-1}^{\nu_{m+1}} \ell_{m-1}^{\tau_{m}}\ell_{m-1}^{d-2} \nonumber\\
    &\qquad \qquad \qquad + \frac{1}{d} g^{\nu_{m+1} \tau_{m+1}}  \Gamma \left( 1 + \frac{d}{2} \right) \Gamma \left( 1 - \frac{d}{2} \right) x^{\frac{d}{2}-1}\left( 1- x \right)^{\frac{d}{2}+1}  \ell_{m-1}^{\nu_{m}} \ell_{m-1}^{\tau_{m}}\ell_{m-1}^{d-2} \nonumber\\
    &\qquad \qquad \qquad +  \Gamma \left( \frac{d}{2} \right) \Gamma \left( 2 - \frac{d}{2} \right) \left( x\left( 1- x \right)  \right)^{\frac{d}{2}} \ell_{m-1}^{\nu_{m+1}} \ell_{m-1}^{\tau_{m+1}} \ell_{m-1}^{\nu_{m}} \ell_{m-1}^{\tau_{m}}\ell_{m-1}^{d-4} \Bigg) \, .
\end{align}
We now use the identity $\Gamma \left( 1 + x \right) = x \Gamma \left( x \right)$ to factor and cancel the gamma functions, and finally evaluate the Feynman integrals, giving the result
\begin{align}
    L_{m, 2} &= \frac{i\Gamma \left( -\frac{d}{2} \right)}{\left( 4\pi \right)^{\frac{d}{2}} \Gamma \left( d + 2 \right) } \Bigg( \frac{ 3\left( 1 + \frac{d}{2} \right) \frac{d}{2}}{d \left( d + 2 \right) }   g^{(\nu_{m+1} \tau_{m+1}} g^{\nu_{m} \tau_{m})}   \Gamma \left( \frac{d}{2}+1 \right)^2 \ell_{m-1}^{d}   \nonumber \\
    &\qquad \qquad \qquad + \frac{\left( \frac{d}{2} \right) \left( -\frac{d}{2} \right)  }{d} g^{\mu_{m} \nu_{m}} \ell_{m-1}^{\nu_{m+1}} \ell_{m-1}^{\tau_{m+1}}  \Gamma \left( \frac{d}{2} + 2 \right) \Gamma \left( \frac{d}{2} \right) \ell_{m-1}^{d-2} \nonumber\\
    &\qquad \qquad \qquad + \frac{\frac{d}{2}\left( -\frac{d}{2} \right) }{d} g^{\nu_{m+1} \nu_{m}} \Gamma \left( \frac{d}{2} + 1 \right)^2 \ell_{m-1}^{\tau_{m+1}} \ell_{m-1}^{\tau_{m}} \ell_{m-1}^{d-2} \nonumber\\
    &\qquad \qquad \qquad + \frac{\frac{d}{2}\left( -\frac{d}{2} \right) }{d} g^{\tau_{m+1} \nu_{m}}   \Gamma \left( \frac{d}{2} + 1 \right)^2 \ell_{m-1}^{\nu_{m+1}} \ell_{m-1}^{\tau_{m}}\ell_{m-1}^{d-2} \nonumber\\
    &\qquad \qquad \qquad + \frac{\frac{d}{2}\left( -\frac{d}{2} \right) }{d} g^{\nu_{m+1} \tau_{m+1}}   \Gamma \left( \frac{d}{2} + 2 \right) \Gamma \left( \frac{d}{2} \right)  \ell_{m-1}^{\nu_{m}} \ell_{m-1}^{\tau_{m}}\ell_{m-1}^{d-2} \nonumber\\
    &\qquad \qquad \qquad +  \left( 1-\frac{d}{2} \right) \left( -\frac{d}{2} \right)   \Gamma \left( \frac{d}{2} + 1 \right)^2 \ell_{m-1}^{\nu_{m+1}} \ell_{m-1}^{\tau_{m+1}} \ell_{m-1}^{\nu_{m}} \ell_{m-1}^{\tau_{m}}\ell_{m-1}^{d-4} \Bigg) \, .
\end{align}
We see that all terms scale as $\ell_{m-1}^{d}$ and the only divergent gamma function is $\Gamma \left( -\frac{d}{2} \right) $. This establishes the result used in the body of the paper, in the text above equation (\ref{final_n_loop_2_vertex_dimreg}), which can then be applied iteratively to evaluate the remaining loop integrals.

\bibliographystyle{utphys}
\bibliography{ref}

\end{document}